\documentclass[prl,aps,reprint,twocolumn,showpacs,amsmath,superscriptaddress,longbibliography,notitlepage]{revtex4-2}
\usepackage{amssymb,mathtools,amsmath}
\usepackage{physics}

\usepackage{mathrsfs}
\usepackage{graphicx}
\usepackage{float}
\usepackage[caption=false]{subfig}
\usepackage{dcolumn}
\usepackage{appendix}
\usepackage{tabularx}
\usepackage{booktabs}
\usepackage{epstopdf}
\usepackage[normalem]{ulem}
\usepackage{verbatim}
\usepackage{bm}
\usepackage{natbib}
\usepackage{multirow}
\usepackage{tikz}
\usepackage{empheq}
\usetikzlibrary{fit}
\usepackage{footnote}
\usepackage{xcolor}
\usepackage{hyperref}
\hypersetup{
	colorlinks,
	citecolor=green,
	linkcolor=magenta,
	urlcolor=cyan}

\usepackage{orcidlink}

\DeclareMathAlphabet\mathbfcal{OMS}{cmsy}{b}{n}

\usepackage{soul}
\begin{document}
\title{Protected chaos in a topological lattice}

	\author{Haydar Sahin\,\orcidlink{0000-0002-6361-8161}}
	\email{sahinhaydar@u.nus.edu}
	\affiliation{Department of Electrical and Computer Engineering, National University of Singapore, Singapore 117583, Republic of Singapore}
	\affiliation{Institute of High Performance Computing (IHPC), Agency for Science, Technology and Research (A*STAR), Singapore 138632, Republic of Singapore}

	\author{Hakan Akg\"un\,\orcidlink{0009-0009-3474-1599}}
	\email{hakan.akgun@ug.bilkent.edu.tr}
	\affiliation{Department of Physics, Bilkent University, Ankara 06800, T\"urkiye}
	
	\author{Zhuo Bin Siu\,\orcidlink{0000-0002-7056-937X}}
	\email{elesiuz@nus.edu.sg}
	\affiliation{Department of Electrical and Computer Engineering, National University of Singapore, Singapore 117583, Republic of Singapore}
	
	\author{S.~M. Rafi-Ul-Islam\,\orcidlink{0000-0002-2275-8889}}
	\email{elesmr@nus.edu.sg}
	\affiliation{Department of Electrical and Computer Engineering, National University of Singapore, Singapore 117583, Republic of Singapore}
	
	\author{Jian\nobreak \space Feng Kong\,\orcidlink{0000-0001-5980-4140}}
	\email{kong\_jian\_feng@ihpc.a-star.edu.sg}
	\affiliation{Institute of High Performance Computing (IHPC), Agency for Science, Technology and Research (A*STAR), Singapore 138632, Republic of Singapore}
	\affiliation{Quantum Innovation Centre (Q.InC), Agency for Science, Technology and Research (A*STAR), 2 Fusionopolis Way, Innovis \#08-03, Singapore 138634, Republic of Singapore}
	
	\author{Mansoor B.~A. Jalil\,\orcidlink{0000-0002-9513-8680}}
	\email{elembaj@nus.edu.sg}
	\affiliation{Department of Electrical and Computer Engineering, National University of Singapore, Singapore 117583, Republic of Singapore}
	
	\author{Ching Hua Lee\,\orcidlink{0000-0003-0690-3238}}
	\email{phylch@nus.edu.sg}
	\affiliation{Department of Physics, National University of Singapore, Singapore 117542, Republic of Singapore}

\begin{abstract}
\noindent \textit{\textbf{Published version}---}\textit{Adv. Sci.} 2025, e03216. \url{https://doi.org/10.1002/advs.202503216}\\
	
\noindent \textit{\textbf{Keywords}---} protected chaos, non-linear topology, topological protection, machine learning\\
	
The erratic nature of chaotic behavior is thought to erode the stability of periodic behavior, including topological oscillations. However, we discover that in the presence of chaos, non-trivial topology not only endures but also provides robust protection to chaotic dynamics within a topological lattice hosting non-linear oscillators. Despite the difficulty in defining topological invariants in non-linear settings, non-trivial topological robustness still persists in the parametric state of chaotic boundary oscillations. We demonstrate this interplay between chaos and topology by incorporating chaotic Chua's circuits into a topological Su-Schrieffer-Heeger (SSH) circuit. By extrapolating from the linear limit to deep into the non-linear regime, we find that distinctive correlations in the bulk and edge scroll dynamics effectively capture the topological origin of the protected chaos. Our findings suggest that topologically protected chaos can be robustly achieved across a broad spectrum of periodically-driven systems, thereby offering new avenues for the design of resilient and adaptable non-linear networks.
\end{abstract}

\maketitle
\section{Introduction}

Chaotic dynamics, despite being notoriously unpredictable and sensitive to initial conditions~\cite{lorenz1963deterministic}, govern the behavior of many complex systems~\cite{may1976simple,shinbrot_chaos_1992} such as networks of coupled oscillators and dynamical non-linear lattices~\cite{watts_collective_1998,hofstrand_discrete_2023,chaunsali_stability_2021,chaunsali_dirac_2023,chen_circuit_2014,sone_topological_2022}. In such coupled dynamical networks, even small perturbations are amplified rapidly, leading to complex global behaviors~\cite{watts_collective_1998,aydiner_chaotic_2018,sugitani_synchronizing_2021}. While synchronization can appear at times, chaos, with its inherent unpredictability, disrupts this order and makes the system decidedly irregular in the long run. However, we discover that this well-established pattern of behavior may no longer hold when a network of coupled chaotic oscillators is arranged in a topological manner. Despite the lack of well-defined band topological invariants in non-linear systems, we find that topological robustness potently persists, even as chaotic behavior introduces randomness that would erode the bulk periodicity and threaten any semblance of topological bulk-boundary correspondence.

Surprisingly, not only can topological robustness be well-preserved in certain chaotic systems, it even \emph{protects} chaotic oscillations from strong perturbations, demonstrating a parametric robustness inherited from band topology. This protection stems from the stability of non-trivial edge oscillations, whose absence would lead to the rapid decay of bulk chaotic attractors, as observed in the case of trivial edge oscillations.

Recent progress in non-linear topological physics has expanded to include models combining non-linear and non-Hermitian mechanisms. Of particular interest are phenomena such as non-Hermitian global synchronization, where the inherent non-orthogonality of eigenmodes facilitates enhanced communication between modes, enabling global synchronization~\cite{zhang_nonhermitian_2024}; size-dependent non-Hermitian topological synchronization, characterized by distinct edge and bulk oscillations, with bulk oscillation frequencies explicitly depending on lattice size~\cite{di_observation_2025}; and a growth blockade arising from modulational instability observed in non-Hermitian periodic Hatano-Nelson lattices~\cite{longhi_modulational_2025}. Additionally, various types of non-linear skin modes have recently been reported~\cite{yuce_nonlinear_2025,many_manda_skin_2024,simonyan_non-hermitian_2024}. Interestingly, while non-linearity is conventionally considered disruptive to topological mechanisms, it is now intentionally utilized to configure these very topological states~\cite{bai_arbitrarily_2024}. Despite these exciting developments, existing studies predominantly consider non-chaotic non-linearities, employing regular oscillator models such as Stuart-Landau or Kerr-type oscillators.

In this work, we choose chaotic Chua circuits among various potential candidates to implement the Lorenz system~\cite{lorenz1963deterministic}, as they offer an experimentally accessible and easily tunable platform for investigating chaotic dynamics~\cite{kennedy_three_1993,lu_network_2022,nekorkin_homoclinic_1995}. Identical Chua circuits are coupled such that they form a one-dimensional topological Su-Schrieffer-Heeger (SSH) circuit array with onsite chaotic Chua oscillators~\cite{lee_topolectrical_2018,sahin_impedance_2023,zhang_anomalous_2023,wang_topologically_2019,kotwal_active_2021,hadad_self-induced_2018,marquie_nonlinear_1995,li_synergetic_2023,bai_observation_2024,sahin_topolectrical_2025,tang_strongly_2023}. In our setup, the degree of non-linearity can be adjusted by tuning the current-voltage characteristics of the Chua diodes, with high non-linearity typically expected to substantially distort linear eigenstates.

In non-linear systems, a key challenge in attributing robustness to band topology is the difficulty in defining bulk topological invariants, which are inherently properties of linear band structures. (Nonetheless, see Refs.~\cite{sone_nonlinearity-induced_2024,zhou_topological_2022,sone_transition_2025,zhou_topological_2024,isobe_bulk-edge_2024,wong_probing_2023} for proposed non-linear topological invariants in lattices with generally Kerr-type non-linearities.) To overcome this, we continuously deform the Chua circuit array from its linear SSH limit by tuning its non-linear Chua diodes. Notably, the topological edge oscillations are preserved remarkably well even deep into the non-linear regime. This allows for the identification of non-linear topological boundaries, which can be made further precise with the aid of machine learning techniques~\cite{gilpin_generative_2024}.

Importantly, we emphasize that although we shall demonstrate topologically protected chaos only with Chua circuits, this novel form of robustness is not specific to the Lorenz equations, and extends generally to chaotic attractors with well-defined bounded oscillations. Furthermore, the physical implementation is applicable to any medium which can host the requisite non-linear dynamics, not just electronic circuit metamaterials. As such, our discovery of topologically protected chaos also opens up new possibilities in fields as diverse as neural networks~\cite{kesgin_implementing_2024,terada_chaotic_2024}, chaotic quantum dynamics~\cite{wienand_emergence_2024} and mechanical metamaterials~\cite{kim_nonlinear_2022}.

\section{Results}
\subsection{Chaotic Chua oscillators as unit cells}

\begin{figure}%[ht!]
	\centering
	\includegraphics[width=\linewidth]{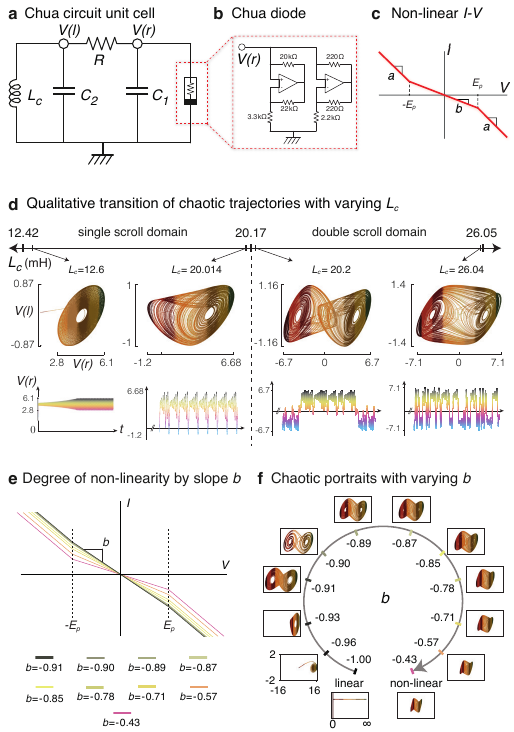}
	\caption{\textbf{Overview of a Chua circuit and its chaotic dynamics.} 
	\textbf{a} The Chua circuit is a two-node ($V(l)$ and $V(r)$) circuit comprising four passive components: two capacitors, one inductor, and one resistor, as well as the all-important active Chua diode.
	\textbf{b} The Chua diode provides the requisite non-linearity for chaotic dynamics, and can be implemented using operational amplifiers in inverting feedback loops. 
	\textbf{c} The Chua diode's non-linear $I$-$V$ characteristic consists of piece-wise linear segments, with slopes $b$ for the middle segment and $a$ for the other segments.	
	\textbf{d} The oscillations construct chaotic phase portraits in the parametric space of the voltage at the two nodes (obtained from Equation~\eqref{eqLorenz}). These portraits are classified based on the total number of equilibrium states, also known as attractors. We illustrate the qualitative transition of the phase portraits from single to double scroll and determine two domains based on the inductance of the inductor (marked on the $L_c$ scale). The lower and upper $L_c$ values of $12.42\,\text{mH}$ and $26.05\,\text{mH}$ indicate the critical thresholds beyond which the oscillations either quickly attenuate or diverge to infinity, respectively. The bottom row of plots shows the corresponding voltage oscillations at node $V(r)$, where the Chua diode is attached. Here, $b=-0.71$. 
	\textbf{e} Representative $I$-$V$ characteristics of the Chua diode with various slopes $b$ of the middle segment (of width $2E_p$). Throughout this study, we set $a=-1.14$ and $E_p=3\,\text{V}$, with variable non-linearity controlled by $b$. 
	\textbf{f} The non-linearity $b$ of the Chua diode profoundly controls the type and amplitude of its chaotic scroll, with limit point behavior crossing over to chaotic scrolls as $b$ is increased from its linear limit. Here, $L_c=24\,\text{mH}$, and other default parameter values are $C_1=10\,\text{nF}$, $C_2=100\,\text{nF}$, and $R=1.85\,\text{k}\Omega$.
	}
	\label{figChua}
\end{figure}

To show how topology can protect chaos in a topological lattice, we first design unit cells that host chaotic oscillations. A key ingredient of chaos is non-integrability, which can be realized in various non-linear systems of differential equations. A particularly stable candidate is the Chua oscillator~\cite{leon_chua_ten_years}, and below we review its physical implementation in an electrical circuit.

Chua circuits are recognized as the first and among the simplest electrical circuits capable of exhibiting chaos~\cite{muthuswamy_simplest_2010}. A Chua circuit consists of four passive linear components, namely an inductor ($L_c$), two capacitors ($C_1$ and $C_2$) and a regular resistor ($R$), as well as an active non-linear resistor known as the `Chua diode', as shown in Figure~\ref{figChua}a. While the circuit behaves as a trivial lossy LC oscillator if only passive linear elements are present, an active non-linear circuit component significantly alters its regular oscillatory behavior. Here, the Chua diode [Figure~\ref{figChua}b], has a piece-wise linear current-voltage ($I$-$V$) characteristic with non-constant negative slope, yielding an overall non-linear negative differential resistance [Figure~\ref{figChua}c]. In practice, the non-linear $I$-$V$ characteristic of the Chua diode can be achieved by employing appropriately connected operational amplifiers (op-amps) within a negative feedback loop~\cite{michael_peter_kennedy_robust_1992}, as detailed in Figure~\ref{figChua}b (In our LTspice simulations, we employ LT1351 op-amps).

Chua circuits form the building blocks of our topological lattice described later, with qualitatively different dynamical behaviors in different parameter regimes. Even though chaotic behavior is inherently stochastic, chaotic trajectories are confined within certain characteristic phase space regimes known as attractors~\cite{tsonis_chaos_1989,kuznetsov_hidden_2023}. Two relevant types of Chua oscillator attractors---the single and the double scroll---are presented in Figure~\ref{figChua}d. They arise from the underlying non-linear Lorenz differential equations~\cite{lorenz1963deterministic} 
\begin{equation}
	\begin{aligned}
		\dv{x}{\tau} &= \alpha(y-x-f_b(x)),\\
		\dv{y}{\tau} &= x-y+z,\\
		\dv{z}{\tau} &= -\beta y.
	\end{aligned}
	\label{eqLorenz}
\end{equation}
where $\tau$ is a dimensionless time variable and $x(\tau),y(\tau),z(\tau)$ represent the dynamical state in 3D space. Of crucial importance is the non-linear function $f_b(x)$, which we choose to be the dimensionless non-linear piece-wise function 
\begin{equation}
f_b(x) = bx +\frac{1}{2}(a-b)(|x+E_p|-|x-E_p|),
\label{fb}
\end{equation}
that replicates the $I$-$V$ characteristic shown in Figure~\ref{figChua}c, where $a$ and $E_p$ are real constants and $b$ is a parameter that controls the extent of non-linearity ($f_b(x)$ is linear when $b=a$). In general, a wide variety of deterministic and chaotic behavior~\cite{kennedy_three_1993,kennedy_three_1993_2} exists in various regimes of $\alpha$, $\beta$ and $b$.

Since this work is focused on the experimentally measurable consequences of topologically protected chaotic behavior, rather than the general classification of the Lorenz dynamics, we shall adapt the quantities in Equation~\eqref{eqLorenz} to particular electrical circuit component properties that best showcase single and double-scroll chaotic behavior (see Methods: Relating an array of Lorenz equations with our Chua-SSH circuit lattice). Explicitly, we identify
\begin{equation}
		 x = \frac{V(r)}{E_p}, \qquad  y = \frac{V(l)}{E_p}, \qquad  z = I_{L_c} \frac{R}{E_p},
	\label{xyzvariablesMainChua}
\end{equation}
where $x$ and $y$ are the dimensionless versions of the right and left node voltages ($V(r)$ and $V(l)$, see Figure~\ref{figChua}a) normalized by the Chua diode's linear voltage range $E_p$, while $z$ is the dimensionless current through the inductor $L_c$ rescaled by $R/E_p$. In addition, we define the dimensionless ratios 
\begin{equation}
	\alpha = \frac{C_2}{C_1}, \qquad 	\beta = \frac{R^2C_2}{L_c}.
	\label{EqAlphaBeta}
\end{equation}
Hence, in terms of the physical circuit quantities, Chua circuit dynamics are fully captured by the time dependence of the pair of voltages $(V(r),V(l))$, which will henceforth be plotted as phase portraits [Figure~\ref{figChua}d].

Importantly, whether the Chua circuit exhibits regular or chaotic oscillations is regulated by $\alpha$ and $\beta$, which can be tuned by varying the relevant circuit components. To reduce the redundancy in this tuning, we will henceforth set the following circuit parameters values as default: $C_1=10\,\text{nF}$, $C_2=100\,\text{nF}$, and $R=1.85\,\text{k}\Omega$, $a=-1.14$, and $E_p=3\,\text{V}$, while keeping $L_c$ and $b$ as parameters to be varied, unless otherwise stated. 

In Figure~\ref{figChua}d, using these default parameters, it is shown that varying the inductance $L_c$ can induce two qualitatively different classes of oscillations in the dynamical phase portraits $(V(r),V(l))$. The voltage oscillations between nodes $V(l)$ and $V(r)$ form a single-scroll chaotic phase portrait when $L_c$ is between $12.42\,\text{mH}$ and $20.17\,\text{mH}$. Beyond that, the Chua circuit exhibits double scroll dynamics in between $20.17\,\text{mH}$ and $26.05\,\text{mH}$. Around the critical value $L_c=20.17\,\text{mH}$ separating these two domains, the phase portrait rapidly deforms and transitions to the other configuration. The oscillations also become unstable at the lower and upper boundaries of these $L_c$ domains (see Section IV in the Supporting Information for detailed analyses of scroll transitions as a function of $L_c$, $C_1$, and $C_2$).

Physically, the single and double scroll scenarios are distinguished by the oscillations of $V(r)$ relative to that of $V(l)$. While the polarity of the voltage at node $V(r)$ remains mostly the same in the single scroll cases, i.e., positive, it strongly oscillates between positive and negative voltages in the double scroll cases, as shown in the time evolution of $V(r)$ in bottom row of Figure~\ref{figChua}d.

Besides the inductance $L_c$, the other key tuning parameter $b$ controls the non-linearity of the Chua diode through the slope of the middle segment of its $I$-$V$ characteristic. As shown in Figure~\ref{figChua}e, the system approaches its linear limit as the parameter $b$ is decreased. The corresponding chaotic phase portraits are depicted in Figure~\ref{figChua}f for the default parameter combination and $L_c=24\,\text{mH}$. While the exactly linear limit is given by $b\rightarrow a=-1.14$ using our default parameters, to facilitate the analysis of chaotic dynamics, the system is already decidedly linear by $b=-1$. As we approach the linear limit (more negative $b$), the portrait trajectories wander and fluctuate more, until they become non-chaotic and attenuated beyond $b = -0.91$. These behaviors remain qualitatively similar had we used other physical parameter combinations, as long as we are still in the double scroll domain.

\subsection{Chaotic topological SSH circuit}

\begin{figure*}[t!]
	\centering
	\includegraphics[width=\textwidth]{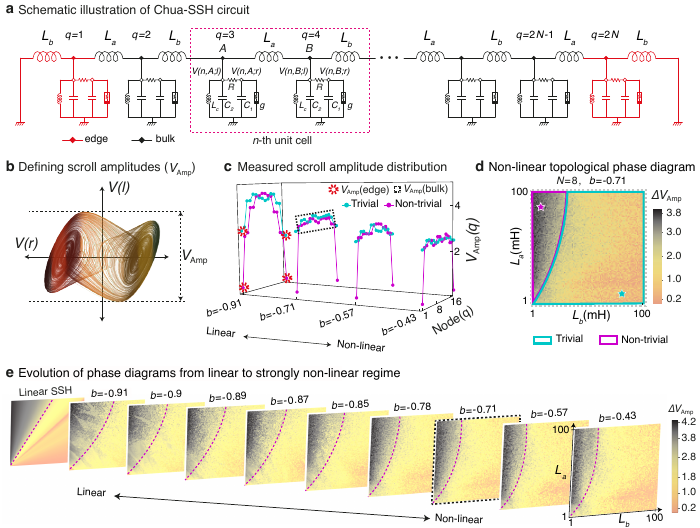}
	\caption{\textbf{The SSH lattice of Chua oscillators and its topological characterization by extrapolating from its linear limit.} 
	\textbf{a} The Chua-SSH circuit is constructed by coupling identical Chua circuits at their left nodes ($l$) with alternating intra-cell and inter-cell coupling inductances of $L_a$ and $L_b$. $A$ and $B$ labels the sublattice nodes of each unit cell. We also alternatively label the nodes from $q = 1$ to $2N$, where $N$ is the total number of unit cells.
	\textbf{b} Dynamical evolution in one Chua oscillator, as characterized by $V_\text{Amp}$, the peak-to-peak amplitude of $V(l)$ fluctuations. 
	\textbf{c} By determining the amplitudes of each scroll in the Chua-SSH lattice, the evolution of the scroll dynamics across the nodes of an $N=8$ circuit can be plotted for different levels of non-linearity. Notably, starting from a topologically non-trivial (magenta with $(L_a,L_b)=(80\,\text{mH}, 8\,\text{mH})$ corresponding to $(\delta_a,\delta_b)=(4.278, 42.78)$) SSH configuration in the linear limit, the edge scrolls are generally significantly suppressed compared to the bulk scrolls. But not so for the topologically-nontrivial (cyan, $(L_a,L_b)=(8\,\text{mH}, 80\,\text{mH})$ corresponding to $(\delta_a,\delta_b)=(42.78, 4.278)$) cases. Here, $\delta_a$
	\textbf{d} Difference in amplitude $V_\text{Amp}$ (Equation~\eqref{EqDeltaVamp}) between the bulk and edge scrolls, with distinct black and yellow regions extrapolated from the topologically non-trivial and trivial regimes in the linear limit.
	\textbf{e} Evolution of the $V_\text{Amp}$ diagram from the linear to deeply non-linear regime, showing how the black region continuously evolves from the exact topologically non-trivial SSH phase in the linear limit (see Methods: Determination of linear SSH phase diagram), manifestly showing how topological edge robustness persists in the non-linear regime. The inherit topological boundaries (dashed) can be rigorously determined via machine learning (see Methods: Topological phase characterizations via unsupervised learning). Circuit parameters used in all simulations are $N=8$, $L_c=24\,\text{mH}$, $C_1=10\,\text{nF}$, $C_2=100\,\text{nF}$, $R=1.85\,\text{k}\Omega$, and $a=-1.14$.}
	\label{figPhasediag}
\end{figure*}

We now show how to construct a topological lattice that admits chaotic oscillations. A paradigmatic 1D topological model is the SSH lattice model, which exhibits robust symmetry-protected boundary states in its topologically non-trivial phase~\cite{hohmann_observation_2023,rafi-ul-islam_type-ii_2022,du_chiral_2024}. The idea is to replace every site of the SSH lattice with a copy of the Chua circuit, whose dynamics are described by Equation~\eqref{eqLorenz}. As shown in Figure~\ref{figPhasediag}a, the two equivalent sites ($A$ and $B$) of each SSH unit cell are thus each grounded with a Chua circuit, and its $l$ node of a Chua circuit is connected that of adjacent sites by inductors of either $L_a$ (intra-cell) or $L_b$ (inter-cell), depending on the even/oddness of the site position. Hence, each unit cell consists of two identical Chua circuits, such that a $N$-unit cell Chua-SSH (cSSH) circuit contains a total of $2N$ Chua circuits. 

To elucidate what behavior can be potentially new in topological chaotic circuits, we need to review known behavior in non-topologically coupled Chua circuits. Firstly, for a pair of coupled Chua circuits (as in the $L_b\rightarrow \infty$ case where unit cells are decoupled), the two Chua circuits eventually become synchronized, a phenomenon known as \textit{chaos synchronization}~\cite{chaos_sync_leo,chai_wah_wu_synchronization_1995}, which has been extensively studied through various coupling components and combinations~\cite{yao_synchronization_2019,kapitaniak_experimental_1997,gamez-guzman_synchronization_2009,feki_adaptive_2003,kolumban_role_1998}. This synchronization occurs because the information exchange leads to the merging of the attractor points of the two chaotic systems~\cite{anishchenko_dynamics_1995,xie_hybrid_2002,moon_chaos_2021}. The time required for synchronization depends on the coupling strength, which essentially determines the amount of shared information.

However, complete synchronization is generally not observed in a 1D topologically trivial array of Chua oscillators~\cite{lu_network_2022,sugitani_synchronizing_2021}. Instead, they are known to exhibit traveling waves~\cite{nekorkin_travelling_1996,nekorkin_chaotic_1996}, homoclinic orbits and solitary waves~\cite{nekorkin_homoclinic_1995}. Yet, complete synchronization is not observed, but only chimera states~\cite{abrams_chimera_2004,omelchenko_robustness_2015}. 

The absence of complete synchronization in a lattice array is due to the chaotic interference between coherent and incoherent groups~\cite{mishra_chimeras_2023,panaggio_chimera_2015}, where some portions of the network become synchronized while the rest remain unsynchronized~\cite{muni_chimera_2020}. These incoherent groups consistently introduce irregularity to the lattice, amplifying chaotic dynamics across the system. While complete synchronization is already quite rare and difficult to attain---even in networks with global couplings where each oscillator interacts with every other---it becomes unachievable in a 1D lattice array with local (nearest-neighbor) couplings due to persistent perturbations in the trajectories~\cite{acebron_kuramoto_2005,strogatz_kuramoto_2000}. Such interference in general leads to the suppression of periodic behavior in non-linear chaotic networks.

However, as described later, it turns out that the introduction of topological modes can fundamentally alter the chaotic dynamics. While the aggregate feedback among a lattice of Chua oscillators may intuitively seem to be overall destructive, compromising the oscillation synchronization at the very least, a lattice with protected \emph{topological} states would turn out to have more \emph{robust} chaotic behavior. We would find that chaotic dynamics enjoy some protection by topological non-trivial-ness, despite the expectation that non-linearity and chaos would erode the topology. 
	
To proceed, we define our cSSH circuit [Figure~\ref{figPhasediag}a] through its equations of motion. Unlike the uncoupled Chua system described in Equation~\eqref{eqLorenz}, each variable in Equation~\eqref{xyzvariablesMainChua} is now labeled by its unit cell number $n$ and its sublattice node label $\mu=A,B$, and its notation is generalized to
\begin{equation}
	\begin{aligned}
		& x(n,\mu) = \frac{V(n,\mu;r)}{E_p}, \qquad  y(n,\mu) = \frac{V(n,\mu;l)}{E_p},\\
		& \qquad \qquad  z(n,\mu) = I_{L_c}(n,\mu) \frac{R(n,\mu)}{E_p},
	\end{aligned}
	\label{xyzvariablesMainCSSH}
\end{equation}
and for the Chua diode non-linearity,
\begin{equation}
	\begin{aligned}
	f_b(x(n,\mu)) &=  b x(n,\mu)\\ & +\frac{1}{2}(a-b)(|x(n,\mu)+E_p|-|x(n,\mu)-E_p|).
	\end{aligned}
	\label{fbSSH}
\end{equation}
Here, we express the node voltages for the cSSH circuit as $V(n,\mu;l)$ and $V(n,\mu;r)$, where $l$ and $r$ label the left and right nodes at position $(n,\mu)$, respectively (previously, for a single Chua circuit, they were just $V(l)$ and $V(r)$). The dynamical evolution on the $A$ sublattice of each Chua circuit is
\begin{equation}
	\begin{aligned}
		& \dv{x(n,A)}{\tau}= \alpha(y(n,A) - x(n,A) -  f_b(x(n,A))),\\
		& \dv{y(n,A)}{\tau} =  x(n,A) - y(n,A) + z(n,A) + u(n) - v(n-1) ,\\
		& \dv{z(n,A)}{\tau} = - \beta y(n,A),
	\end{aligned}
		\label{fullequationsA}
\end{equation}		
while for sublattice $B$ they are 
\begin{equation}
	\begin{aligned}
		& \dv{x(n,B)}{\tau} = \alpha(y(n,B) - x(n,B) -  f_b(x(n,B))),\\
		& \dv{y(n,B)}{\tau} = x(n,B) - y(n,B) + z(n,B) -u(n) + v(n),\\
		& \dv{z(n,B)}{\tau} = - \beta y(n,B).
	\end{aligned}
		\label{fullequationsB}
\end{equation}
Notice that Equations~\eqref{fullequationsA} and \eqref{fullequationsB} involve two additional dimensionless state variables, i.e., $u$ and $v$, representing currents through intra-cell and inter-cell couplings $L_a$ and $L_b$ respectively [Figure~\ref{figPhasediag}a]:
\begin{equation}
	u(n) = I_{L_a}(n) \frac{R(n,\mu)}{E_p}, \qquad  v(n) = I_{L_b}(n) \frac{R(n,\mu)}{E_p}.
	\label{uvvariablesMain}
\end{equation}
The temporal evolution of these currents is described by the dimensionless differential voltage as
\begin{equation}
	\begin{aligned}		
		& \dv{u(n)}{\tau} = \delta_a (y(n,B) - y(n,A)),\\
		& \dv{v(n)}{\tau} = \delta_b (y(n+1,A) - y(n,B)),
	\end{aligned}
	\label{fullequationsC}
\end{equation}
where the dimensionless $\delta_a,\delta_b$ depend on inductances $L_a,L_b$ via
\begin{equation}
	\delta_{a} = \frac{C_2 R^2}{L_{a}}, \qquad \delta_{b} = \frac{C_2 R^2}{L_{b}}.
	\label{EqDelta}
\end{equation}
Importantly, $\delta_a$ and $\delta_b$ allows for the tuning of the topological phase (alongside the tunable non-linearity $b$), with other dimensionless coefficients fixed at $\alpha = 10$ and $\beta = 14.26$, as calculated from Equation~\eqref{EqAlphaBeta} with fixed $L_c=24\,\text{mH}$ and other default circuit parameters. For the coupling inductance range that we shall use in this work, between $1$ mH and $100$ mH, $\delta_a$ and $\delta_b$ can range from $342.25$ to $3.4225$.

By combining Equations~\eqref{fullequationsA} and \eqref{fullequationsB} with Equation~\eqref{fullequationsC}, we obtain the equations governing the dynamical evolution of our cSSH circuit. Open boundary conditions are implemented by imposing $\dv{v(0)}{\tau} = \delta_b y(1,A)$ and setting $y(n+1,A) = 0$. For dynamical simulations, the initial conditions are set as $x(n,A)\bigr\rvert_{\tau = 0}=0.01$, while all other variables are initialized to zero. For the explicit derivation of these equations of motion and the conversions for each dimensionless variable introduced earlier, please refer to Methods: Relating an array of Lorenz equations with our Chua-SSH circuit lattice. Note that for ease of presentation, each node in our simulation results would be labeled with a single numeric index $q$ for simplicity. Here $q$ ranges from 1 to $2N$, starting from the left edge, with the corresponding voltages denoted as $V(q;l)$ and $V(q;r)$.

\subsection{Topological robustness in the non-linear regime}

Even though topological invariants are rigorously defined only for linear systems~\cite{haldane1988model,kawabata_symmetry_2019,gong_topological_2018,yao_edge_2018,jiang2023dimensional, soluyanov_computing_2011}, topological signatures in our system interestingly remain robust well into the non-linear regime. Specifically, we observe that drastic differences between bulk and edge chaotic oscillations persist as our Chua diode is interpolated from the linear to strongly non-linear regime.

In the linear regime where the slope of the middle segment of the Chua diode's $I$-$V$ characteristic is $b\approx a$, Equations~\eqref{fullequationsA},~\eqref{fullequationsB}, and~\eqref{fullequationsC} reduce to the linear relations
\begin{equation}
		\begin{aligned}
&	 \lambda_J V(n,A) = \frac{1}{L_a} V(n,B) + \frac{1}{L_b} V(n-1,B),\\
&	 \lambda_J V(n,B) = \frac{1}{L_a} V(n,A) + \frac{1}{L_b} V(n+1,A),
		\end{aligned}
\end{equation}
giving rise to a circuit Laplacian $J$ of SSH-type~\cite{sahin_impedance_2023,yang_circuit_2024}. In momentum space ($k$)-space, it takes the form 
\begin{equation}
	\begin{aligned}
		J(k) = \begin{pmatrix}
			 \lambda_J & -\frac{1}{j \omega L_a} - \frac{1}{j \omega L_b} e^{-jk} \\ -\frac{1}{j \omega L_a} - \frac{1}{j \omega L_b} e^{jk} & \lambda_J
		\end{pmatrix},
		\label{LapSSHmomentum}
	\end{aligned}
\end{equation}
where $\lambda_J = j \omega C_2 + \frac{1}{R} + \frac{1}{j \omega L_a} + \frac{1}{j \omega L_b} + \frac{1}{j \omega L_c} $ represents the onsite components, $j$ is the imaginary unit and $\omega$ is the angular frequency. It is well-known that the SSH model is topologically non-trivial when $L_a > L_b$, and trivial otherwise~\cite{lee_topolectrical_2018,chen_engineering_2025,hu_observation_2024,rafi-ul-islam_interfacial_2022,zou_experimental_2024}; below, we shall see how this phase boundary is modified in the non-linear context. 

In our Chua circuit lattice, the robustness of the topological edge states is manifested in the dramatically different amplitudes of the voltage oscillations in boundary and bulk sites. As shown in the phase portrait of an isolated Chua oscillator in Figure~\ref{figPhasediag}b, oscillations of voltage $V(r)$ at node $r$ of a chosen Chua oscillator site serves to discriminate between scroll types, while $V(l)$ at node $l$ remains relatively constant as non-linearity is varied [Figure~\ref{figChua}f]. As such, for the purpose of detecting the persistence of topological signatures as non-linearity is introduced, we shall focus on $V_\text{Amp}(q)$, the peak-to-peak amplitude of $V(q;l)$. Thus, $V_\text{Amp}(q)$ omits the Chua node label, as we consistently refer to $V_\text{Amp}$ by the voltage amplitude of the $l$ node at the $q$-th site.

Figure~\ref{figPhasediag}c shows the qualitatively different profiles of $V_\text{Amp}$ between topological and non-topological scenarios, across all the circuit nodes for $N=8$ unit cells ($q=16$ nodes). In the linear limit, the cases $(L_a, L_b) = (8\,\text{mH}, 80\,\text{mH})$ (cyan) and $(L_a, L_b) = (80\,\text{mH}, 8\,\text{mH})$ (magenta) correspond to the topologically trivial and non-trivial scenarios, respectively. Note that, the $V(l)$ amplitude $V_\text{Amp}$ for the trivial case (cyan) is relatively constant across all the sites, apart from slight suppressions at the edge unit cells $n=1$ and $8$ due to edge effects. However, for the topologically non-trivial case (magenta), the edge $V_\text{Amp}$ is almost 4 volts lower than the $V_\text{Amp}$ within the bulk, due to the presence of topological edge states. Yet, in both topological and non-topological scenarios, the oscillation amplitudes are very similar across any $V_{\text{Amp}}(\text{bulk})$ (i.e., any bulk node). Hence we take this dramatically suppressed voltage oscillation amplitude at the edge nodes as the signature of topological edge states:
\begin{equation}
	\Delta V_\text{Amp} = V_{\text{Amp}}(\text{bulk}) - V_{\text{Amp}}(\text{edge}).
	\label{EqDeltaVamp}
\end{equation}

Interestingly, the topological edge suppression of $V_\text{Amp}$ persists even as the Chua diode non-linearity $b$ is tuned into the strongly non-linear regime, up to $b=-0.43$ [Figure~\ref{figChua}e]. As apparent in Figure~\ref{figPhasediag}c, the edge $V_\text{Amp}$ of the topological scenarios (magenta) remain strongly suppressed even as the circuit becomes significantly non-linear, such that $V_\text{Amp}$ pertains to the amplitude of the chaotic scrolls $V(l)$. Even though we have deliberately chosen topological and non-topological parameters such that either the inter- or intra-cell inductance is much larger, i.e. with $L_a/L_b=10$, i.e., close to the dimerized limit, our observed topological robust edge behavior contrasts strongly with the synchronized oscillations~\cite{anishchenko_dynamics_1995,lahav_synchronization_2018,lahav_topological_2022} observed in a fully dimerized cSSH.

Importantly, this persistence of the large edge amplitude suppression $\Delta V_\text{Amp}$ for non-linear chaotic oscillations allows one to define topological phase boundaries even under highly non-linear conditions. Shown in Figure~\ref{figPhasediag}d is a phase diagram capturing the difference between the amplitudes of bulk and edge portraits $\Delta V_\text{Amp}$ in the parameter space of $L_a$ and $L_b$, for an illustrative non-linearity parameter $b=-0.71$. 
Two distinct regions, namely black and yellow are clearly visible, separated by a rather abrupt gradient in $\Delta V_\text{Amp}$. The presence of this rather sharp demarcation parameter regions of high and low $\Delta V_\text{Amp}$ provides further support that topological robustness indeed extends into the non-linear regime, allowing for the continued identification of topological (black) and non-topological (yellow) regions.

\begin{figure*}[ht!]
	\centering
	\includegraphics[width=\textwidth]{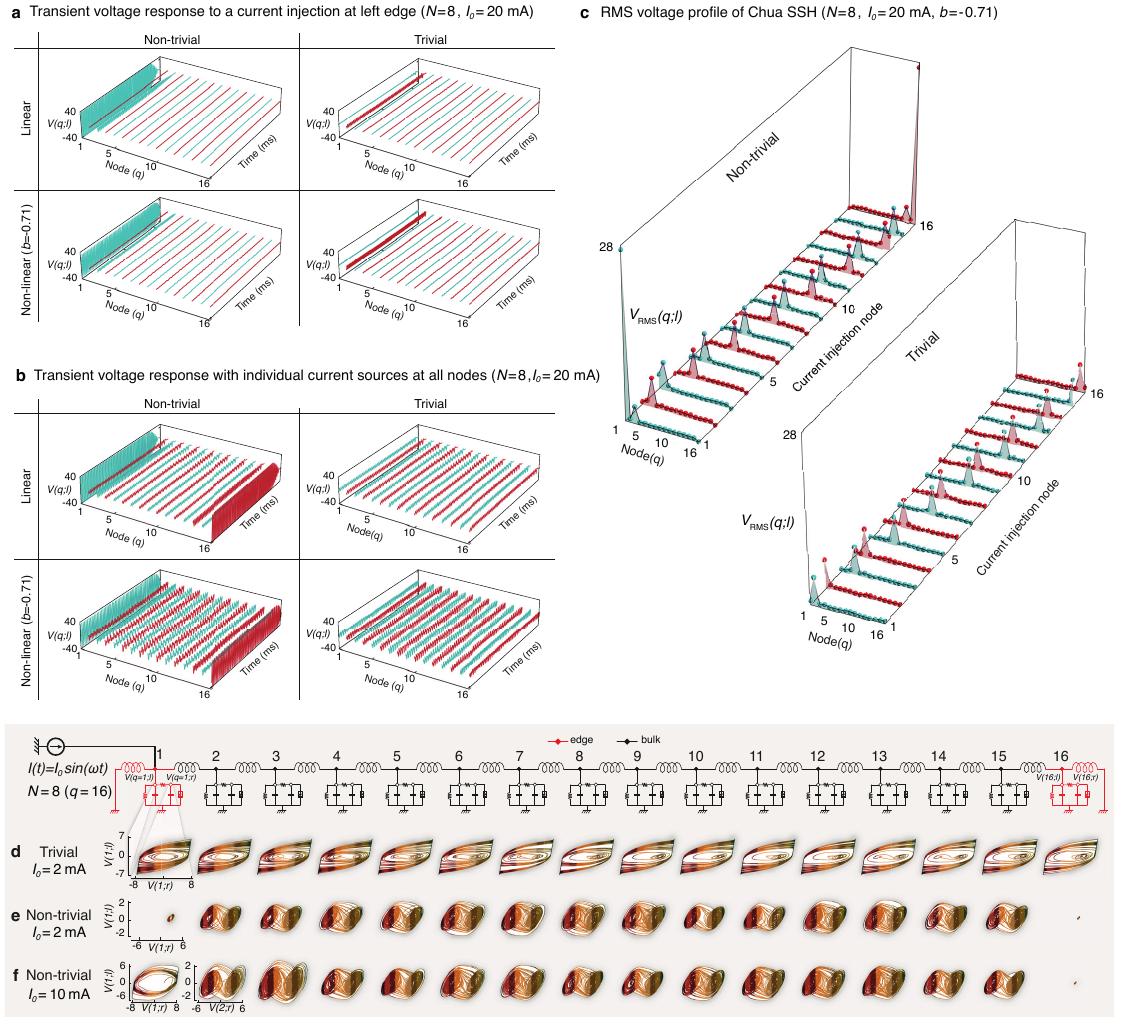}
	\caption{\textbf{Topological protection of chaotic dynamics under perturbations, analyzed by contrasting topological and non-linear phases via LTspice simulations.} 
	\textbf{a} Transient voltage response is tested by applying an AC current with a magnitude of $I_0=20\,\text{mA}$ to the left edge (node 1) of an $N=8$ Chua-SSH circuit. The linear and non-linear regimes shows similar profile behaviors, with strong edge oscillations occurring at the left edge in the non-trivial phase of both regimes. 
	\textbf{b} Pronounced transient oscillations are observed only at the edges when a current source is applied to each node simultaneously and individually across the circuit. \textbf{c} RMS voltages are plotted for each node in the topologically non-trivial and trivial phases, showing clear edge localization in the non-trivial phase compared to the trivial phase. 
	\textbf{d} In the topologically trivial case, a relatively small current of $2\,\text{mA}$ injected at the left edge (red) transforms chaotic oscillations into limit cycle oscillations shortly after injection, even though each node initially exhibits double-scroll chaotic behavior without an stimuli. 
	\textbf{e} By contrast, with topological parameters, chaotic oscillations are robust under the same current injection, i.e., $2\,\text{mA}$, demonstrating the topological protection of chaos in the lattice. 
	\textbf{f} Further demonstration of the robustness of topological protection, with bulk chaotic portraits remaining intact even under a large current injection of $10\,\text{mA}$. In all cases, for the non-trivial non-linear phase $(L_a,L_b,f_r,b)=(80\,\text{mH}, 8\,\text{mH}, 6.42\,\text{kHz},-0.71)$ were used. The values of $L_a$ and $L_b$ are switched for the trivial phase.}
	\label{figExternalProtect}
\end{figure*}

To explicitly identify the two distinct non-linear parameter space regions with the rigorously defined topological and non-topological regions in the linearized effective cSSH system, we present the evolution of $\Delta V_\text{Amp}$ phase diagrams under various non-linearities in Figure~\ref{figPhasediag}e. By varying the non-linearity parameter $b$ [Figure~\ref{figChua}e], we obtain an interpolation from the linear limit ($b=-0.91$, beyond which the chaotic oscillations break down, as elaborated in Methods: Determination of linear SSH phase diagram) to the highly non-linear regime (up to $b=-0.43$). Indeed, across various degrees of non-linearity $b$, the phase diagram exhibits a demarcation into two regions (black and yellow) that varies gradually with $b$. Note that the jump in $\Delta V_\text{Amp}$ is intrinsically not perfectly sharp even in the linear limit with sharply defined topological boundary $L_a=L_b$, since bulk states also contribute to the voltage dynamics in all cases.

Henceforth, we shall call the black region the `topological' regime. The omnipresence of the distinctive black regions in the phase diagrams of Figure~\ref{figPhasediag}e, even in the strongly non-linear limit ($b=-0.43$), showcases that qualitatively topological edge behavior remains robust for chaotic dynamics. The determination of the non-linear phase boundaries (purple dashed), which differs from $L_a=L_b$ of the linear limit, can be made rigorous with a machine learning model, as elaborated in Methods: Non-linear to linear phase boundary extrapolation. (These phase boundaries remain unchanged for sufficiently long lattices; see Methods: The Effect of Lattice Size on Phase Boundaries, and under structural disorders; see Section III in the Supporting Information.)

\subsection{Topology protects chaos}

Contrary to the intuitive expectation that chaotic dynamics would erode the topological properties of a system, we find that, surprisingly, it is however the topology that protects the chaotic dynamics. To investigate the topological robustness of chaotic edge oscillations in an experimentally realistic setting, we performed LTspice circuit simulations of the voltage dynamics driven by an external alternating current (AC) denoted by $I(t)=I_0 \sin(2\pi f_r t)$, where $I_0$ represents the current magnitude and $f_r$ is the AC frequency. 

Recalling our default parameter values used in Equations~\eqref{fullequationsA}, \eqref{fullequationsB}, \eqref{fullequationsC} and the coupling inductance values indicated by the stars in Figure~\ref{figPhasediag}d (i.e., $\{L_a,L_b\}=\{8\,\text{mH}, 80\,\text{mH}\}$), we find the resonant frequency as $|\text{Re}(f_r)|=6.42\,\text{kHz}$ (see Methods: Determination of resonant frequency). For the non-trivial phase, we set $L_a = 80\,\text{mH}$ and $L_b = 8\,\text{mH}$, with these values switched for the trivial phase. Given that the topological response of topolectrical circuits manifests itself when the signal frequency matches with the resonant frequency~\cite{lee_topolectrical_2018,sahin_impedance_2023,rafi-ul-islam_system_2022,shang_observation_2024,rafi-ul-islam_unconventional_2022,rafi-ul-islam_chiral_2024,zou2024experimental}, we maintain the signal frequency at $6.42\,\text{kHz}$ in all simulations.

The response of our lattice under an external current stimulus provides insights into the robustness of topological dynamics in the presence of circuit non-linearity, as presented in Figure~\ref{figExternalProtect}. In Figures~\ref{figExternalProtect}a and b, we contrast the dynamics between linear and non-linear scenarios, as well as between topologically trivial and non-trivial parameter regions. To isolate the protection from topological edge modes, we also compared the effects of injecting the current in three different ways: In Figure~\ref{figExternalProtect}a, an AC current is injected at the left edge node (node 1) 
where topological modes reside. The voltage response $V(q;l)$ clearly does not decay with time only at the left node, only when the system has topologically non-trivial parameters. Most saliently, the oscillations at the left node remain just as robust deep in the non-linear regime ($b=-0.71$), beyond the scope of usual topological invariants. The prominent edge oscillations, even in the non-linear regime, indicate the persistence of topological zero modes, which can only be excited by the resonant frequency, set at $6.42\,\text{kHz}$ in our system.

Alternatively, in Figure~\ref{figExternalProtect}b, rather than using a single current source, we attach dedicated multiple current sources $I_0$ to each node simultaneously and display the voltage oscillations across all nodes to see the global response. As before, the oscillations are consistently much smaller on the bulk nodes ($q=2$ to 15) at the gapped topological frequency. However, now the oscillations survive at both the left and right edges, further confirming the origin of the robustness from the topological edge modes, given that the current injection occurs democratically at all nodes. Importantly, the voltage dynamics remain qualitatively unchanged as circuit non-linearity is introduced, like in Figure~\ref{figExternalProtect}b, a testimony of continued topologically protected robustness.

\begin{figure*}[ht!]
	\centering
	\includegraphics[width=\textwidth]{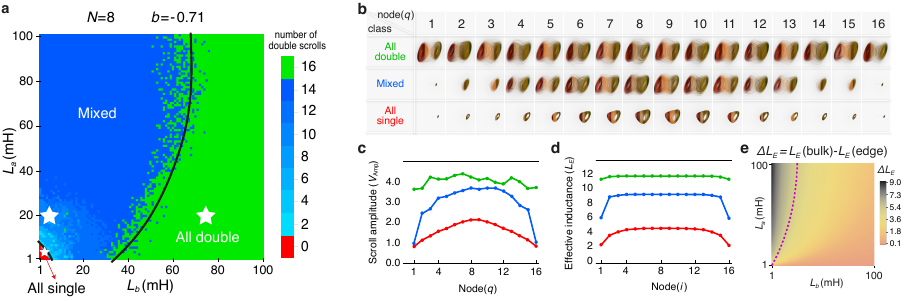}
	\caption{\textbf{Classification of topological chaos based on edge scrolls attenuation, and its comparison with effective inductance predictions.} 
	\textbf{a} Classification of chaotic lattice dynamics in the parameter space of $(L_a,L_b)$ based on the number of double and single chaotic scrolls out of $2N=16$ Chua oscillators. These scenarios can be categorized into three main classes: i) All-single (red), where the entire portraits are single scroll, ii) Mixed (shades of blue), where single and double scroll portraits coexist in the circuit, and iii) All-double (green). The boundaries of these three regimes are defined by a machine learning model involving a wider parameter space (see Methods: Finding explicit parametric equations of scroll space through machine learning). 
	\textbf{b} Scroll profiles of each Chua oscillator, for the three representative cases indicated by the white stars in `a'. Saliently, as the edge oscillations are attenuated, some double scrolls are also converted into single scrolls. 
	\textbf{c} The scroll amplitude profiles for these three examples (green, blue, red) are plotted, with the edge scrolls significantly suppressed in blue but not in green, while the bulk scrolls are suppressed toward the red class. 
	\textbf{d} The effective inductance $L_E(i)$ (Equation~\eqref{eqtotalindMain}) profiles corresponding to the same three examples, all which exhibit a qualitative correlation with the scroll amplitudes (also see Section V in the Supporting Information).
	\textbf{e} This close qualitative agreement is evident across the entire parameter space, which is mapped by measuring the difference between bulk and edge effective inductances, i.e., $\Delta L_E = L_E(\text{bulk})-L_E(\text{edge})$, similar to $\Delta V_\text{Amp}$ diagram. The purple dashed line indicates the approximate phase boundary. Here $b=-0.71$. }
	\label{figScrollMap}
\end{figure*}

To further demonstrate that non-linear oscillations exhibit the same type of edge dependence as topological modes, in Figure~\ref{figExternalProtect}c, we investigate the circuit response from injecting current one node at a time, for each node, and measuring the resultant root-mean-square voltage ($V_{\text{RMS}}$) at all nodes. Evidently, the edge nodes in the topological case [Figure~\ref{figExternalProtect}c] exhibit far larger $V_{\text{RMS}}$ than the bulk nodes, as well as the $V_{\text{RMS}}$ of all nodes from the topologically trivial case. While the significantly enhanced edge dynamics in the non-trivial phase and the absence of boundary modes in the trivial phase are well-known and expected phenomena in the linear regime, the surprising aspect is the persistence of these topological features even in the strongly non-linear regime.

For a concrete demonstration of how topology protects chaos, we now turn to Figure~\ref{figExternalProtect}d, which shows the phase portraits $V(q;l)$ vs. $V(q;r)$ for every node $q$ in the circuit. It is known that chaotic oscillations (which arise in isolated, self-driven setups) can in general be significantly destroyed by external perturbing drives. This is exemplified in the topologically trivial case in Figure~\ref{figExternalProtect}d, which shows that the voltages on all nodes exhibit limit cycles instead of chaotic dynamics when an external current of amplitude $I_0 = 2\,\text{mA}$ is introduced at the left edge (node 1). However, the same current $I_0 = 2\,\text{mA}$ fails to penetrate the circuit bulk in the topologically non-trivial case, and the double scrolls saliently survive at all bulk nodes (refer to Figure~\ref{figExternalProtect}e). Notably, even with a much larger current magnitude, such as $I_0 = 10\,\text{mA}$, as shown in Figure~\ref{figExternalProtect}f, the strong topological edge localization continues to protect the chaotic double scroll oscillations from the perturbative inputs.

Thus, topological protection remarkably enables chaos to persist even under substantial perturbations that would otherwise push the chaotic trajectories to saturate at the op-amps' output limit (can be even tens of volts) in an isolated Chua circuit. If that were to happen, external perturbations would be amplified by the gain feedback, ultimately leading to limit cycle oscillations. However, this is suppressed by the very robust edge states inherited from linear band structure topology, thereby constituting a topological protective mechanism for the Chua oscillations. (See Sections II and III in the Supporting Information for experimentally realizable protected chaos under consideration of disorder, uncertainties, and inevitable loss.)

\subsection{Phase boundary determination from edge scroll attenuation and effective impedance}

In the following, we describe other observables that provide alternative signs of topological robustness in chaotic lattices. 

\subsubsection{Edge scroll attenuation and total number of double scrolls}

Saliently, topological signatures in Chua oscillators lattices also manifest in the scroll pattern of the oscillator dynamics. Since each isolated Chua oscillator exhibits double scroll dynamics by default, the number of double scrolls across all $2N$ oscillators would signify the extent of oscillator independence i.e. lack of correlation. Plotted in Figure~\ref{figScrollMap}a is the number of double scrolls in a chain with $2N=16$ Chua oscillators, in the parameter space of inductance values $L_a$ and $L_b$. We identify three main classes of behavior: the all-double scroll class (green region), the all-single scroll class (red region), as well as the transitional `mixed' class (various shades of blue) with both double and single scrolls coexisting. Interestingly, the double/mixed class boundary in this parameter space plot almost coincides with the topological boundary shown earlier in Figure~\ref{figPhasediag}, although the very small all single (red) parameter region represents distinct new behavior.

To understand the suggestive topological origin of the scroll patterns, we provide a visual illustration of the scrolls in Figure~\ref{figScrollMap}b. Plotted are the chaotic portraits (defined in Figure~\ref{figPhasediag}b) for all the $2N=16$ oscillators, for representative parameters in each class (green, blue, red), as indicated by the white stars in Figure~\ref{figScrollMap}a. Notably, as we reduce $L_b$ and $L_a$, the double scrolls near the ends of the lattice generally tend to attenuate significantly, such that they become single scrolls. The extent of single scroll penetration increases as $L_a,L_b\rightarrow 0$ towards the red region, which is also deep inside the topological region found earlier. (For additional examples, see Section VI in the Supporting Information.)

In Figure~\ref{figScrollMap}c, we quantitatively establish the connection between the classification by number of double scrolls [Figure~\ref{figScrollMap}a] with $\Delta V_\text{Amp}$ (defined in Equation~\eqref{EqDeltaVamp}) which determines the inherited topological regions. Plotted in Figure~\ref{figScrollMap}c, is the $V_\text{Amp}$ spatial profile for the abovementioned representative examples in each class (green, blue, red).
While edge suppression is less pronounced in the all-double and all-single examples compared to the bulk, the edge scroll amplitude is significantly lower than that of the bulk scrolls in the mixed class. Indeed, the blue and red cases with also exhibits significant edge effects, unlike in the purportedly topologically trivial green region where large $L_b$ gives rise to nearly decoupled unit cells.

\subsubsection{Node effective inductance}

Next, we furthermore reveal that the edge sensitivity of the scroll amplitude profiles in Figure~\ref{figScrollMap}c can be qualitatively predicted from \emph{effective inductance}
($L_E$) profile, which is a lumped-element concept for \emph{linear} systems. Even though we have a highly non-linear dynamics, the inspiration is that the network connectivity of the circuit, in particular the elements that give rise to SSH band topology, all stem from linear inductors. 
 
The effective inductance $L_E(i)$ at a bulk node $i$ is given recursively by
\begin{equation}
	L_E(i)=\left(\frac{1}{L_c}+\frac{1}{L_E^\text{left}}+\frac{1}{L_E^\text{right}}\right)^{-1},
	\label{eqtotalindMain}
\end{equation}
where $L_E^\text{left}$ and $L_E^\text{right}$ represent the effective lumped inductances on the left and right sides of the $i$th node, respectively. The effective inductance method assumes that chaotic scrolls across different coupling inductances evolve due to the $L_E$ at each node, which results from the cumulative contributions of $L_c$, $L_a$, and $L_b$. (For the detailed derivation and explanation, see Methods: Derivation of effective inductance formulae and Methods: Qualitative proportionality between the effective inductance and scroll amplitude.)

Figure~\ref{figScrollMap}d shows the $L_E(i)$ profiles for each representative example (green, blue, red) from the three regimes in Figure~\ref{figScrollMap}a. The effective inductance profile of each example exhibits a very similar trend across the nodes in the chain as the voltage amplitude [Figure~\ref{figScrollMap}c]. Based on this qualitative correspondence, we proceed to evaluate whether this relationship holds across the entire parameter space. Analogous to the $\Delta V_\text{Amp}$ diagram that defines the inherited band topology, in Figure~\ref{figScrollMap}e, we plot the effective inductance difference between a bulk node and an edge node, defined as $\Delta L_E = L_E(\text{bulk}) - L_E(\text{edge})$. The two distinct regions, which closely resemble the $\Delta V_\text{Amp}$ diagrams in Figure~\ref{figPhasediag}e, provide strong evidence of a qualitative link between $\Delta V_\text{Amp}$ and $\Delta L_E$, and also an alternative justification of our non-linear topological characterization. 

Remarkably, the effective inductance approaches can be generalized to multi-band models. Beyond the dimer SSH, we present an example of a trimer Chua-SSH circuit, in which the close relationship between effective inductance and scroll amplitudes prominently holds (see Methods: Derivation of effective inductance formulae: Trimer Chua-SSH circuit).

\section{Discussion}
The realization and characterization of chaotic topological lattices require careful and new approaches, as existing topological classifications are based on linear systems. Here, we have demonstrated topologically protected chaos in a one-dimensional topological circuit array of inductively coupled identical chaotic Chua circuits. Counterintuitively, incorporating chaotic dynamics into the periodic structure of the topological SSH circuit results in protected chaos rather than eroding the circuit's topological behavior. In other words, the topological non-trivial phase of the SSH lattice can protect the bulk dynamical scrolls of the Chua circuit elements from external perturbations, as shown in Figure~\ref{figExternalProtect}d-f. The topological protection of chaos is significant for preserving chaotic circuits to be potentially used in encryption and secure communications~\cite{kolumban_role_1998,feki_adaptive_2003,moon_chaos_2021,guan_chaos-based_2005,xiong_simplest_2021}, where even minor variations in the initial conditions can lead to significant deviations. Moreover, topologically protected chaos might provide pivotal phenomenological stability for potential chaotic processors in analog computing, which enhances machine learning~\cite{kesgin_implementing_2024}. Ensuring these processors are isolated from their environment is essential within a network.

One of the main challenges in such non-linear topological systems is the inability to evaluate the topological invariants directly. In this work, we utilized an analogous way to evaluate the dynamics of our system inspired by topological bulk-edge correspondence. Protected topological edge dynamics alters the chaotic phase portraits at each node. Therefore, we topologically characterized our system by measuring the scroll amplitudes for different non-linearities and coupling strengths, as presented in Figure~\ref{figPhasediag}e. The phase diagram obtained from the difference between the bulk and edge scrolls plotted as a function of coupling parameters becomes more pronounced and resembles that of the linear counterpart as non-linearity is tuned down.

To further investigate the interplay between chaos and topology, we developed an alternative characterization based on the node effective impedance. This approach stems from our observation that the said topology arises from the values of the linear inductors, and that the chaotic dynamics of Chua circuits at each node do not solely rely on individual component values, but rather on the cumulative effective impedance due to other circuit elements at that node. This behavior enabled us to examine the dynamical evolution of both chaotic and topological phenomena in our circuit. Our methodology simplifies the circuit analysis and successfully captures the non-linear behavior from an approximate linear model. By evaluating the difference between the bulk and edge effective inductance, we observe a similar qualitative phase demarcation as the scroll amplitude evolution [Figure~\ref{figScrollMap}e]. Furthermore, we provide analytical expressions for the effective inductance calculations. From the scroll amplitude and the effective inductance methods, we construct a comprehensive picture of topologically protected chaos in the circuit.

In addition, our study constitutes an exemplary study of the coexistence of topological phenomena and chaotic dynamics. This work may pave the way for future research into the interplay between topology and multi-scroll chaotic systems~\cite{liu_generation_2016,chen_circuit_2014}, where chaotic oscillations extend beyond single or double attractor points. Further studies can examine the effects of using Chua circuit components directly as intra-cell and inter-cell couplings, or employing non-oscillatory couplings such as resistors, diodes, or memristors~\cite{zheng_analysis_2018,yu_dynamic_2015,yu_new_2017}. Extending our model to include these elements is expected to reveal protected chaos due to the strong persistence of topology, potentially enriching the observations derived from this study.

Finally, we emphasize that our approach is generic and applicable to a variety of platforms, such as photonics~\cite{spitz_private_2021}, acoustics~\cite{tamang_dynamical_2021}, and optical cavities~\cite{fan2021real}, since they are based on mathematical non-linear differential equations rather than a specific circuit model. The framework and methods presented in this study can further offer new insights into possible topological robustness in diverse chaotic systems such as ecological networks, biological rhythms, and disease dynamics, thus highlighting the relevance of this study to numerous scientific fields.

\section{Methods}

\subsection{Determination of linear SSH phase diagram}
\begin{figure}[ht!]
	\centering
	\includegraphics[width=\linewidth]{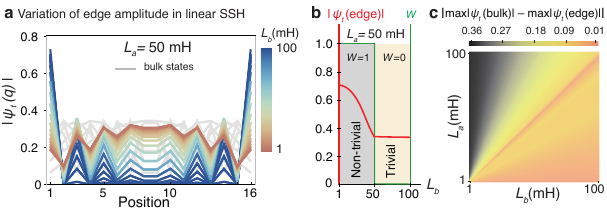}
	\caption{\textbf{Methodology for determining the linear SSH phase diagram from measurable quantities.}
	\textbf{a} The SSH circuit in the linear limit exhibits distinct edge states, shown in color as a function of the inter-cell coupling, with $L_a=50\,\text{mH}$.
	\textbf{b} Although the topological invariant changes abruptly at the phase boundary, the edge state amplitudes increase gradually. The red solid line represents the maximum amplitude of the edge states as a function of $L_b$ across both topological phases. \textbf{c} Analogous to the scroll amplitude diagrams, the phase diagram of the linear SSH can be obtained by analyzing the difference between the edge and bulk eigenstates for each combination of $L_a$ and $L_b$. In the effective circuit Laplacian, the non-linear Chua diode and the capacitor $C_1$ are omitted, as the regular resistor strongly dampens their contribution to the effective impedance. Here, $|\psi|$ denotes the eigenmodes of the effective linear SSH Laplacian corresponding to the fully linearized cSSH. The parameters used are $(L_c,C_2,R)=(24\,\text{mH}, 100\,\text{nF}, 1.85\,\text{k}\Omega)$.}
	\label{figVampAndEdgeAmplitude}
\end{figure}

Here, we describe the process of obtaining the linear SSH amplitude diagram in relation to the voltage $\Delta V_\text{Amp}$ diagrams, where distinct regions represent two different behavioral regimes. To compare this qualitative behavior with the topological edge amplitudes across $L_a$ and $L_b$, we examine the SSH circuit in the linear limit, where the non-linear Chua diode is disregarded as it blocks the AC signal in this regime. Consequently, the SSH Laplacian effectively consists only of the onsite components $L_c$, $C_2$, and $R$ in series with $C_1$ grounded. The topological phases are characterized by the topological invariant---the winding number $W$---for this circuit, calculated as
\begin{equation}
	W=\frac{1}{2\pi j}\int_{-\pi}^{\pi} dk \frac{d}{dk} \log h(k), 
	\label{eqWinding}
\end{equation}
where $h(k) = - \frac{1}{j \omega L_a} - \frac{1}{j \omega L_b} e^{j k}$ encodes the topological phase information. Our effective Laplacian, given in Equation~\eqref{LapSSHmomentum}, describes the linear limit of the Chua-SSH circuit and is non-trivial with a non-zero winding number when $L_a/L_b > 1$, and trivial otherwise.

However, while this topological invariant can be calculated for the linear limit Laplacian, it cannot be directly applied once non-linearity is introduced. Therefore, we instead rely on the consequences of the topological robustness to trace the topological properties of our non-linear Chua-SSH circuit. The key idea is to evaluate the difference in bulk and edge state amplitudes as the coupling inductances are varied.

In Figure~\ref{figVampAndEdgeAmplitude}a, we show how the edge amplitude changes as the inter-cell inductance $L_b$ is varied. Despite the quantized nature of the topological invariants, the edge amplitude increases gradually, transitioning from faded-red to deep-blue. To highlight this behavior, we plot the maximum edge amplitude in Figure~\ref{figVampAndEdgeAmplitude}b. The parameter regimes highlighted by the gray and yellow-shaded backgrounds correspond to the non-trivial and trivial phases, respectively, as defined by Equation~\eqref{eqWinding}. While the winding number (green) abruptly drops from $W=1$ to $W=0$ at $L_a = L_b$, the maximum edge amplitude (red) begins to decrease gradually as approaching $L_b = 50\,\text{mH}$, contrasting with the discrete behavior of the topological invariant.

Similar to our $\Delta V_\text{Amp}$ approach, we initially calculate the dynamical state amplitudes using
\begin{equation}
	\ket{\psi_t(q)} = e^{-iHt}  \ket{\psi_{t_0}(q)}, 
\end{equation}
where $H = J/{j\omega}$ is the effective Hamiltonian, and $\psi_{t_0}(q)$ is an initial state. In Figure~\ref{figVampAndEdgeAmplitude}c, we illustrate the difference between the maximum left edge and bulk amplitudes across all $L_a$ and $L_b$ values. The yellow/orange region represents a very small (or negligible) amplitude difference, while the difference increases rapidly, though not suddenly, for smaller $L_b$. This approach, based on directly measurable quantities reflecting topological effects, enables us to evaluate the topological phase diagram without relying on the topological invariant, but instead on measurable values. The strong resemblance between Figure~\ref{figVampAndEdgeAmplitude}c and the $\Delta V_\text{Amp}$ diagrams shown in Figure~\ref{figPhasediag}e as the system approaches the linear limit provides further evidence of the topological origin of the two regions, black and yellow, in our non-linear phase diagrams.

\subsection{Determination of resonant frequency}
At the linear limit, the Chua diode can be treated as a regular resistor with a constant value, allowing the calculation of the AC frequency of the circuit without accounting for the complex structure of the diode. This approximation is valid because, with a linear $I$-$V$ relationship, the Chua diode no longer contributes to the non-linear feedback that drives the complex dynamics of the system. In our simulations, at the linear limit, the output voltage of the op-amps diverged to the saturation value, introducing only a constant DC voltage. Therefore, it becomes feasible to determine the resonant frequency by ignoring the Chua diodes while including all other components in the circuit. The resonant frequency of the circuit is then given by
\begin{equation}
	f_r=\frac{1 + \sqrt{1-4 R^2 (C_1 + C_2)(L_c^{-1} + L_a^{-1} + L_b^{-1})}}{j 4 \pi R (C_1 + C_2)}.
	\label{resfreq}
\end{equation}
The resonant frequency is indeed complex, as our Chua-SSH circuit includes the onsite resistors which are part of the Chua circuits. While the imaginary part represents decay, the real part corresponds to the actual resonant frequency. Using the default parameters, we obtain the resonant frequency as $|\text{Re}(f_r)| = 6.42\,\text{kHz}$. This calculated resonant frequency is further confirmed by the fast Fourier transform, which shows a harmonic peak at this calculated frequency.

\subsection{Relating an array of Lorenz equations with our Chua-SSH circuit lattice}
\begin{figure}[b!]
	\centering
	\includegraphics[width=8.7cm]{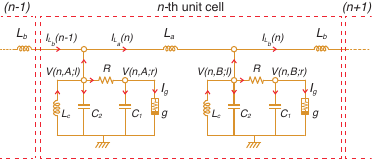}
	\caption{\textbf{The schematic of a unit cell of our Chua-SSH circuit.} Two sublattice nodes within a unit cell, $A$ and $B$, are connected by a coupling inductor with inductance $L_a$, while adjacent unit cells are connected by an inductor with inductance $L_b$. Each unit cell is labeled by $n$, with the nodes on the left and right sides of the conventional resistor $R$ denoted as $V(n,\mu;l)$ and $V(n,\mu;r)$, respectively, where $\mu=\{A,B\}$. The red arrows indicate the considered direction of current flow. The non-linear Chua diodes are labeled as $g$.}
	\label{figunit}
\end{figure}

The equations of motion for our cSSH circuit are defined by a set of eight dimensionless equations for each unit cell, as illustrated in Figure~\ref{figunit}. These equations govern the temporal variation of the current passing through each energy storage component. They are expressed as a function of the unit cell index $n$, as shown in Equations~\eqref{fullequationsA}, \eqref{fullequationsB} and \eqref{fullequationsC}. Our circuit's equation set includes two new dimensionless parameters, $\delta_a$ and $\delta_b$. These parameters represent the energy exchange between the nearest neighboring circuits and cells, respectively. The introduction of these parameters allows us to adjust the circuit's topological phase, owing to their dependence on the intra-cell and inter-cell coupling inductances. To derive the system's dimensionless Lorenz equations, which describe our coupled Chua's circuits, we write the currents passing through the components connected to the $(n,A,l)$ and $(n,B,l)$ sublattice nodes as
\begin{equation}
	\begin{aligned}
		&  I_{C_2}(n,\mu) = I_{L_c}(n,\mu) - I_{R}(n,\mu) \pm I_{L_b}(n-1) \mp I_{L_a}(n),\\
		&  I_{C_1}(n,\mu) = I_{R}(n,\mu) - I_{g}(n,\mu) ,\\
		&  L_c(n,\mu) \dv{I_{L_c}(n,\mu)}{t} = - V(n,\mu;l),
	\end{aligned}
	\label{currentsA}
\end{equation}
where $n$ and $\mu$ denote the unit cell number and the sublattice nodes $A$ and $B$, respectively, that is, $\mu \in {A,B}$. The signs of the currents $I_{L_a}$ and $I_{L_b}$ depend on the sublattice node, as will be further detailed in the following derivation. We note that, while this notation is more suitable for deriving the equations of motion, we use $q$ in place of $(n,\mu)$ to indicate position in the figures and discussions. Above, $I_{C_1}(n,\mu)$, $I_{C_2}(n,\mu)$, $I_{R}(n,\mu)$, and $I_{L_c}(n,\mu)$ represent the currents across the capacitors $C_1$ and $C_2$, the regular resistor $R$, and the inductor $L_c$ in each of Chua's circuits, respectively. Since we couple two of Chua's circuits using $L_a$ and the neighboring unit cells with $L_b$, we write the currents across these two additional inductors as follows:
\begin{equation}
	\begin{aligned}
		& L_a(n) \dv{I_{L_a}(n)}{t} = V(n,A;l) - V(n,B;l),\\
		& L_b(n) \dv{I_{L_b}(n)}{t} = V(n,B;l) - V(n+1,A;l),
	\end{aligned}
	\label{iLs}
\end{equation}
where $L_a$ and $L_b$ represent the coupling inductances of the intra-cell and inter-cell couplings, respectively, and $V$ denotes the voltage at the node indicated by its subscript. We proceed by employing the relationship $I_i = C_i \frac{\mathrm{d}V_i}{\mathrm{d}t}$ for capacitors. For inductors, the voltage across each inductor is described by the temporal evolution of the current passing through it, as $V_i = L_i \frac{\mathrm{d}I_i}{\mathrm{d}t}$. To transform the abovementioned equations into Lorenz equations, we introduce the dimensionless variables
\begin{equation}
	\begin{aligned}
		& x(n,\mu) = \frac{V(n,\mu;r)}{E_p}, \qquad  y(n,\mu) = \frac{V(n,\mu;l)}{E_p},\\
		& \qquad \qquad  z(n,\mu) = I_{L_c}(n,\mu) \frac{R(n,\mu)}{E_p},\\
		& u(n) = I_{L_a}(n) \frac{R(n,\mu)}{E_p}, \qquad  v(n) = I_{L_b}(n) \frac{R(n,\mu)}{E_p},
	\end{aligned}
	\label{xyzvariables}
\end{equation}
where, in addition to the dimensionless variables $x, y, z$, we introduce $u$ and $v$. The term $E_p$ represents the breakpoint voltage, which defines the extent of the slope of the middle segment of the $I$-$V$ characteristics of the Chua diode. We further introduce the dimensionless parameters
\begin{equation}
	\begin{aligned}
		& \alpha = \frac{C_2(n,\mu)}{C_1(n,\mu)}, \qquad  \beta= R(n,\mu)^2 \frac{C_2(n,\mu)}{L_c(n,\mu)},\\
		& \qquad \qquad \tau = \frac{t}{R(n,\mu) C_2(n,\mu)},\\
		& \delta_a= \frac{C_2(n,\mu)}{L_a(n)} R(n,\mu)^2, \qquad \delta_b= \frac{C_2(n,\mu)}{L_b(n)} R(n,\mu)^2.
		\label{dimensvar}
	\end{aligned}
\end{equation}

\subsubsection{Dimensionless non-linear function}
The non-linear current-voltage characteristic of Chua diode is expressed as a piece-wise function comprising three segments defined by the dimensional $m_0$ and $m_1$ slopes, given by
\begin{equation}
	\begin{aligned}
	&I_{g}(n,\mu) =\frac{1}{2}(m_{1}-m_{0})\\ &\times \left(|V(n,\mu;r)+E_p|-|V(n,\mu;r)-E_p| \right)+m_{0} V(n,\mu;r),
	\end{aligned}
	\label{g1}
\end{equation}
where $E_p$ is the breakpoint voltage of the non-linear $I$-$V$, which practically defines the extend of the voltage oscillations at the two terminals of the Chua circuits. $V(n,\mu;r)$ denotes the voltage at node $V(r)$ of the $\mu$-sublattice node ($A$ or $B$) of the $n$-th unitcell. We now introduce the dimensionless parameters for the non-linear resistor as
\begin{equation}
	\begin{aligned}
		& 	f(x(n,\mu)) = I_{g}(n,\mu) \frac{R(n,\mu)}{E_p},\\
		&	a=m_1 R(n,\mu) , \quad b=m_0 R(n,\mu),
	\end{aligned}
	\label{nonlinfunc}
\end{equation}
where $I_{g}(n,\mu)$ represents the non-linear current across the Chua diode. The slopes of the three segments, defined by $m_0$ and $m_1$, are crucial parameters that enable tuning of the dimensionless non-linearity factors $a$ and $b$, expressed in Equation~\eqref{fb} and \eqref{fbSSH}.

Utilizing the dimensionless slope parameters specified in Equation~\eqref{nonlinfunc}, the dimensionless non-linear function is expressed as
\begin{equation}
	\begin{aligned}
	 f_b(x(n,\mu)) &= b x(n,\mu) \\ &+\frac{1}{2}(a-b)(|x(n,\mu)+E_p|-|x(n,\mu)-E_p|).
	 \end{aligned}
	\label{diodef}
\end{equation}
In our simulations, we specifically set $E_p = 3\,\mathrm{V}$, and $a = -1.14$, while varying $b$ to implement the non-linear function described above. The subscript $b$ in $f_b$ denotes that the non-linear function depends on $b$.

\subsubsection{Detailed derivations of dimensionless equations}
We express the current for the left node (i.e., $(n,\mu;l)$) of each Chua's circuit in our chaotic SSH circuit as a function of the unit cell $n$ and sublattice node $\mu$, as follows
\begin{equation}
	I_{C_2}(n,\mu) = I_{L_c}(n,\mu) - I_{R}(n,\mu) \pm I_{L_a}(n) \mp I_{L_b}(n-1) .
	\label{derc2_1}
\end{equation}
Here, unlike Equation~\eqref{currentsA}, the sign of $I_{L_a}(n)$ and $I_{L_b}(n-1)$ varies depending on the sublattice node. For example, when $\mu=A$, $I_{L_a}(n)$ and $-I_{L_b}(n-1)$, but when $\mu=B$, $-I_{L_a}(n)$ and $I_{L_b}(n-1)$. The above equation can be reformulated to express it in terms of the voltage across the capacitors $C_2$ and the resistors $R$ as
\begin{equation}
	\begin{aligned}
	C_2(n,\mu) \dv{V(n,\mu;l)}{t} =&  I_{L_c}(n,\mu) \pm I_{L_a}(n) \mp I_{L_b}(n-1) \\
	& - \frac{1}{R(n,\mu)} \left(V(n,\mu;l)-V(n,\mu;r)\right).
	\end{aligned}
	\label{derc2_2}
\end{equation}
Now, we substitute the dimensionless variables specified in Equation~\eqref{xyzvariables} and Equation~\eqref{dimensvar} into the above equation and rewrite it as
\begin{equation}
	\begin{aligned}
	\dv{y(n,\mu)}{\tau} =& x(n,\mu) - y(n,\mu) \\ &+ \frac{R(n,\mu)}{E_p} \left( I_{L_c}(n,\mu) \pm I_{L_a}(n,\mu) \mp I_{L_b}(n-1) \right).
	\end{aligned}
	\label{derc2_3}
\end{equation}
We then rearrange Equation~\eqref{xyzvariables} and derive one of the dimensionless system equations as
\begin{empheq}[box=\fbox]{equation}
	\dv{y(n,\mu)}{\tau} = x(n,\mu) - y(n,\mu) + z(n,\mu) \pm u(n) \mp v(n-1) .
	\label{derc2_4}
\end{empheq}
Note that the sign of the dimensionless variables $u(n)$ and $v(n-1)$ changes such that $\dv{y(n,A)}{\tau}=\dots + u(n)-v(n-1)$ and $\dv{y(n,B)}{\tau}=\dots - u(n)+v(n-1)$. Similarly, we express the current for the right node (i.e., $(n,\mu;r)$) of each Chua's circuit as
\begin{equation}
	I_{C_1}(n,\mu) = I_{R}(n,\mu) - I_{g}(n,\mu),
	\label{derc1_1}
\end{equation}
and
\begin{equation}
	\begin{aligned}
	C_1(n,\mu) \dv{V(n,\mu;r)}{t} = &\frac{1}{R(n,\mu)} \left(V(n,\mu;l)-V(n,\mu;r)\right)\\ & - I_{g}(n,\mu).
	\end{aligned}
	\label{derc1_2}
\end{equation}
We apply the dimensionless variables from Equation~\eqref{xyzvariables} and Equation~\eqref{dimensvar}, and rewrite the above equation as
\begin{equation}
	\frac{C_1(n,\mu) }{C_2(n,\mu)} \dv{x(n,\mu)}{\tau} = \left(y(n,\mu) - x(n,\mu)\right) - I_{g}(n,\mu) \frac{R(n,\mu)}{E_p},
	\label{derc1_3}
\end{equation}
and finally we have
\begin{empheq}[box=\fbox]{equation}
	\dv{x(n,\mu)}{\tau} = \alpha \big( y(n,\mu) - x(n,\mu) - f(x(n,\mu))\big).
	\label{derc1_4}
\end{empheq}

To derive the third equation in the Lorenz system, we write the current across the Chua inductor as
\begin{equation}
	L_c(n,\mu) \dv{I_{L_c}(n,\mu)}{t} = - V(n,\mu;l),
	\label{iLc1}
\end{equation}
explicitly write
\begin{equation}
	\dv{I_{L_c}(n)}{\tau} \frac{L_c}{R(n,\mu) C_2(n,\mu)} = - E_p y(n,\mu),
	\label{iLc2}
\end{equation}
and with $\beta = R(n,\mu)^2 C_2(n,\mu) / L_c(n,\mu)$, we obtain
\begin{empheq}[box=\fbox]{equation}
	\dv{z(n,\mu)}{\tau} = - \beta y(n,\mu).
	\label{iLc4}
\end{empheq}

We now derive the dimensionless form of the intra-cell coupling, denoted as $L_a(n)$, within a unit cell. The current for the inductors $L_a(n)$ is expressed as
\begin{equation}
	L_a \dv{I_{L_a}(n)}{t} = V(n,B;l)-V(n,A;l).
	\label{iL11}
\end{equation}
We incorporate the dimensionless parameters from Equation~\eqref{xyzvariables} and Equation~\eqref{dimensvar} and reformulate the above equation as
\begin{equation}
	\dv{I_{L_a}(n)}{\tau} \frac{L_a}{R(n,\mu) C_2(n,\mu)} = E_p \left(y(n,B)-y(n,A)\right),
	\label{iL12}
\end{equation}
and
\begin{equation}
	\frac{R(n,\mu)}{E_p} \dv{I_{L_a}(n)}{\tau}\frac{L_a(n)}{R(n,\mu)^2 C_2(n,A)} =y(n,B) - y(n,A).
	\label{iL13}
\end{equation}
We now employ the dimensionless new variables $u(n)$ and $\delta_a$, as explicitly defined in Equation~\eqref{xyzvariables} and Equation~\eqref{dimensvar}, to further simplify the above equation as
\begin{empheq}[box=\fbox]{equation}
	\dv{u(n)}{\tau} = \delta_a \big( y(n,B)-y(n,A) \big).
	\label{iL14}
\end{empheq}
Similarly, we can now derive the dimensionless form of the inter-cell coupling, denoted as $L_b(n)$, within a unit cell. The current for the inductors $L_b(n)$ is expressed as
\begin{equation}
	L_b(n) \dv{I_{L_b}(n)}{t} = V(n+1,A;l) - V(n,B;l).
	\label{iL21}
\end{equation}
By following similar steps and utilizing the new dimensionless variables $v(n)$ and $\delta_b$, we obtain
\begin{empheq}[box=\fbox]{equation}
	\dv{v(n)}{\tau} = \delta_b \big(y(n+1,A) - y(n,B) \big).
	\label{iL22}
\end{empheq}

We have derived the dimensionless system equations in terms of the unit cell number $n$ and the sublattice node $\mu$, which can be either $A$ or $B$. While it is possible to treat each component as an individual entity and assign specific values to each, we standardized the components across all unit cells in our cSSH for periodicity. Consequently, we set $L_a=L_a(n)$, $L_b=L_b(n)$, $L_c=L_c(n,\mu)$, $C_1=C_1(n,\mu)$, $C_2=C_2(n,\mu)$, and $R=R(n,\mu)$ across the entire lattice.

\subsection{Derivation of effective inductance formulae}

Here, we derive the effective inductance experienced at any node, and show how it may affect the local topological behavior of its attached chaotic oscillator. In our cSSH circuit, we couple Chua circuits at their $l$ nodes via inductors with inductances $L_a$ and $L_b$, forming an SSH circuit with onsite Chua circuits. Coupling identical Chua circuits results in a breakpoint of the identity of Chua circuits because each Chua circuit experiences effectively different inductance at the coupling nodes due to the open boundary conditions. For example, while the total impedance at the $l$ nodes is $L_c + L_a + L_b + C_2 + R$, the effective impedance varies depending on the position of the node. Building on this behavior, we assess the circuit's chaotic and topological dynamics using the effective inductance approach, as inductance is a common parameter in both the Chua circuits and the SSH circuit. In the following, we describe how we derive the effective inductance in our circuit by considering an approximate circuit consisting solely of inductive elements.

\begin{figure}[b!]
	\centering
	\includegraphics[width=8.7cm]{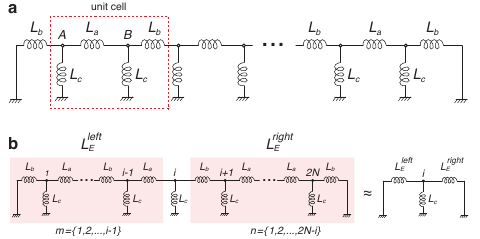}
	\caption{\textbf{The approximate inductive SSH circuit model with open boundary conditions.} \textbf{a} The chaotic SSH circuit can be approximated to a linear circuit to which only the same type of components are considered. This circuit corresponds to a fully inductive linear SSH circuit with the intra-cell and inter-cell inductances $L_a$ and $L_b$ and onsite inductor with inductance $L_c$. \textbf{b} The effective inductance at each node can be calculated analogous to the lumped-element method, which considers the impedance of each direction as a single equivalent impedance. This method requires lumping all the elements along each direction and allows us to determine the impedance at a node (inductance for our circuit) by considering each side as a single entity, i.e., $L_E^\text{left}$ and $L_E^\text{right}$. The impedance of the lumped elements can be determined starting from the edge elements. $m$ and $n$ represent the indices of each iteration. }
	\label{figCircuitSchematic}
\end{figure}

We first derive the effective inductance for the bulk nodes of the circuit depicted in Figure~\ref{figCircuitSchematic}a. The effective inductance at a node is not simply the sum of the inductances of the connected inductors (i.e., $L_a + L_b + L_c$); instead, it is determined by the contributions of all the inductors on both sides of the node across the boundary. To calculate the effective node inductance, we lump all the nodes on both sides of the node, as illustrated in Figure~\ref{figCircuitSchematic}b. This concept corresponds to the lumped-element model, where the circuit elements are concentrated at a single node at which the effective inductance can be calculated as

\begin{figure}[t!]
	\centering
	\includegraphics[width=8.7cm]{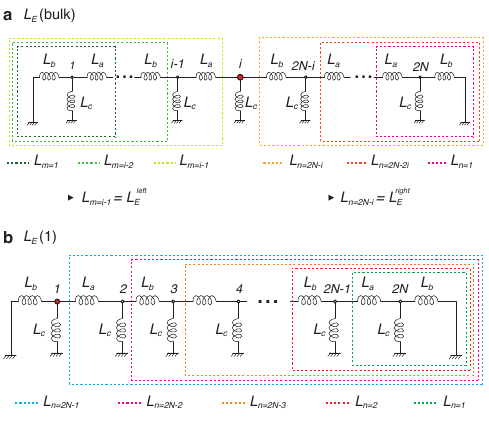}
	\caption{\textbf{Examples for the derivation of recursive effective inductance for the edge and bulk nodes.} \textbf{a} The effective inductance of a bulk node is obtained by lumping the elements found both sides of the circuit. The effective inductances of the left and right lumps are obtained when $L_E^{\text{left}} = L_{m=i-1}$ and $L_E^{\text{right}} = L_{n=2N-i}$, respectively. Once $L_E^{\text{left}}$ and $L_E^{\text{right}}$ are obtained, $L_E(i)$ is given by Equation~\eqref{eqtotalind}. \textbf{b} The effective inductance of an edge node, e.g., $L_E(1)$, is obtained by starting to lump the inductors from the right side of the circuit. As we proceed to the opposite edge, we label each summation with the recursion index $n$. $n=2N-1$ gives us the effective inductance of the right side of the circuit. $N$ denotes the total number of unit cells.}
	\label{figRecursiveSchematic}
\end{figure}

\begin{equation}
	L_E(i)=\left(\frac{1}{L_c}+\frac{1}{L_E^\text{left}}+\frac{1}{L_E^\text{right}}\right)^{-1},
	\label{eqtotalind}
\end{equation}
where $L_E(i)$ represents the effective inductance at the $i$th node, and $L_E^\text{left}$ and $L_E^\text{right}$ are the effective lumped inductances of the left and right sides of node $i$, respectively. 

Now, to determine $L_E^\text{left}$ and $L_E^\text{right}$, we begin from each boundary and sum the inductances towards node $i$. We recursively label the summation of the left and right sides as $L_m$ and $L_n$, where $m$ and $n$ are indices that range from $1$ to $i-1$ and 1 to $2N-i$, respectively. For example, at both edges (refer to Figure~\ref{figRecursiveSchematic}a), inductors $L_b$ and $L_c$ are in parallel, and they are in series with $L_a$ in the first iteration. This leads to $L_{m,n=1}=(\frac{1}{L_c} + \frac{1}{L_b})^{-1} + L_a$ for the first iterations. This implies that $L_{m,n=1}$ effectively reduces the inductance of the three inductors at the edge node to a single equivalent inductor between the edge node and ground. The first iteration corresponds to the dark green and magenta dashed rectangles in Figure~\ref{figRecursiveSchematic}a. Following the first iteration, the second iteration (i.e., $\{m,n\}=2$) results in $L_{m=2}=(\frac{1}{L_c} + \frac{1}{L_{m=1}})^{-1} + L_b$ and $L_{n=2}=(\frac{1}{L_c} + \frac{1}{L_{n=1}})^{-1} + L_b$. As we progress to node $i$, we obtain the recursive relation
\begin{equation}
	\begin{aligned}
		L_{m} =& \left(\frac{1}{L_c}+\frac{1}{L_{m-1}}\right)^{-1}+L_{a,b} ,\\
		L_{n} =& \left(\frac{1}{L_c}+\frac{1}{L_{n-1}}\right)^{-1}+L_{a,b} ,
	\end{aligned}
		\label{recursiveRelations}
\end{equation}
where $L_0=L_b$ and $L_{a,b}$ alternates between $L_a$ and $L_b$ depending on the parity of the indices $m$ and $n$. Specifically, for the odd values of $m$ and $n$, $L_{a,b}$ assumes the value of $L_a$, while for even values, it takes on the value of $L_b$. The above recursive relations provide us with the effective inductance of each lump when $m=i-1$ and $n=2N-i$ implying that $L_{m=i-1} = L_E^\text{left}$ and $L_{n=2N-i} = L_E^\text{right}$. Once the effective inductances on both sides of the $i$th node are determined, the effective inductance at the $i$th node is given by Equation~\eqref{eqtotalind}. 

In the case of edge node effective inductances, i.e., $L_E(i=1)$ (left) and $L_E(i=2N)$ (right), the recursive indices become $m=0$ and $n=2N-1$, or $m=2N-1$ and $n=0$, respectively. We illustrate the model for the left edge calculation in Figure~\ref{figRecursiveSchematic}b. In the edge $L_E$ calculations, we have a single lump covering the remaining nodes, which is the case corresponding to $m=0$ or $n=0$.

\begin{figure}[b!]
	\centering
	\includegraphics[width=\linewidth]{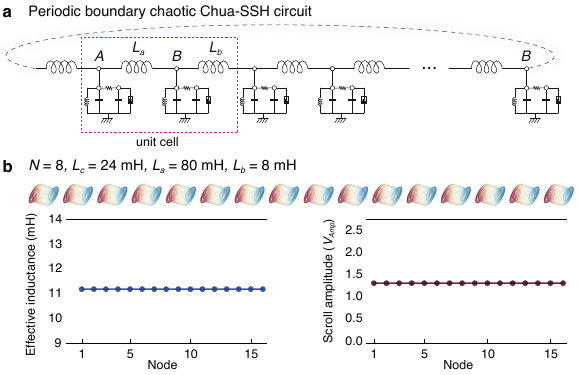}
	\caption{\textbf{Chua-SSH circuit under periodic boundary conditions (PBC) and its effective inductance and scroll amplitude profiles.} \textbf{a} The periodic boundary cSSH is achieved by connecting the two edge nodes with $L_b$. The bulk of the circuit under OBC is completely identical to that of the PBC cSSH. \textbf{b} The demonstration of chaotic portraits, effective inductance, and scroll amplitude profiles exhibits a consistent profile throughout the entire circuit attributed to the implementation of PBC.}
	\label{figPBCSSH}
\end{figure}

The effective inductance profile is significantly altered depending on the boundary conditions. To illustrate this, we consider a semi-infinite circuit where $N$ is large enough. We recall Equation~\eqref{recursiveRelations} and express it for a bulk node as
\begin{equation}
	\scriptstyle
	\begin{aligned}
	& L_E^{\text{left}}(\text{bulk}) \approxeq L_E^{\text{right}}(\text{bulk})  =\\ &  L_{a,b} + \cfrac{1}{\frac{1}{L_c} + \cfrac{1}{L_{a,b} + \cfrac{1}{\frac{1}{L_c} + \cfrac{1}{L_{a,b}  + \cfrac{1}{\frac{1}{L_c} + \cfrac{1}{L_{a,b} + \cfrac{1}{\ddots}}}}}}},
	\end{aligned}
	\label{recursionEquation}
\end{equation}
where $L_{a,b}$ takes on value of $L_a$ or $L_b$ depending on the parity of the node $i$.
The above equation implies that the effective inductance of bulk nodes is approximately equivalent. This is because $L_E^\text{left}$ and $L_E^\text{right}$ have nearly equal weight due to the long-depth of the fractions.

However, as we approach one of the edges, the depth of one recursion in Equation~\eqref{recursionEquation} decreases while the other increases, corresponding to $n \ll m$ or $m \ll n$. A decrease (increase) in the recursion depth leads to an overall increase (decrease) in $L_E^\text{left}$ or $L_E^\text{right}$. This results in a rapid change in the effective inductance at the nodes near the boundaries. From Equation~\eqref{eqtotalind}, since the effective inductances appear in the denominator, an increase in the inductance of the lumps leads to a decrease in the overall node inductance as we approach the edges, i.e., $L_E(i \rightarrow 1)$ or $L_E(i \rightarrow 2N)$.

Whereas, in the case of periodic boundary conditions [Figure~\ref{figPBCSSH}a], the effective inductance across the entire circuit is expected to remain uniform. This is because $m$ and $n$ are approximately equal to $N$, making the contributions of $L_E^\text{left}$ and $L_E^\text{right}$ always roughly equal. As a result, $L_E(i)$ remains constant for all $i$, regardless of its parity. For example, in Figure~\ref{figPBCSSH}b, we present the effective inductance profile of the periodic Chua-SSH circuit alongside the scroll amplitude profile, both of which exhibit uniform amplitudes. Remarkably, the scroll amplitudes exhibit qualitatively the same profile as the effective inductance, further supporting the presence of topologically protected chaotic dynamics in the open boundary Chua-SSH circuit.

\subsubsection{Trimer Chua-SSH circuit}

\begin{figure}[h!]
	\centering
	\includegraphics[width=\linewidth]{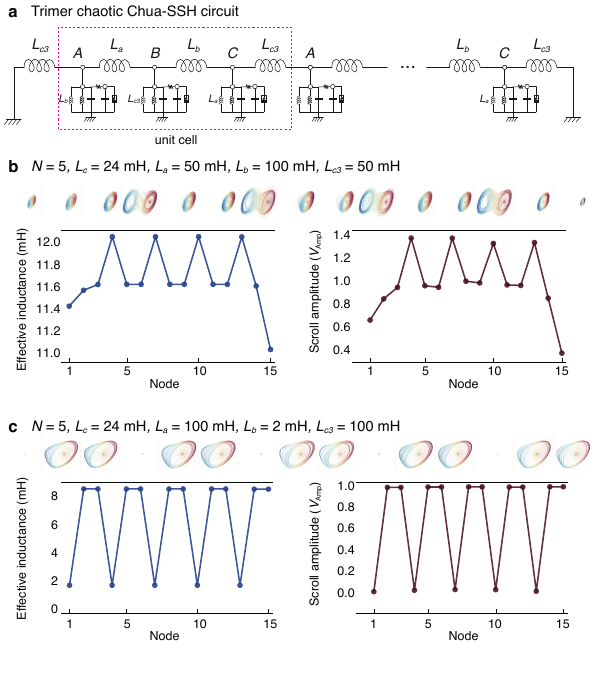}
	\caption{\textbf{Trimer chaotic Chua-SSH circuit and examples of its chaotic phase portraits, effective inductance, and scroll amplitude profiles.} \textbf{a} The trimer Chua-SSH circuit comprises three sublattice nodes—$A$, $B$, and $C$—and three types of inductors with inductances $L_a$, $L_b$, and $L_{c3}$. Additional inductors are connected between each node and ground to ensure uniform node conductance across the circuit. \textbf{b} and \textbf{c} Two examples of chaotic phase portraits, effective inductance, and scroll amplitude profiles are presented, demonstrating the effectiveness of the effective inductance approach and its correlation with scroll amplitude in generic circuits. The parameters besides those indicated are the default parameters.}
	\label{figTrimerSSH}
\end{figure}

To examine the generality of the qualitative correspondence between the effective inductance and the scroll amplitudes, we present a trimer Chua-SSH circuit. The inclusion of an additional sublattice node ($C$) in each unit cell requires modifying the circuit's grounding. As shown in Figure~\ref{figTrimerSSH}a, we ground each node with an additional inductor whose inductance corresponds to the coupling inductor that is not connected to the considered node. This arrangement ensures that each node experiences the same conductance. Following this structure, the effective inductance for the trimer chaotic SSH circuit involving the new additional inductance $L_{c3}$ is given by
\begin{equation}
	L_E^\text{TCSSH}(i)=\left(\frac{1}{L_c}+\frac{1}{L_{a,b,c3}(i)}+\frac{1}{L_E^\text{left}}+\frac{1}{L_E^\text{right}}\right)^{-1},
	\label{eqtotalindTrimer}
\end{equation}
where 
\begin{equation}
	L_{a,b,c3}(i) = 
	\begin{cases} 
		L_b, & \text{if } i \mod3 = 1, \\
		L_{c3}, & \text{if } i \mod3 = 2, \\
		L_a, & \text{if } i \mod3 = 0. 
	\end{cases}
\end{equation}
The effective inductance of the left and right lumped elements, i.e., $L_E^\text{left}$ and $L_E^\text{right}$, are determined by
\begin{equation}
	\begin{aligned}
		L_{m} =& \left(\frac{1}{L_c} + \frac{1}{L_{b,c3,a}} + \frac{1}{L_{m-1}}\right)^{-1}+L_{a,b,c3} ,\\
		L_{n} =& \left(\frac{1}{L_c} + \frac{1}{L_{a,c3,b}} + \frac{1}{L_{n-1}}\right)^{-1}+L_{b,a,c3}.
	\end{aligned}
	\label{recursiveRelationsTrimer}
\end{equation}
where each inductance term, such as $L_{a,b,c3}$ or $L_{b,c3,a}$, represents one of $L_a$, $L_b$, or $L_{c3}$ depending on $m \mod 3 $ and $n \mod 3 $. For example, when $m \mod 3 = 1$, $L_{b,c3,a}$ becomes $L_b$; when $m \mod 3 = 2$, it becomes $L_{c3}$; and when $m \mod 3 = 0$, it becomes $L_a$. This same cyclic logic applies to the additional terms outside the parentheses. The effective inductance of the left and right lumps are achieved when $m=i-1$ and $n=2N-i$, implying that $L_{m=i-1} = L_E^\text{left}$ and $L_{n=2N-i} = L_E^\text{right}$.

Significantly, as can be seen from the two examples presented in Figures~\ref{figTrimerSSH}b and \ref{figTrimerSSH}c, the correlation between the effective inductance profile and scroll amplitude is a generic property that can be applied to more complex or higher dimensional non-linear lattice models to examine the topological properties in non-linear circuits. Our novel approach offers a valuable approximation for evaluating non-linear circuits using linear models. We will discuss this by providing analytical solutions of these numerical approaches and demonstrating their general agreement with the scroll amplitude profiles in the following subsections.

\subsection{Analytical effective inductance formulae}
Utilizing the recursive structure of $L_E(i)$, it is possible to obtain the analytical expressions for the edge and bulk nodes. The approach relies on assessing each continued fraction representation by expressing the nested dependencies in terms of repeated sequences. The recursive nature gives rise to an infinite continued fraction, which can be simplified by expressing it in terms of two interdependent variables. These variables are defined recursively, with each relying on the other, leading to a nested solution. Solving these coupled equations yields closed-form expressions for both edge and bulk node inductances. We provide a detailed derivation in Section I in the Supporting Information. Considering a semi-infinite circuit, we obtain the analytical form of $L_E(1)$ as
\begin{equation}
		\begin{aligned}
	&L_E(1) = \\ &\left( \frac{1}{L_c}+\frac{1}{L_b} + \frac{2}{L_a + \sqrt{\frac{\left(L_a+2 L_c\right) \left(2 L_c \left(L_a+L_b\right)+L_a L_b\right)}{L_b+2 L_c}}}\right)^{-1}.
		\end{aligned}
	\label{LE1closedform3}
\end{equation}
Similarly, the analytical expression for the second node from the left edge is given by
\begin{equation}
	\begin{aligned}
		L_E(2) =& \left( \frac{1}{L_c}+\frac{1}{L_a+\frac{1}{\frac{1}{L_c}+\frac{1}{L_b}}} \right. \\ &\left. \quad +\frac{2}{L_b+\sqrt{\frac{\left(L_b+2 L_c\right) \left(2 L_c \left(L_a+L_b\right)+L_a L_b\right)}{L_a+2 L_c}}} \right)^{-1}.
	\end{aligned}
	\label{LE2closedform3}
\end{equation}
We provide a closed-form expression for the third node in Section I in the Supporting Information. The bulk closed-form effective inductance expression is also obtained as
\begin{equation}
	\begin{aligned}
		L_E(&\text{bulk}) = \\ &\frac{L_a L_c \left(L_b+L_c\right)+L_b L_c^2}{\sqrt{\left(L_a+2 L_c\right) \left(L_b+2 L_c\right) \left(2 L_c \left(L_a+L_b\right)+L_a L_b\right)}}.
	\end{aligned}
	\label{LEBulkclosedform3}
\end{equation}

\begin{figure}[ht!]
	\centering
	\includegraphics[width=8.7cm]{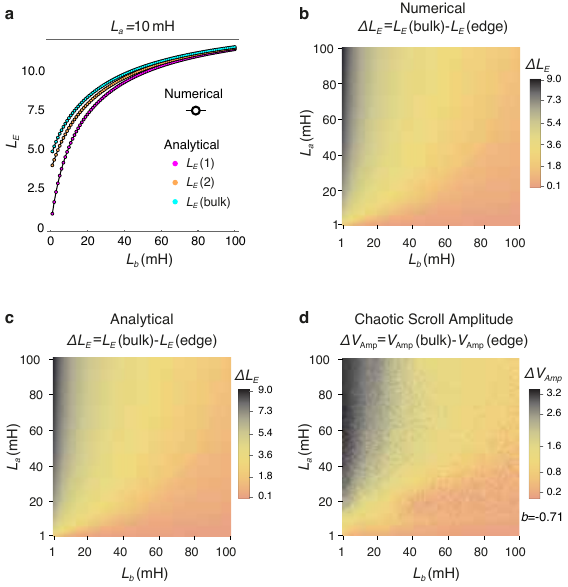}
	\caption{\textbf{The comparison of the numerical and analytical effective inductance diagrams and the correspondence between the effective inductance and scroll amplitudes.} \textbf{a} The numerical effective inductances of the edge node, second node, and a bulk node, as calculated from Equation~\eqref{eqtotalind}, show good agreement with the analytical effective inductances of the edge node, second node, and a bulk node calculated from Equations~\eqref{LE1closedform3}, \eqref{LE2closedform3}, and \eqref{LEBulkclosedform3}. The parameters used are $L_a=10\,\text{mH}$. \textbf{b} The numerical effective inductance diagram across $L_a$ and $L_b$ is obtained from Equation~\eqref{eqtotalind}. The diagram display the $L_E$ difference between bulk and edge nodes. \textbf{c} The effective inductance diagram is obtained from the analytical closed-form expressions for the edge and bulk nodes, i.e., Equations~\eqref{LE1closedform3}, and \eqref{LEBulkclosedform3}. As provided a specific $L_a$ example in \textbf{a}, both the numerical and analytical effective inductance expressions show good agreement across the entire range of the parameter space. \textbf{d} The chaotic scroll amplitudes were obtained through numerical simulations of our circuit with $N=8$. We measured the peak-to-peak amplitude of the voltage oscillations at the node $l$s and calculated the difference between the bulk and edge node scroll amplitudes. The resulting density plot of $\Delta V_\text{Amp}$ reveals two distinct regions in parallel to the effective inductance diagrams showing their direct correspondence. The noisy appearance of the density plot is attributable to the dynamic characteristics of the chaotic scrolls. For each panel here, we consistently set $L_c=24\,\text{mH}$.
	}
	\label{figPhasedensityplots}
\end{figure}

These analytical expressions we derived for the first two edge nodes and a bulk node align well with the numerical effective inductances. In Figure~\ref{figPhasedensityplots}a, we present the analytical and numerical $L_E(1)$, $L_E(2)$, and $L_E(\text{bulk})$ results, highlighting significant differences between the bulk and edge inductance for small $L_b$ values. Figure~\ref{figPhasedensityplots}b demonstrates this behavior across a wide range of intra- and inter-cell coupling inductances. In Figure~\ref{figPhasedensityplots}c, the analytical $\Delta L_E$ results align well with the numerical ones across all parameter spaces.

The qualitative correspondence between the effective inductance and the scroll amplitude exhibits very similar behavior across the same parameter space. For example, $\Delta V_{\text{amp}}$ shown in Figure~\ref{figPhasedensityplots}d exhibits similar dynamics where the difference between the bulk and edge quantities rapidly increases as the parameters approach the black region, which is the core argument to define the topological phases in our non-linear lattice. Importantly, this close agreement between the two different quantities allows us to study the non-linear behavior using the linear effective inductance model.

\subsection{Qualitative proportionality between the effective inductance and scroll amplitude}
To explain the correlation between the effective inductance and scroll amplitudes, we consider a bulk node in Figure~\ref{figCircuitSchematic}b. The current flowing into the bulk node is given by
\begin{equation}
	I_{L_c} = I_\text{right} - I_\text{left}.
\end{equation}
where $I_\text{right}$ and $I_\text{left}$ represent the currents flowing through the right and left inductor lumps, respectively. Considering the current flow directions shown in Figure~\ref{figunit}, we substitute the current integral expressions into the above relation as
\begin{equation}
		\frac{1}{L_c} \int V(i) \,dt =  - \frac{1}{L_E^\text{left}} \int V(i) \,dt - \frac{1}{L_E^\text{right}} \int V(i) \,dt,
\end{equation}
and rearranging this relation, we arrive at
\begin{equation}
	\left( \frac{1}{L_c} + \frac{1}{L_E^\text{left}} + \frac{1}{L_E^\text{right}} \right) \int V(i) \,dt = 0.
\end{equation}
By substituting Equation~\eqref{eqtotalind}, we obtain
\begin{equation}
	\frac{1}{L_E(i)} \int V(i) \,dt = 0.
	\label{curVoltIntegral}
\end{equation}
The above equation suggests that if $L_E(i)$ is larger, the node can sustain larger oscillations in voltage (since inductors resist changes in current, causing voltage to rise). In other words, the system allows higher voltage swings when the inductance is high because the inductor slows down current changes. As such the voltage at the node must increase to maintain the same rate of change in current. Conversely, if $L_E(i)$ is smaller, the voltage needed to drive the same current change is smaller. The integral reflects the scaling of voltage oscillations over time by the effective inductance. For instance, a larger inductance value $L_E(i)$ allows more energy to be stored, facilitating larger voltage amplitudes at the node, i.e.,
\begin{equation}
	V(i) \propto L_E(i).
\end{equation}

Taking the derivative of both sides of Equation~\eqref{curVoltIntegral} also shows that the rate of change of current through the inductor is inversely proportional to the inductance. The current flows primarily through the inductors connected to the ground, which offer lower impedance. A low inductance at the edge nodes increases the rate of change of current, compressing the dynamics into a more constrained oscillatory state. Conversely, high inductance at the bulk nodes results in slower voltage evolution. The double-scroll attractors therefore persist because the state can still alternate between two stable oscillatory regions.

In the case of an external AC current injection, as shown in Figure~\ref{figExternalProtect}, it is intuitive that the current flows through the ground, as the end nodes are connected to the ground with $L_b$, which is smaller in the non-trivial phase, corresponding to the black region in Figure~\ref{figPhasediag}d.

\subsection{The effect of lattice size on phase boundaries}

\begin{figure}[ht!]
	\centering
	\includegraphics[width=0.9\linewidth]{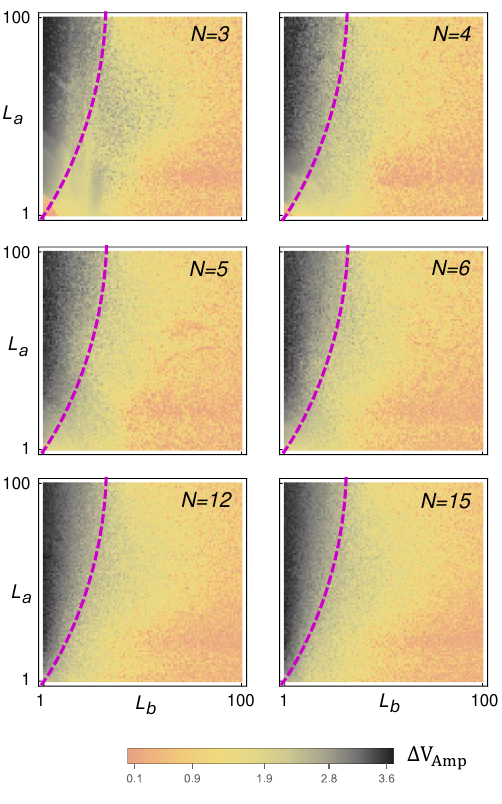}
	\caption{\textbf{Evaluation of non-linear phase boundaries under varying lattice sizes.} The phase diagrams are obtained from the amplitude difference between bulk and edge scrolls, defined by Equation~\eqref{EqDeltaVamp}. The magenta dashed lines correspond to the reference phase boundary from Figure~\ref{figPhasediag}d (for $N=8$), facilitating visual comparison. The non-linear phase boundaries remain qualitatively unchanged for lattice sizes $N\geq5$. Circuit parameters are fixed at $(L_c,C_1,C_2,R,b)=(24\,\text{mH}, 10\,\text{nF}, 100\,\text{nF}, 1.85\,\text{k}\Omega,-0.71)$.}
	\label{figSize}
\end{figure}

We now discuss the influence of the system size on the non-linear phase boundaries. In Figure~\ref{figSize}, we present non-linear phase diagrams obtained for various system sizes ($N=3,4,5,6,12$, and $15$), keeping constant other circuit parameters, including the non-linearity ($b=-0.71$) and Chua circuit parameters. The magenta dashed lines correspond to the phase boundary from the reference diagram in Figure~\ref{figPhasediag}d (for $N=8$), enabling an easy comparison to observe how the phase boundaries evolve with changing lattice size.

For smaller lattice sizes ($N=3,4$), significant deviations from the reference boundary (magenta dashed line) appear, particularly in regions where both $L_a$ and $L_b$ are small. In contrast, for lattice sizes $N \geq 5$, the non-linear phase boundaries remain qualitatively consistent across almost the entire parameter space.

Specifically, for lattices smaller than $N=5$, the lower-left regions of the diagrams in Figure~\ref{figSize} predominantly exhibit regular or damped oscillations, indicated by relatively uniform regions compared to chaotic regions, which typically appear noisy. This finding aligns with expectations, as topological phenomena inherently rely on sufficient periodicity within the bulk lattice structure. Hence, a lattice size of at least $N=5$ unit cells is determined to be sufficient to achieve clearly defined topological protection across the investigated parameter range.

Nevertheless, even for smaller lattices ($N=3,4$), when the ratio of $L_a$ to $L_b$ is sufficiently large (e.g., $(L_a,L_b)=(80\,\text{mH}, 8\,\text{mH})$), prominent suppression of edge modes in the non-trivial phase persists. We illustrate this behavior by providing examples of the spatial distribution of chaotic scrolls in Section VI in the Supporting Information.

\subsection{Finding explicit parametric equations of scroll space through machine learning}

We present how we identify the explicit parametric equations for determining the three main scroll classes, i.e., all single, mixed and all double. The black lines in Figure~\ref{figScrollMap}a defines the boundaries between the classes. We determine these boundaries by employing linear discriminant analysis (LDA) and support vector machines (SVM). LDA serves as a dimensionality reduction tool that distinguishes classes by maximizing the ratio of between-class variance to within-class variance~\cite{fisher_use_1936}. Meanwhile, SVM classifies data by identifying optimal hyperplanes that maximize the margin between classes~\cite{cortes_support_vector_1995}. We utilize the LDA and SVM accuracy results to evaluate linear separability. 

\begin{figure}[t!]
	\centering
	\includegraphics[width=.8\linewidth]{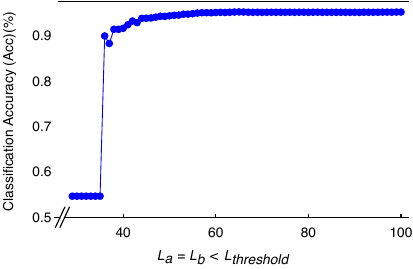}
	\caption{\textbf{Accuracy evolution over trained parameter region.} This figure demonstrates how accuracy across the entire state space changes with respect to a given training space. The training space is defined by a threshold inductance value, i.e., $L_a < L_{\text{threshold}}$ and $L_b < L_{\text{threshold}}$, where the x-axis corresponds to $L_{\text{threshold}}$. The high accuracy (Acc - Equation~\eqref{eqAccuracy}) remaining beyond \(L_{\text{threshold}} = 50\) indicates that all information defining the state space can be acquired from the region \(L_a < 50\) and \(L_b < 50\). This suggests that the found equations continue to govern the state space over a much larger extended region.} 
	\label{classification_accuracy_over_N}
\end{figure}

The state space is likely to be linearly separable if distinct clusters are uncovered by LDA or high accuracy is achieved by SVM. Linear separability suggests the feasibility of using linear equations to determine the boundaries between classes. In our study, we transform from a non-linear state space $(L_{a}, L_{b})$ in Figure~\ref{figScrollMap}a to a higher-dimensional state space $(L_a, L_b, L_E(\text{bulk}), \Delta L_E)$. In contrast to the results obtained in the two-dimensional non-linear state space, applying LDA and SVM to the newly defined four-dimensional state space results in separated clusters in the LDA-reduced space and high classification accuracy (Acc) with SVMs (up to 95\% as shown in Figure~\ref{classification_accuracy_over_N}). The accuracy is defined as:
\begin{equation}
	\text{Acc} = \frac{1}{M} \sum_{k=1}^{M} \mathbb{I}(o_k = \hat{o}_k).
	\label{eqAccuracy}
\end{equation}
Here, $M$ is the total number of instances, $o_k$ is the true label of the \( k \)-th instance, \( \hat{o}_k \) is the predicted label of the \( k \)-th instance, and \( \mathbb{I}(\cdot) \) is the indicator function. The parametric equations found by SVM define the black lines in Figure~\ref{figScrollMap}a, which can be obtained from
\begin{equation}
	\begin{cases} 
		7.4698 L_{E_{\text{bulk}}} - 30.8352 < \Delta L_E & \text{All single}, \\
		4.4354 L_{E_{\text{bulk}}} - 5.7499 > L_b & \text{Mixed}, \\
		\text{otherwise} & \text{All double}.
	\end{cases}
\end{equation}

Note that the conditions are evaluated sequentially: if condition 1 is satisfied, the data point belongs to `All double'; if condition 1 is not satisfied but condition 2 is, the data point belongs to `Mixed'; otherwise, the data point belongs to `All single.'

Remarkably, even though the classes can be determined using only data points with $L_a < 50$ and $L_b < 50$, yet $95\%$ accuracy is achieved in classifying the extended region $L_a < 100$ and $L_b < 100$. By analyzing the accuracy trend across the trained parameter region (Figure~\ref{classification_accuracy_over_N}), it becomes clear that beyond $L_a < 50$ and $L_b < 50$, the system provides no additional information, and the accuracy stabilizes. Consequently, these inequalities accurately describe a broader range of unsimulated parameter regions, highlighting their strong generalizability (see Sections VII and VIII in the Supporting Information). 

This approach reduces the need for computationally intensive simulations across entire phase spaces and allows for explicit theoretical formulations. Furthermore, this approach can be extended to identify non-linear phase boundaries using the SVM kernel trick.

\begin{figure*}[ht!]
	\centering
	\includegraphics[width=\textwidth]{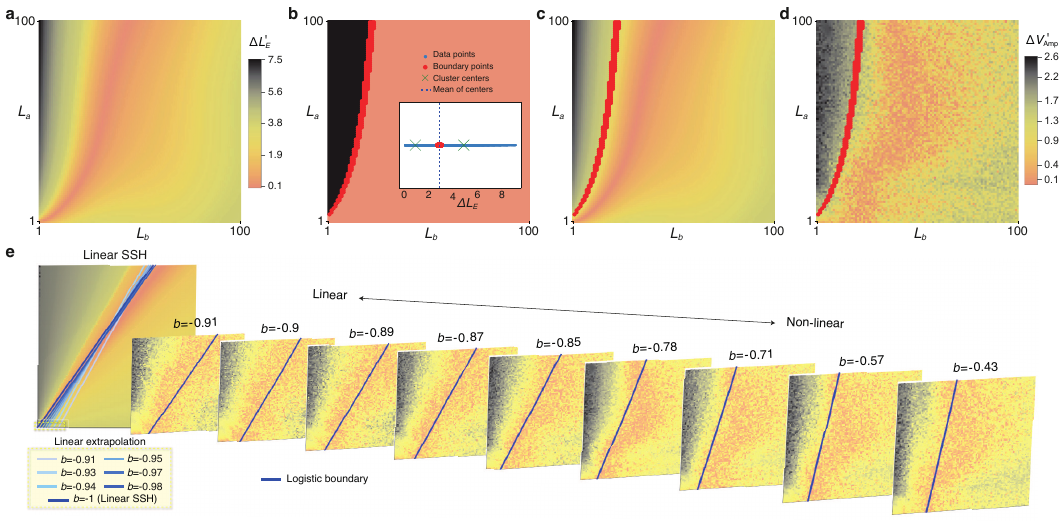}
	\caption{\textbf{Phase boundary detection and evolution through machine learning} \textbf{a} High-pass filtering of $\Delta L_E$ around the corresponding k-means boundary yields $\Delta L'_E$, allowing us to examine phase transitions in real space with enhanced noise resilience. \textbf{b} The quantification of the phase boundary is provided by the $k$-means clustering algorithm, where phase boundaries are defined as equidistant points from the cluster centers. \textbf{c} Combining these results for $\Delta L_E$, the overlap between high-pass filtering and $k$-means clustering reinforces confidence in the identified phase boundaries. \textbf{d} High-pass filtering of $\Delta V_\text{Amp}$ around its k-means boundary yields $\Delta V^\prime_\text{Amp}$ where $b=-0.71$, extracting a noise-resilient phase boundary. Overlaying the $k$-means boundary of $\Delta L_E$ onto this reveals a close alignment, suggesting both phase spaces share the same topological nature. \textbf{e} We use the $k$-means algorithm to effectively identify clusters and apply logistic regression to determine the logistic boundaries, shown as blue lines. As chaotic patterns disappear beyond $b = -0.91$, we apply linear regression using the slope and intercept from the previous nine samples' lines to continue tracking the evolution of the phase boundary. The extrapolated non-linear boundary directly overlaps with the linear SSH boundary, reinforcing the idea that non-linear boundaries are reminiscent of the linear topological boundary.}
	\label{KMeans_high-pass}
\end{figure*}

\subsection{Topological phase characterizations via unsupervised learning} 
Here, we outline how we define the phases in the $\Delta V_\text{Amp}$ and $\Delta L_E$ diagrams of the main text. The non-linear topological phase is defined by sudden changes in voltage amplitude difference, i.e., $\Delta V_\text{Amp}$, and effective inductance difference, i.e., $\Delta L_E$. Beyond studying the qualitative correlation between $\Delta V_\text{Amp}$ and $\Delta L_E$, we further analyze their phase boundaries to investigate this correlation quantitatively. The $k$ means clustering unsupervised learning algorithm and high-pass Fourier filtering signal processing technique are utilized for this investigation. The $k$-means algorithm provides the phase transition regions from the statistical distribution of the quantitative difference in the phase diagrams, while Fourier filtering reduces uncertainty near the phase space edges owing to its improved noise resilience.

Since our non-linear phase diagram is defined by the difference between the bulk and edge magnitudes, the high-frequency components directly define the fast-changing data points. In this context, Fourier filtering, with its strong noise resilience, emerges as the natural choice. This approach is particularly effective for the $\Delta V_\text{Amp}$ phase boundary, given its susceptibility to noise caused by the inherent randomness in its phase space. Utilizing Gaussian high-pass filtering, we remove slow-changing points in the state space by selecting high-frequency components around the identified $k$-means boundaries. Starting from real space, we apply the Fourier transform to access frequency space, then filter out low frequencies using the following mask function:

\begin{equation}
	F = 1 - e^{-\frac{\|\mathbf{u} - \mathbf{u}_0\|^2}{2\sigma^2}},
	\label{MaskFiltering}
\end{equation}
where \( \mathbf{u} \) represents the frequency space vector corresponding to the real-space variables \( (L_a, L_b) \), \( \mathbf{u}_0 \) is the center of the Gaussian function, and \( \sigma \) is its standard deviation. After setting \( \sigma = 0.5 \) and masking frequency space, we apply the inverse Fourier transform to recover real space with the two distinguished phases, as shown for \( \Delta L_E \) in Figure~\ref{KMeans_high-pass}a.

While high-pass filtering offers noise-resilient insights by isolating fast-changing components, $k$-means clustering provides a statistically rigorous method for identifying clusters and their boundaries in the data. It partitions the data into \(k\) clusters by iteratively minimizing the sum of squared distances between data points and their corresponding cluster centroids. This corresponds to solving the optimization problem:
\begin{equation}
	\min_{\{\eta_k\}} \sum_{k=1}^{k=2} \sum_{q\in {\{\mathcal{C}_1, \mathcal{C}_2}\}} \|q - \eta_k\|^2,
	\label{KMeans_high-passEq}
\end{equation}
where $\eta_k$ is the centroid of cluster \(\mathcal{C}_i\) and \(q\) denotes a data point. The topologically non-trivial (\(\mathcal{C}_2\)) and trivial regions (\(\mathcal{C}_1\)) correspond to different cluster types (black and orange regions in Figure~\ref{KMeans_high-pass}b), and the phase boundary is acquired by the equidistant points from the centroids ($\eta_k$), as detailed in the inset of Figure~\ref{KMeans_high-pass}b. 

Using this methodology, we acquire the phase boundary of $\Delta L_E$ presented in Figure~\ref{KMeans_high-pass}b. We showcase $k$-means phase boundary of $\Delta L_E$ determined in Figure~\ref{KMeans_high-pass}b on top of high-pass filtered diagrams of $\Delta L_E$ [Figure~\ref{KMeans_high-pass}c] and $\Delta V_\text{Amp}$ [Figure~\ref{KMeans_high-pass}d]. This combined analysis of $k$-means and high-pass filtering reveals a remarkable similarity between the phase boundaries of $\Delta L_E$ and $\Delta V_\text{Amp}$, indicating that both phase spaces are governed by the same topological nature.

\subsection{Non-linear to linear phase boundary extrapolation}
Chaotic boundary portraits begin to vanish beyond $b=-0.91$ due to the highly linearized $I$-$V$ characteristics. For instance, the partial regular regions emerge as $b$ decreases, as shown in Figure~\ref{figPhasediag}e, particularly in the small $L_a$ and $L_b$ parameter regime. To estimate whether the phase boundaries determined using the methodology in Figure~\ref{KMeans_high-pass}a-d approximate the linear SSH phase boundary beyond $b = -0.91$, we apply cluster identification with the $k$-means algorithm within the region $b < -0.91$, which enables us to track the evolution of phases effectively. By modeling these boundaries linearly for simplicity and extrapolating their behavior to \(b > -0.91\), we find that the extrapolated boundaries in the non-linear regime closely align with those in the linear regime, demonstrating the continuity in the system’s behavior despite the transition.

To acquire the linear boundaries for different $b$ settings, we apply logistic regression to the identified $k$-means clusters. In our setting, logistic regression models the probability of a sample belonging to a phase as:
\begin{equation}
	P(q \in \mathcal{C}_1 | L_a, L_b) = \frac{1}{1 + e^{-(\lambda_{J_0} + \lambda_{J_1} L_a + \lambda_{J_2} L_b)}},
\end{equation}
where $\lambda_{J_0}$, $\lambda_{J_1}$, and $\lambda_{J_2}$ are the model coefficients, and \(q \in \mathcal{C}_1\) corresponds to the sample being part of the topologically trivial cluster. The coefficients $\lambda_{J_0}$, $\lambda_{J_1}$, and $\lambda_{J_2}$ are determined by maximizing the log-likelihood function, given by:
\begin{equation}
	\begin{split}
		\mathscr{L}(\lambda_J) = \sum_{q\in {\{\mathcal{C}_1, \mathcal{C}_2}\}} \Big[ \mathbb{I}(q \in \mathcal{C}_1) \log P(q \in \mathcal{C}_1 | L_{a}, L_{b}) \\
		+ \mathbb{I}(q \in \mathcal{C}_2) \log P(q \in \mathcal{C}_2 | L_{a}, L_{b}) \Big] .
	\end{split}
\end{equation}
Then the decision boundary is found when the probability is 0.5, which corresponds to:
\begin{equation}
	\lambda_{J_0} + \lambda_{J_1} L_a + \lambda_{J_2} L_b = 0 .
\end{equation}
This equation represents a logistic boundary separating the classes in the phase space denoted in Figure~\ref{KMeans_high-pass}e with blue lines. These lines linearly approximate the visual purple boundaries shown in Figure~\ref{figPhasediag}e by using k-means clusters to demonstrate the direct quantitative evolution from linear to non-linear.

As chaotic portraits vanish beyond $b=-0.91$, we use linear regression to perform extrapolation based on the previous nine sample's slope ($b=-0.43$ to $b=-0.91$). The extrapolated logistic boundaries are shown in varying blue tones in the first panel of Figure~\ref{KMeans_high-pass}e. Additionally, to increase the fidelity of the regression across the fully linear SSH, we used the intercept data ranging from $b=-0.85$ to $b=-0.91$, with an interval of $0.05$. See Section IX in the Supporting Information for extraction of extrapolated linear boundaries for the mentioned data points.

This provides a quantitative demonstration of the evolution from the non-linear to the linear regime, with the overlap of the extrapolated boundary and the SSH linear phase boundary indicating they share the same origin.

\section{Acknowledgments}
This work was supported by the Ministry of Education (MOE) Tier-II Grant No. MOE-T2EP50121-0014 (NUS Grant No. A-8000086-01-00) and MOE Tier-II Grant No. MOE-T2EP50222-0003, andMOE Tier-I FRC Grant (NUS Grant No. A-8000195-01-00), and MOE Tier-I FRC Grant (NUS Grant No. A-8002656-00-00). H.S. would like to thank the Agency for Science, Technology and Research (A*STAR) for their support of our research through the SINGA fellowship program.

\subsection{Data and code availability}
The datasets and necessary simulation codes generated and/or analyzed during the current study are available in the Zenodo repository at \url{http://doi.org/10.5281/zenodo.15392779}.

\subsection{Competing interests}
The authors declare no competing interests.

\subsection{Author contributions}
C.H.L. proposed the idea, initiated and supervised the project, and provided extensive revisions to the manuscript. H.S. conducted the initial research, developed the theoretical framework, performed numerical and LTspice simulations, and drafted the manuscript. H.A. implemented the machine learning applications and authored the corresponding sections. M.B.A.J. co-supervised the project. Z.B.S., S.M.R., J.F.K and M.B.A.J. reviewed and contributed to all aspects of the manuscript. All authors analyzed the results. The manuscript reflects the contributions of all authors.

\bibliography{TopoChaos_mainAccepted}

\end{document}

\title{Supporting Information for ``Protected chaos in a topological lattice''}

	\author{Haydar Sahin\,\orcidlink{0000-0002-6361-8161}}
	\email{sahinhaydar@u.nus.edu}
	\affiliation{Department of Electrical and Computer Engineering, National University of Singapore, Singapore 117583, Republic of Singapore}
	\affiliation{Institute of High Performance Computing (IHPC), Agency for Science, Technology and Research (A*STAR), Singapore 138632, Republic of Singapore}

	\author{Hakan Akg\"un\,\orcidlink{0009-0009-3474-1599}}
	\affiliation{Department of Physics, Bilkent University, Ankara 06800, T\"urkiye}
	
	\author{Zhuo Bin Siu\,\orcidlink{0000-0002-7056-937X}}
	\affiliation{Department of Electrical and Computer Engineering, National University of Singapore, Singapore 117583, Republic of Singapore}
	
	\author{S.~M. Rafi-Ul-Islam\,\orcidlink{0000-0002-2275-8889}}
	\affiliation{Department of Electrical and Computer Engineering, National University of Singapore, Singapore 117583, Republic of Singapore}
	
	\author{Jian\nobreak \space Feng Kong\,\orcidlink{0000-0001-5980-4140}}
	\affiliation{Institute of High Performance Computing (IHPC), Agency for Science, Technology and Research (A*STAR), Singapore 138632, Republic of Singapore}
	\affiliation{Quantum Innovation Centre (Q.InC), Agency for Science, Technology and Research (A*STAR), 2 Fusionopolis Way, Innovis \#08-03, Singapore 138634, Republic of Singapore}
	
	\author{Mansoor B.~A. Jalil\,\orcidlink{0000-0002-9513-8680}}
	\affiliation{Department of Electrical and Computer Engineering, National University of Singapore, Singapore 117583, Republic of Singapore}
	
	\author{Ching Hua Lee\,\orcidlink{0000-0003-0690-3238}}
	\email{phylch@nus.edu.sg}
	\affiliation{Department of Physics, National University of Singapore, Singapore 117542, Republic of Singapore}

\maketitle

%\begin{center}
%	\textbf{\Large Supplementary Materials for}\\[6pt]
%	\textbf{\large ``Protected chaos in a topological lattice''}\\[10pt]
	
%	Haydar Sahin$^{1,2,*}$, Hakan Akgün$^{3,\dagger}$, Zhuo Bin Siu$^{1,\ddagger}$, S. M. Rafi-Ul-Islam$^{1,\S}$,\\
%	Jian Feng Kong$^{2,4,\mathparagraph}$, Mansoor B. A. Jalil$^{1,\ast\ast}$, and Ching Hua Lee$^{5,\dagger\dagger}$\\[10pt]
	
%	\textit{\small
%		$^{1}$Department of Electrical and Computer Engineering, National University of Singapore, Singapore 117583, Republic of Singapore\\
%		$^{2}$Institute of High Performance Computing (IHPC), Agency for Science, Technology and Research (A*STAR), Singapore 138632, Republic of Singapore\\
%		$^{3}$Department of Physics, Bilkent University, Ankara 06800, Türkiye\\
%		$^{4}$Quantum Innovation Centre (Q.InC), Agency for Science, Technology and Research (A*STAR), 2 Fusionopolis Way, Innovis \#08-03, Singapore 138634, Republic of Singapore\\
%		$^{5}$Department of Physics, National University of Singapore, Singapore 117542, Republic of Singapore
%	}
%\end{center}

%\vspace{1em}

%\noindent\rule{\textwidth}{0.5pt}

\renewcommand{\theequation}{S\arabic{equation}}
\setcounter{equation}{0}
\renewcommand{\thefigure}{S\arabic{figure}}
\setcounter{figure}{0}	
\setcounter{table}{0}

\onecolumngrid
\tableofcontents

\newpage
\section{Detailed derivations of analytical effective inductance formulae}
We begin by deriving the closed-form analytical expressions for the edge nodes, i.e., $L_E(i=1)$ and $L_E(i=2N)$, which involve continued fractions due to recursive relations. Given that our circuit exhibits reflectional symmetry about node $N$, the effective inductance at both edge nodes is equivalent, i.e., $L_E(1) = L_E(2N)$. Therefore, we initially focus on deriving a closed-form expression for the left edge node under the assumption that the right side of the circuit extends semi-infinitely. The effective inductance for the left edge node, where $i=1$, can be ascertained by substituting Equation~(44) into Equation~(43). This yields
\begin{equation}
	\scriptstyle
	L_E(1) =  \cfrac{1}{\frac{1}{L_c} + \frac{1}{L_b}+ \cfrac{1}{\fcolorbox{red}{white}{$\displaystyle L_{a} + \cfrac{1}{\frac{1}{L_c} + \cfrac{1}{L_{b}  + \cfrac{1}{\frac{1}{L_c} +  \cfrac{1}{L_{a} + \cfrac{1}{\ddots}}}}} $} }}.
	\label{LE1recursive}
\end{equation}
The above equation realizes Equation~(43) of the main text when $m = 0$ and $n = \propto$. The red box represents $L_E^\text{right}$. Since the red box involves a continued fraction ($L_E^\text{right}$) and we are dealing with a nested structure, we can express the two repeated sequences as
\begin{equation}
	A = L_a + \left( \frac{1}{L_c}+\frac{1}{B}\right)^{-1},
\end{equation}
where $A$ is defined in terms of $B$, which means that to solve for $A$, we express $B$ in a form that eventually leads back to $A$ itself as
\begin{equation}
	B = L_b + \left( \frac{1}{L_c}+\frac{1}{A}\right)^{-1}.
\end{equation}
The continued fraction represented by $A$ is, in essence, an infinite nested structure that, due to its recursive nature, can be described by a finite formula. Breaking down the recursive dependency typically involves substituting one expression into the other and solving for the variable of interest. Therefore, we solve the above two relations simultaneously and find them as
\begin{equation}
	A = \frac{1}{2} \left(L_a+\sqrt{\frac{\left(L_a+2 L_c\right) \left(2 L_c \left(L_a+L_b\right)+L_a L_b\right)}{L_b+2 L_c}}\right),
	\label{Aexplicit}
\end{equation}
and 
\begin{equation}
	B = \frac{1}{2} \left(L_b + \sqrt{\frac{\left(L_b+2 L_c\right) \left(2 L_c \left(L_a+L_b\right)+L_a L_b\right)}{L_a+2 L_c}}\right).
	\label{Bexplicit}
\end{equation}
Now, by rewriting Equation~\eqref{LE1recursive} in terms of $A$, 
\begin{equation}
	L_E(1) =  \left( \frac{1}{L_c} + \frac{1}{L_b}+ \frac{1}{A} \right)^{-1},
	\label{LE1closedform2}
\end{equation}
and substituting $A$ into the above equation, we obtain the closed-form expression for the left edge node as
\begin{equation}
	L_E(1) = \left( \frac{1}{L_c}+\frac{1}{L_b} + \frac{2}{L_a + \sqrt{\frac{\left(L_a+2 L_c\right) \left(2 L_c \left(L_a+L_b\right)+L_a L_b\right)}{L_b+2 L_c}}}\right)^{-1}.
	\label{LE1closedform3Supp}
\end{equation}
The above closed-form expression provides an effective inductance for the left edge node, assuming the right side of the circuit is semi-infinite. This expression aligns well with the numerically derived $L_E(1)$, obtained from recursive relations, as can be seen from Figure~11a. It is important to note that $L_E(1) = L_E(2N)$, reflecting the translational symmetry of our circuit. Following a similar approach, we can derive an analytical expression for the second left-edge node (where $i=2$) by considering the right side of node 2 as a semi-infinite circuit. Applying Equation~(43) with $m=1$ and $n=\infty$ leads to the following numerical recursive relation
\begin{equation}
	L_E(2) =  \cfrac{1}{\frac{1}{L_c} + \cfrac{1}{L_a + \cfrac{1}{\frac{1}{L_c}+\frac{1}{L_b}}}+ \cfrac{1}{L_{b} + \cfrac{1}{\frac{1}{L_c} + \cfrac{1}{L_{a}  + \cfrac{1}{\frac{1}{L_c} + \cfrac{1}{\ddots}}}}}}.
	\label{LE2closedform}
\end{equation}
In the equation above, the right fraction has a continued form due to the semi-infinite approximation, through which we can rewrite it as
\begin{equation}
	L_E(2) =  \cfrac{1}{\frac{1}{L_c} + \cfrac{1}{L_a + \cfrac{1}{\frac{1}{L_c}+\frac{1}{L_b}}}+ \cfrac{1}{B}}.
	\label{LE2closedform2}
\end{equation}
We can explicitly obtain $L_E(2)$ by substituting $B$ in the above equation as
\begin{equation}
	\begin{aligned}
		L_E(2) = \left( \frac{1}{L_c}+\frac{1}{L_a+\frac{1}{\frac{1}{L_c}+\frac{1}{L_b}}} +\frac{2}{L_b+\sqrt{\frac{\left(L_b+2 L_c\right) \left(2 L_c \left(L_a+L_b\right)+L_a L_b\right)}{L_a+2 L_c}}} \right)^{-1}.
	\end{aligned}
	\label{LE2closedform3Supp}
\end{equation}
Using the same approximation methods, it is possible to obtain an analytical expression for $L_E(3)$ by initially writing the recursive expression down as

\begin{equation}
		L_E(3) =  \cfrac{1}{\frac{1}{L_c} + \cfrac{1}{L_b+\frac{1}{\frac{1}{L_c}+\frac{1}{L_a+\frac{1}{\frac{1}{L_c}+\frac{1}{L_b}}}}}+ \cfrac{1}{L_{a} + \cfrac{1}{\frac{1}{L_c} + \cfrac{1}{L_{b}  + \cfrac{1}{\frac{1}{L_c} + \cfrac{1}{\ddots}}}}}}.
		\label{LE3closedform}
\end{equation}
We now replace the continued fraction with $A$ and rewrite the above recursive relation as
\begin{equation}
	L_E(3) =  \cfrac{1}{\frac{1}{L_c} + \cfrac{1}{L_b+\frac{1}{\frac{1}{L_c}+\frac{1}{L_a+\frac{1}{\frac{1}{L_c}+\frac{1}{L_b}}}}}+ \cfrac{1}{A}}.
	\label{LE3closedform2}
\end{equation}
The explicit analytical expression for $L_E(3)$ is obtained when $A$ given in Equation~\eqref{Aexplicit} is substituted in Equation~\eqref{LE3closedform2} as

\begin{equation}
	\begin{aligned}
		L_E(3) =  \left(\frac{1}{L_c} + \cfrac{1}{L_b+\frac{1}{\frac{1}{L_c}+\frac{1}{L_a+\frac{1}{\frac{1}{L_c}+\frac{1}{L_b}}}}} +\frac{2}{L_a+\sqrt{\frac{\left(L_a+2 L_c\right) \left(2 L_c \left(L_a+L_b\right)+L_a L_b\right)}{L_b+2 L_c}}} \right)^{-1}.
	\end{aligned}
	\label{LE3closedform3}
\end{equation}
We have derived the closed-form analytical expressions for the first three edge nodes. Since the effective inductance rapidly converges to the same value as the bulk nodes, this is sufficient to obtain the behaviors of the effective inductance at the boundaries.

We now proceed to derive analytical closed-form expressions for the bulk nodes, employing the same methodology. In contrast to the edge nodes, we are considering a bulk node where both sides are considered to be semi-infinitely large. The recursive relation for such a bulk node is given as
\begin{equation}
	L_E(\text{bulk}) =  \cfrac{1}{\frac{1}{L_c} + \cfrac{1}{L_a + \cfrac{1}{\frac{1}{L_c}+\frac{1}{L_b+\frac{1}{\ddots}}}}+ \cfrac{1}{L_b + \cfrac{1}{\frac{1}{L_c}+\frac{1}{L_a + \frac{1}{\ddots}}}}}.
	\label{LEBulkclosedform}
\end{equation}
Here, since both sides of the bulk node are semi-infinitely large, there are two continued fractions in the denominator of $L_E(\text{bulk})$. Therefore, we can rewrite the above recursive relation as
\begin{equation}
	L_E(\text{bulk}) =  \left( \frac{1}{L_c} + \frac{1}{A}+ \frac{1}{B} \right)^{-1}.
	\label{LEBulkclosedform2}
\end{equation}
We obtain the explicit analytical expression for the bulk nodes by substituting $A$ and $B$ into Equation~\eqref{LEBulkclosedform2} and simplifying the result as
\begin{equation}
	\begin{aligned}
		L_E(\text{bulk}) = \frac{L_a L_c \left(L_b+L_c\right)+L_b L_c^2}{\sqrt{\left(L_a+2 L_c\right) \left(L_b+2 L_c\right) \left(2 L_c \left(L_a+L_b\right)+L_a L_b\right)}}.
	\end{aligned}
	\label{LEBulkclosedform3Supp}
\end{equation}
The closed-form expressions we have derived for the first three edge nodes and bulk nodes (i.e., Equations~\eqref{LE1closedform3Supp}, \eqref{LE2closedform3Supp}, \eqref{LE3closedform3}, and \eqref{LEBulkclosedform3Supp}) are sufficient to obtain the effective inductance profile of our circuit.

\section{Monte Carlo-based LTspice simulations of the Chua-SSH circuit under realistic experimental constraints}

\begin{table}[b]
	\centering
	\begin{tabular}{@{}ccccc@{}}
		\toprule
		\textbf{Component} & \textbf{~~~~Value~~~~} & \textbf{~~~~Manufacturer~~~~} & \textbf{~~~Manufacturer No~~~} & \textbf{Tolerance} \\ \midrule
		$C_1$ & 10\,nF & Murata Electronics & GCM188R72A103KA37D & $\pm$10\% \\
		$C_2$ & 100\,nF & Murata Electronics & GCM21BR72A104KA37K & $\pm$10\% \\
		$L_{a,b}$ & 8.2\,mH & Bourns Inc. & 1140-822K-RC & $\pm$10\% \\
		$L_{a,b}$ & 82\,mH & Abracon LLC & ASPI-0403S-822M & $\pm$20\% \\
		$L_{c,1}$ & 22\,mH & Delevan & 5500R-226J & $\pm$5\% \\
		$L_{c,2}$ & 2\,mH & Pulse Electronics & APCI00121280202KY0 & $\pm$10\% \\
		$R$ & 1.84\,k$\Omega$ & Yageo & RT0603BRD071K84L & $\pm$0.1\% \\
		$R_1$ & 20\,k$\Omega$ & Panasonic & ERA-6ARW203V & $\pm$0.05\% \\
		$R_2$ & 22\,k$\Omega$ & Panasonic & ERA-3ARW223V & $\pm$0.05\% \\
		$R_3$ & 2.2\,k$\Omega$ & Panasonic & ERA-3ARW222V & $\pm$0.05\% \\
		$R_4$ & 3.3\,k$\Omega$ & Panasonic & ERA-3VRW3301V & $\pm$0.05\% \\
		$R_5$ & 220\,$\Omega$ & Panasonic & ERA-3AEB221V & $\pm$0.1\% \\
		Op-Amp & N.A. & Analog Devices & LT1351CS8 & N.A. \\
		\bottomrule
	\end{tabular}
	\caption{
		\textbf{Component specifications used in Monte Carlo-based LTspice simulations.} The table lists the nominal values, manufacturers, part numbers, tolerances of the components used in the LTspice simulations. Here, $L_{c,2}$ is an additional inductor connected in parallel with $L_{c,1}$ to achieve an effective inductance of 24\,mH. The selected components are widely available and suitable for experimental implementation. While the nominal tolerances range from 5\% to 20\% depending on the component type, these can be significantly reduced by preselecting components using standard measurement tools such as an LCR meter or impedance analyzer.}
	\label{TableComponents}
\end{table}

To evaluate the experimental feasibility of protected chaos, we incorporate realistic factors such as component tolerances and parasitic resistances into our LTspice simulations. For this purpose, we use commercially available components selected for their low losses and tight tolerances, as listed in Table~\ref{TableComponents}. Among all components, inductors represent the dominant source of loss, and we model them with their specified series resistances. Another important factor in practical implementations is the deviation in component values due to manufacturing tolerances. Although such variations can be reduced by pre-selecting matched components, we systematically explore their impact by performing Monte Carlo tolerance analyses in LTspice across various tolerance levels.

\begin{figure}
	\centering
	\includegraphics[width=0.9\linewidth]{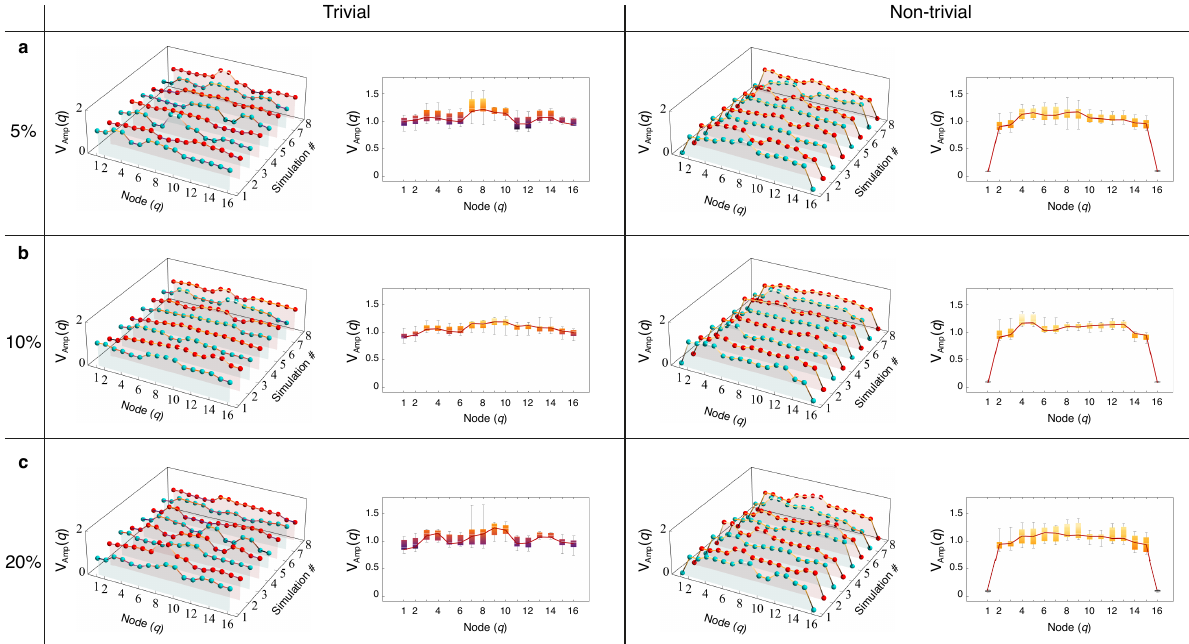}
	\caption{\textbf{Monte Carlo–based LTspice simulations of protected chaos under component tolerances.} Scroll amplitudes $V_\text{Amp}(q)$ are shown for trivial and non-trivial phases across 8 Monte Carlo runs for each tolerance level. Left plots in each panel show individual simulation results; right plots display \textit{box-and-whisker} diagrams representing the variation in scroll amplitudes across nodes.
			\textbf{a} 5\% tolerance applied to all energy-storage components.
			\textbf{b} 10\% tolerance, consistent with typical manufacturer specifications.
			\textbf{c} 20\% tolerance as an upper-bound scenario.
			The protected suppression of edge scrolls in the non-trivial phase remains clearly visible across all cases, indicating strong robustness. Parameters used: $N=8$, $L_c=24\,\text{mH}$, $C_1=10\,\text{nF}$, $C_2=100\,\text{nF}$, $R=1.85\,\text{k}\Omega$, and $b=-0.71$.}
	\label{figrealSim}
\end{figure}

To perform the Monte Carlo worst-case circuit analysis in LTspice, we replace the component values with the command $x \rightarrow {\text{mc}(x,tol)}$, where $x$ is the nominal value of the component and \textit{tol} denotes the maximum deviation from this nominal value. All other properties from the datasheet are kept unchanged. The function ${\text{mc}(x,\text{tol})}$ assigns a randomly generated tolerance value to each component for each simulation run. In our analysis, we perform at least eight independent simulations for each case to ensure adequate sampling of possible variations within the specified tolerance window.

The simulations are based on the parameter set extensively studied in the main text: $N=8$, $L_c = 24\,\text{mH}$, $C_1 = 10\,\text{nF}$, $C_2 = 100\,\text{nF}$, $R = 1.85\,\text{k}\Omega$, and $b = -0.71$. For the trivial and topological phases, we use $(L_a, L_b) = (8.2\,\text{mH}, 82\,\text{mH})$ and $(L_a, L_b) = (82\,\text{mH}, 8.2\,\text{mH})$, respectively. In these simulations, we apply a fixed 5\% tolerance to $C_1$ and $C_2$, and vary the tolerance for the inductors $L_a$, $L_b$, and $L_c$ using 5\%, 10\%, and 20\% deviation windows. The resistors are not perturbed in these runs, as their specified tolerances are already below 0.1\%, as listed in Table~\ref{TableComponents}. Wire resistances are also included by setting the series resistances to the inductors as follows: $25\,\text{m}\Omega$ for the $24\,\text{mH}$ inductor, $75\,\text{m}\Omega$ for the $8.2\,\text{mH}$ inductor, and $150\,\text{m}\Omega$ for the $82\,\text{mH}$ inductor. This setting effectively reflects an experimental circuit setup.

\begin{figure}
	\centering
	\includegraphics[width=0.95\linewidth]{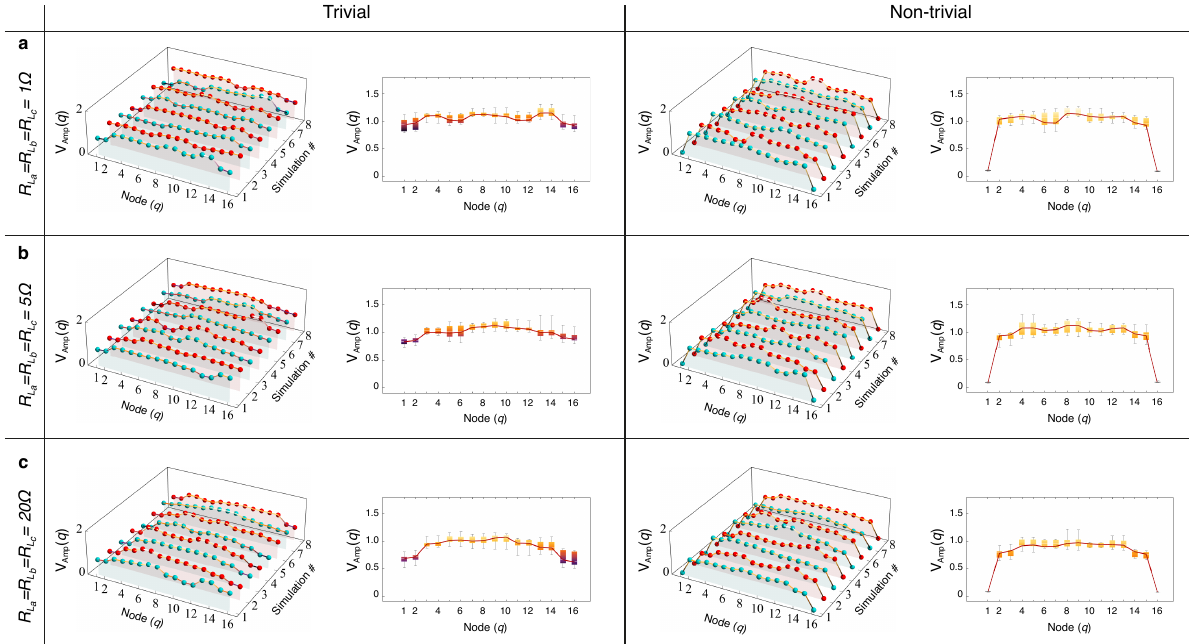}
	\caption{\textbf{Monte Carlo–based LTspice simulations of protected chaos under increasing losses.} 
			Scroll amplitudes $V_\text{Amp}(q)$ are shown for simulations including fixed tolerances and increasing series resistance in all inductors. Left plots show individual simulation results; right plots show node-wise box-and-whisker distributions.
			\textbf{a} Series resistance set to $1\,\Omega$.
			\textbf{b} Increased to $5\,\Omega$.
			\textbf{c} Increased further to $20\,\Omega$.
			The qualitative features of protected chaos, including edge-mode suppression in the non-trivial phase, persist across all resistance values. Parameters used: $C_1\pm2\%$, $C_2\pm2\%$, $L_a$, $L_b$, $L_c\pm5\%$, and other values same as in Figure~\ref{figrealSim}.}
	\label{figrealSimResistance}
\end{figure}

We first consider a scenario in which all energy-storage components---$C_1$, $C_2$, $L_a$, $L_b$, and $L_c$---are assigned a maximum deviation of 5\%. In practice, such deviations can be minimized by preselecting components using measurement tools such as an LCR meter or an impedance analyzer. Each simulation is run for a sufficiently long duration (a few hundred milliseconds), and we record the voltage oscillations at the left nodes of each Chua circuit, denoted by $V(q;l)$, which determine the scroll amplitudes. In experiments, this voltage amplitude can be directly accessed by probing the left node of a Chua circuit with an oscilloscope, where the peak-to-peak amplitude can be extracted using standard oscilloscope functions.

We repeat the simulations eight times and plot the scroll amplitudes, $V_\text{Amp}(q)$, at each node $q$ for all runs. The results for both the trivial and non-trivial phases are shown in the left column of each panel in Figure~\ref{figrealSim}. In the corresponding right column of each panel, we present the node voltage amplitude distributions using \textit{box-and-whisker} plots. Each \textit{whisker} indicates the minimum and maximum $V_\text{Amp}(q)$ values observed across all eight simulations, while each \textit{box} represents the median value. For visual clarity, we also include a red curve connecting the medians across nodes to highlight the overall scroll amplitude distribution trend.

In Figure~\ref{figrealSim}, we present the simulation results of the Chua-SSH circuit under 5\%, 10\%, and 20\% tolerance levels. Figure~\ref{figrealSim}a corresponds to simulations with 5\% deviations, representing experiments with pre-selected components. Figure~\ref{figrealSim}b shows results for 10\% tolerances, which reflect the typical tolerance window specified by the manufacturer. In Figure~\ref{figrealSim}c, we examine an extreme case in which the inductor values vary by up to 20\% from their nominal values. Remarkably, even under 20\% tolerance, the edge scrolls in the non-trivial phase remain nearly unaffected. For instance, while bulk scroll amplitudes may vary by as much as 0.25\,V in some cases, the edge scrolls show only minimal deviation. This result highlights the robustness of the protected chaotic state, demonstrating its resilience even in the presence of significant parameter uncertainties.

We now examine the Chua-SSH circuit under large loss due to the components, wire and contact resistances. By setting the tolerance window constant as $C_1\pm2\%$, $C_2\pm2\%$, $L_a\pm5\%$, $L_b\pm5\%$, and $L_c\pm5\%$, which is easily achievable for an experimental circuit, we first increase the series resistance of the inductors as $R_{L_a}=R_{L_b}=R_{L_c}=1\,\Omega$. We observe the qualitative behavior is mostly preserved, as seen in Figure~\ref{figrealSimResistance}a. Increasing the series resistances to $R_{L_a}=R_{L_b}=R_{L_c}=5\,\Omega$, in Figure~\ref{figrealSimResistance}b, we observe slightly more suppressed voltages, however, importantly, the edge scrolls are still distinct in the non-trivial phase. To further examine our circuit with large resistive parasitics, we set $R_{L_a}=R_{L_b}=R_{L_c}=20\,\Omega$ and observe the same qualitative behavior in Figure~\ref{figrealSimResistance}c.

We now examine the Chua-SSH circuit under conditions of increased loss due to component parasitics, wire resistances, and contact resistances. Keeping the tolerance window fixed at $C_1\pm2\%$, $C_2\pm2\%$, $L_a\pm5\%$, $L_b\pm5\%$, and $L_c\pm5\%$---a range that is easily achievable in practical experimental setups---we vary the series resistances of the inductors to simulate different loss scenarios. First, we set the inductor series resistances to $R_{L_a} = R_{L_b} = R_{L_c} = 1\,\Omega$, and observe that the qualitative behavior of the scroll amplitudes remains largely preserved, as shown in Figure~\ref{figrealSimResistance}a. Increasing the resistance further to $5\,\Omega$ in Figure~\ref{figrealSimResistance}b results in slightly more suppressed voltage amplitudes; however, the edge scrolls in the non-trivial phase remain clearly distinguishable. Finally, when the resistance is increased to $20\,\Omega$ (Figure~\ref{figrealSimResistance}c), the circuit still maintains the same qualitative features, with the protected edge dynamics remaining robust despite the significant parasitic losses.

Although these resistance values exceed the parasitic resistances specified by the manufacturers of the components used in our realistic simulations, the qualitative behavior of the circuit remains robust. In fact, the deviations shown in the right-hand plots of each panel in Figure~\ref{figrealSimResistance} become even smaller as the series resistance increases. Overall, our simulations---which incorporate all relevant real-world experimental effects---demonstrate that protected chaos is highly feasible in practical implementations, with the key features remaining well-preserved even under large tolerance windows and significant parasitic losses.

\section{The effect of structural disorder}

\begin{figure}
	\centering
	\includegraphics[width=14cm]{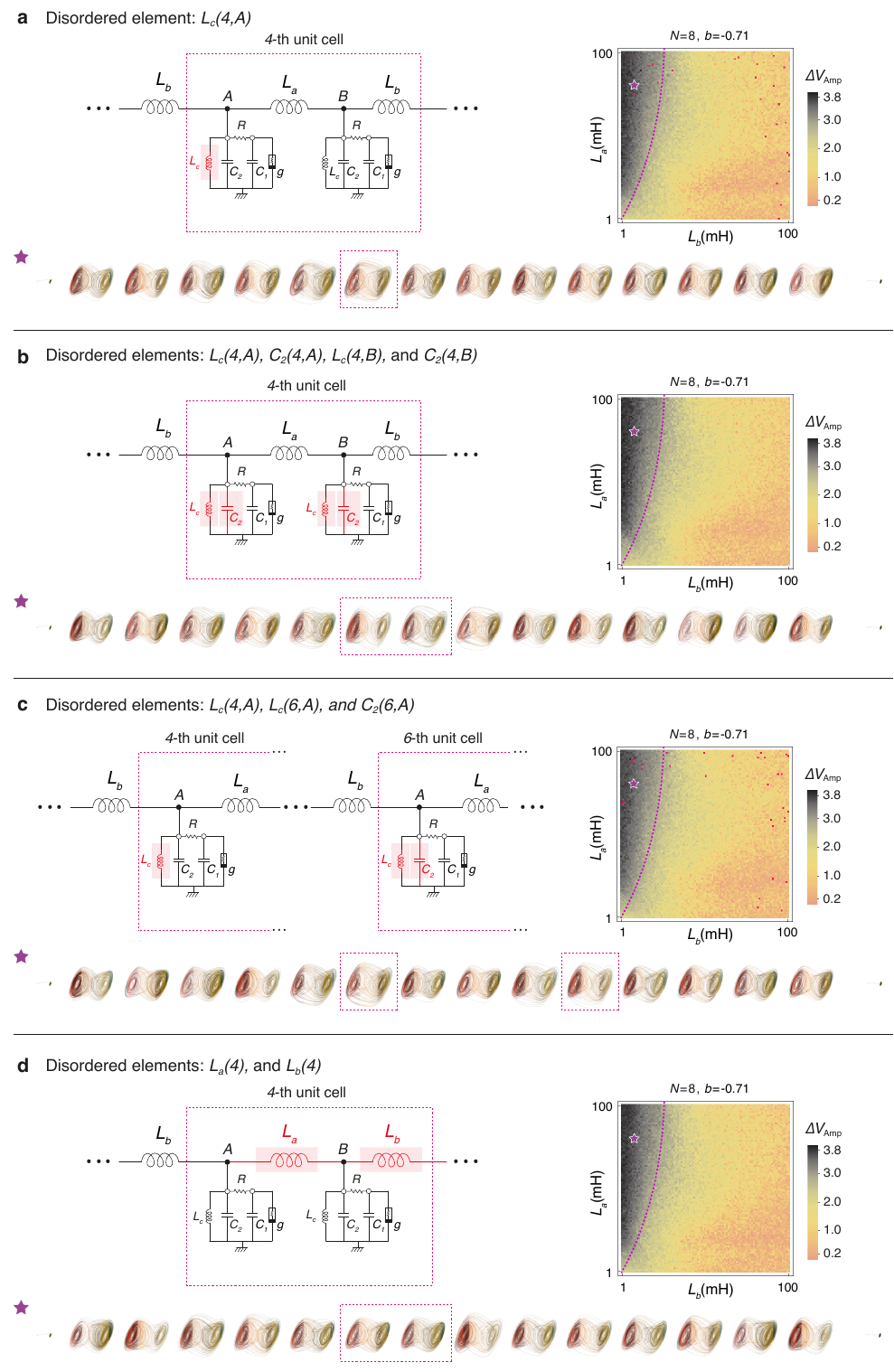}
	\caption{\textbf{Disordered Chua-SSH circuit and chaotic dynamics.} \textbf{a-d} Various scenarios demonstrating structural disorders introduced into the bulk Chua circuits and topological couplings. The altered/disordered elements are highlighted in red within the dashed boxes. Corresponding phase diagrams (right column) illustrate the amplitude difference between bulk and edge scrolls ($\Delta V_{\text{amp}}$). Magenta stars indicate the parameters $(L_a,L_b)=(80\,\text{mH},8\,\text{mH})$ selected for visualization of scroll profiles (bottom rows). The magenta dashed lines in the phase diagrams are reference phase boundaries derived from disorder-free simulations, indicating minimal impact from structural disorders. The parameters used throughout are fixed at $(L_c,C_1,C_2,R,b)=(24\,\text{mH},10\,\text{nF},100\,\text{nF},1.85\,\text{k}\Omega,-0.71)$.}
	\label{figDisorderSuppMat}
\end{figure}

To investigate the robustness of chaotic dynamics induced by topology against structural disorders, we deliberately introduce disorders into bulk Chua circuits and topological coupling elements. As demonstrated in Figure~1d of the main text, chaotic trajectories transition to limit cycles when the inductance ($L_c$) of an isolated Chua circuit exceeds $26.06\,\text{mH}$, with all other parameters fixed as in the main text. In the following scenarios, we analyze how similar perturbations influence the collective dynamics in an $N=8$ Chua-SSH lattice.

First, we introduce a localized structural disorder by altering the inductance $L_c$ of a single Chua circuit at sublattice node $A$ of the 4th unit cell from the default $24\,\text{mH}$ to $30\,\text{mH}$, leaving all other parameters unchanged (Figure~\ref{figDisorderSuppMat}a). Remarkably, this localized perturbation has negligible impact on the scroll profiles and the overall non-linear phase boundary, as illustrated visually and through the phase diagram simulation.

Next, we introduce stronger disorder by simultaneously altering both the inductors $L_c$ and capacitors $C_2$ within the two Chua circuits attached to sublattice nodes $A$ and $B$ of the 4th unit cell. Specifically, we set $L_c(4,A)=27\,\text{mH}$, $C_2(4,A)=80\,\text{nF}$, $L_c(4,B)=30\,\text{mH}$, and $C_2(4,B)=90\,\text{nF}$ (Figure~\ref{figDisorderSuppMat}b). While such deviations are significant enough to alter chaotic trajectories in isolated Chua circuits, the topological bulk effectively suppresses these strong disorders. The visual scroll profiles (taken at parameters corresponding to the magenta star in the associated phase diagram) remain robust, and the non-linear phase boundary, indicated by the magenta dashed line inherited from the disorder-free lattice, remains largely preserved with only minor expansion toward smaller parameter regimes.

In a third scenario, we examine multiple structural disorders at two distinct unit cells by setting $L_c(4,A)=30\,\text{mH}$, $L_c(6,A)=28\,\text{mH}$, and $C_2(6,A)=90\,\text{nF}$ (Figure~\ref{figDisorderSuppMat}c). Although the scroll amplitudes at the disordered nodes slightly increase compared to the other bulk nodes, the overall non-linear phase boundary remains almost identical to that of the disorder-free system.

Finally, we assess the impact of disorders in the topological couplings at the 4th unit cell. We perturb the coupling inductances by setting $L_a(4) = L_a + L_a/20$ and $L_b(4) = L_b - L_b/20$. Remarkably, the phase boundary remains nearly unaffected throughout the entire parameter space. Additionally, for the visualization of the scroll distributions, we consider an extreme case by exchanging coupling inductances, replacing $L_a(4)$ with $L_b$ and $L_b(4)$ with $L_a$, effectively breaking the periodicity of the bulk couplings. Despite this drastic perturbation, as illustrated in Figure~\ref{figDisorderSuppMat}d, the periodicity and character of chaotic scroll profiles remain essentially unaffected.

Collectively, these four distinct disorder scenarios strongly suggest that our Chua-SSH lattice exhibits exceptional robustness against structural disorders. This inherent robustness ensures that protected chaos is readily observable experimentally, providing a promising pathway toward reliable experimental realization.

\newpage
\section{Chaotic phase transitions of an isolated Chua circuit}

\begin{figure}[h]
	\centering
	\includegraphics[width=1\linewidth]{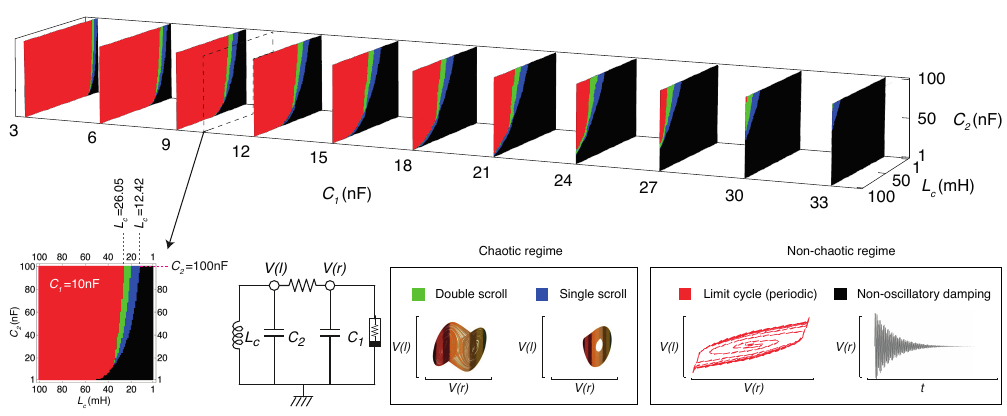}
	\caption{\textbf{Parameter space analysis of chaotic transitions in an isolated Chua circuit.} Systematic classification of dynamical behaviors showing how variations in the energy-storage components ($C_1$, $C_2$, and $L_c$) trigger transitions between chaotic and non-chaotic regimes. Four characteristic regimes identified numerically are indicated by colors: double-scroll chaos (green), single-scroll chaos (blue), periodic limit-cycle (red), and non-oscillatory damping (black). The lower-left inset shows a detailed cross-section for the fixed default capacitance ($C_1=10\,\text{nF}$), clearly marking the chaotic region boundaries at $L_c=12.42\,\text{mH}$ and $L_c=26.05\,\text{mH}$. These boundaries correspond precisely to the chaotic regime limits discussed in Figure~1d. The other parameters used are $(R,b)=(1.85\,\text{k}\Omega,-0.71)$.}
	\label{figSuppParamDiag}
\end{figure}

As topological and chaotic phases in our study are investigated primarily based on variations in the inductances of intra-cell and inter-cell inductors, we previously illustrated chaotic phase transitions by varying the inductance $L_c$ in an isolated Chua circuit while keeping other parameters fixed (Figure~1d). Here, we expand this analysis by systematically examining parametric transitions of chaotic scroll behaviors across all three passive energy storage elements ($L_c$, $C_1$, and $C_2$) of an isolated Chua circuit.

To comprehensively classify chaotic and non-chaotic dynamical regimes, we define four characteristic dynamical behaviors: double-scroll chaos, single-scroll chaos, periodic limit-cycle oscillations, and non-oscillatory damping. By numerically simulating an isolated Chua circuit across a wide parameter space spanned by $L_c$, $C_1$, and $C_2$, we identify these four distinct behavior. In the resulting parameter diagrams (Figure~\ref{figSuppParamDiag}), the double-scroll regime is marked green, the single-scroll regime is marked blue, the periodic limit-cycle regime is red, and the non-oscillatory damping regime is black.

As shown in Figure~\ref{figSuppParamDiag} in the main text, chaotic oscillations---either single scroll (blue) or double scroll (green)---occur between non-oscillatory (black) and periodic limit-cycle (red) regions. Although the parameter space supporting chaotic behavior appears relatively narrow, suitable parameter configurations for robust chaotic dynamics can still be clearly found. For instance, the analysis presented in Figure~1 demonstrated scroll transitions at fixed capacitor values ($C_1=10\,\text{nF}$, and $C_2=100\,\text{nF}$) while varying the inductance $L_c$. There, the chaotic regime was bounded $L_c=12.42\,\text{mH}$ (lower bound) and $L_c=26.05\,\text{mH}$ (upper bound). These boundaries correspond exactly to the cross-sectional diagram at $C_1=10\,\text{nF}$ depicted in the lower-left inset of Figure~\ref{figSuppParamDiag}. Thus, one can readily determine alternative parameter settings to realize protected chaos experimentally.

\subsection{Triggering chaos through circuit parameters}
To elucidate the numerical phase diagrams shown in Figure~\ref{figSuppParamDiag}, we provide a compact analytical explanation of how variations in the passive elements---$L_c$, $C_1$, and $C_2$---lead to different dynamical behaviors in a single Chua circuit. Our aim here is to explain the sequence of transitions observed numerically. We consider the linearized dynamics around an equilibrium point $(x_e, 0, -x_e)$, which yields the Jacobian~\cite{harko_jacobi_2015}
\begin{equation}
J = \begin{pmatrix}
	-\alpha(1 + m) & \alpha & 0 \\
	1 & -1 & 1 \\
	0 & -\beta & 0
\end{pmatrix}, \qquad
\alpha = \frac{C_2}{C_1}, \quad \beta = \frac{R^2 C_2}{L_c}, \quad m = f_b'(x_e) \in \{a, b\}.
\end{equation}
Here, $m$ corresponds to the local slope of the piecewise-linear Chua diode, which switches between $a$ and $b$ depending on whether the state is inside or outside the breakpoint region $|x_e| < E_p$. The eigenvalues of this matrix are determined by the characteristic polynomial
\begin{equation}
\lambda^3 + A \lambda^2 + B \lambda + \Gamma = 0,
\end{equation}
with
\begin{equation}
A = \alpha(1 + m) + 1, \quad B = \alpha(1 + m) + \alpha\beta, \quad \Gamma = \alpha\beta(1 + m).
\end{equation}
The stability of the equilibrium is then governed by the Routh–Hurwitz conditions~\cite{al-azzawi_stability_2012,chu_novel_2018,belouerghi_routh_2023,fowler_complex_1982}. When $A > 0$, $\Gamma > 0$, and $AB - \Gamma > 0$ are all satisfied, the fixed point is stable and all trajectories decay without oscillation, which corresponds to the black region in Figure~\ref{figSuppParamDiag}. The Hopf bifurcation occurs at the threshold $AB - \Gamma = 0$, where two complex-conjugate eigenvalues cross the imaginary axis. This marks the onset of a small-amplitude oscillation around the fixed point. As $\alpha$ or $\beta$ is increased further—for example, by increasing $C_2$ or $L_c$—the amplitude of this limit cycle grows and eventually reaches the nonlinear region of the diode. Once the orbit touches one of the breakpoints, the dynamics become asymmetric and start folding back on themselves, leading to a single-scroll attractor.
A further increase in parameters allows the orbit to span both breakpoints within each cycle, and the system transitions into a double-scroll attractor. This process corresponds to the emergence of more complex dynamics and marks the chaotic regime. The discriminant of the cubic equation,
\begin{equation}
\Delta = 18AB\Gamma - 4A^3\Gamma + A^2B^2 - 4B^3 - 27\Gamma^2,
\end{equation}
provides further insight. When $\Delta < 0$, the eigenvalue structure always includes one real and two complex-conjugate roots, which is necessary for oscillatory behavior. Thus, the region bounded by $\Delta < 0$ and $AB - \Gamma < 0$ supports the growth of oscillations and, ultimately, chaos.
In short, increasing $C_2$ or $L_c$, or decreasing $C_1$, increases the energy stored in the system relative to dissipation. This change first destabilizes the fixed point via a Hopf bifurcation, then enlarges the orbit into the nonlinear regime, where the piecewise structure of the Chua diode generates either single- or double-scroll chaos. These steps follow the color-coded transitions shown in Figure~\ref{figSuppParamDiag} and provide a basic yet useful explanation of how chaos emerges in this circuit by tuning passive components.

\section{Relating the scroll transitions with effective inductance}

\begin{figure}
	\centering
	\includegraphics[width=0.95\textwidth]{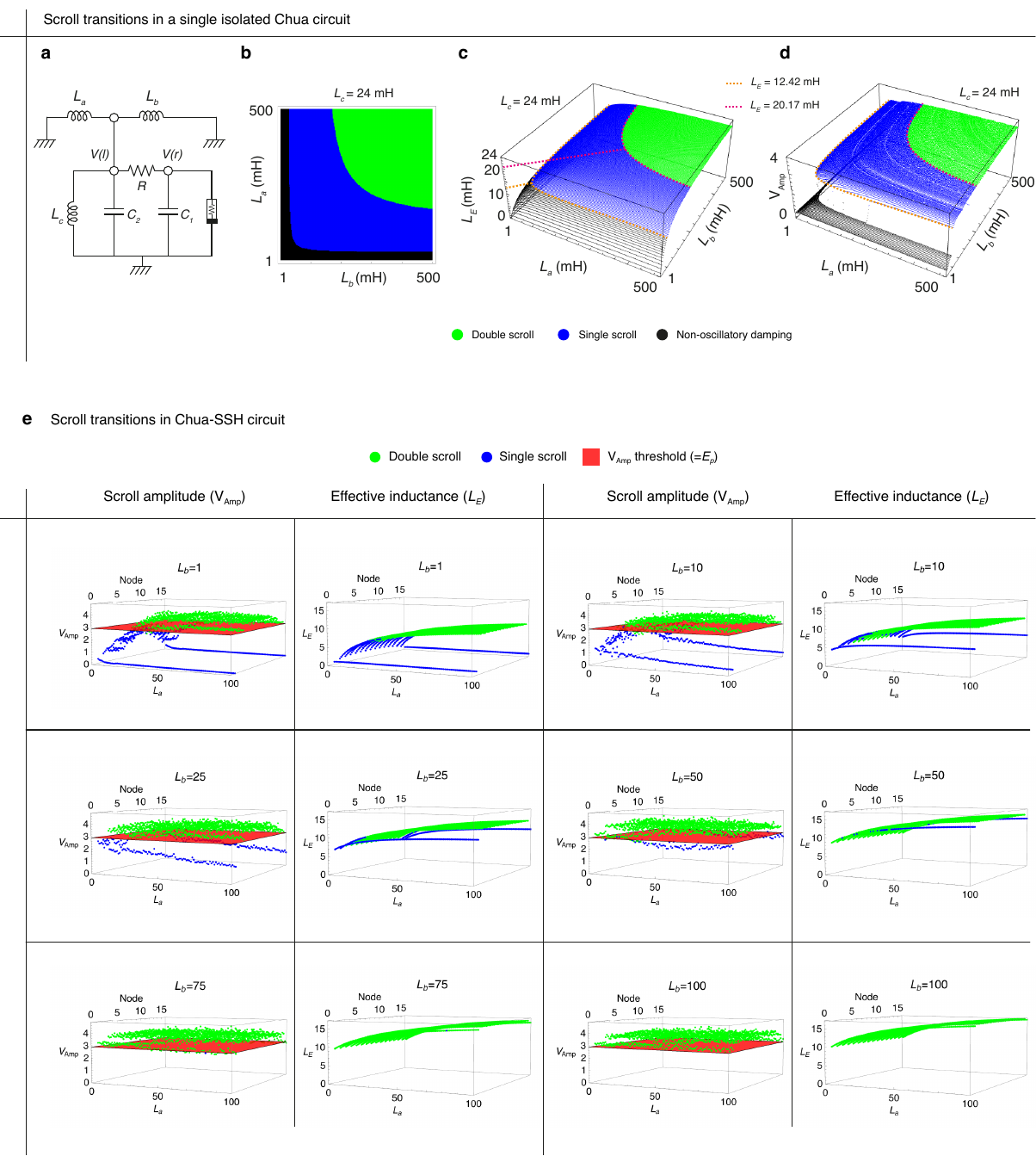}
	\caption{\textbf{Scroll transitions in an isolated Chua circuit and the Chua-SSH array.} 
			\textbf{a} Circuit diagram of the Chua circuit with three inductors $L_a$, $L_b$, and $L_c$ in parallel at the left node $V(l)$ and a single inductor $L_c$ at the right node $V(r)$.
			\textbf{b} Scroll state diagram showing the scroll type as a function of $L_a$ and $L_b$ for a single isolated Chua circuit with fixed $L_c = 24\,\text{mH}$. Blue and green regions respectively indicate single- and double-scroll dynamics. Black region is the non-oscillatory damping regime.
			\textbf{c} Scroll-type transition given in `b' explicitly plotted as a function of the effective node inductance $L_E$. The transitions from non-oscillatory to single-scroll (orange dashed line) and single-scroll to double-scroll (magenta dashed line) occurs precisely at $L_E=12.42\,\text{mH}$ and $L_E=20.17\,\text{mH}$, identical to the single-inductor scenario ($L_c$ only) demonstrated in Figure~1d.
			\textbf{d} Corresponding voltage amplitude $V_\text{Amp}$ illustrating the strong correlation between the effective inductance and node voltage amplitude. 
			\textbf{e} Scroll transitions in a Chua-SSH circuit with $N=8$, evaluated for six different values of $L_b$ ($1$ to $100\,\text{mH}$). For each case, we show the scroll amplitude $V_\text{Amp}$ (left) and the corresponding effective inductance $L_E$ (right) across the nodes for varying $L_a$. Scroll type is color-coded (blue for single-scroll, green for double-scroll). The red transparent plane indicates the threshold amplitude $V_\text{Amp} = E_p = 3\,\text{V}$, which separates the two scroll regimes. A clear correlation between the spatial profile of $V_\text{Amp}$ and $L_E$ emerges, reflecting the impact of effective inductance on local energy confinement and scroll behavior.
			The parameters used throughout are $(L_c,C_1,C_2,R,b)=(24\,\text{mH},10\,\text{nF},100\,\text{nF},1.85\,\text{k}\Omega,-0.71)$.}
	\label{figVampLESuppMat}
\end{figure}

In an isolated Chua circuit, the voltage oscillations at the two nodes are mutually coupled, such that the evolution of one influences the other. This interplay enables an approximate characterization of the scroll type based on the amplitude of voltage oscillations at one of the nodes. In particular, under a reasonable approximation, the scroll type can be inferred from the voltage amplitude at the left node. If we recall our non-linear function (Equation~(2) in the main text):
\begin{equation}
	f_b(x) = bx +\frac{1}{2}(a-b)(|x+E_p|-|x-E_p|),
	\label{Eqfb}
\end{equation}
where $E_p$ is the breakpoint voltage of the Chua diode, which determines the transition point between the inner and outer slopes in the piecewise-linear $I$–$V$ characteristic. As shown in Figure~1c and e, the slope of the nonlinear response changes at these transition points $\pm E_p$. A necessary condition for double-scroll chaotic behavior is that the voltage oscillations at the right node span both nonlinear regions, i.e., $\max|V(r)| > E_p > \min|V(r)|$, so that the system experiences alternating segments of the $I$–$V$ characteristic. In this work, we consistently set $E_p = 3\,\text{V}$ and use a dimensionless outer slope of $a = -1.14$. Once the amplitude of $V(l)$ is sufficiently large, it drives the right node oscillations beyond the negative breakpoint, enabling a transition from single-scroll to double-scroll dynamics. Remarkably, we find that a sufficient condition for the onset of double-scroll behavior is that $V(l)$ exceeds $E_p$. This condition allows us to infer the scroll type from the left-node voltage amplitude, whose qualitative profile reflects the effective inductance at the node.

\subsection{Scroll transitions in a single isolated Chua circuit}
To elaborate on how the effective inductance influences the scroll type, we analyze a single isolated Chua circuit in which two additional inductors, representing the intra-cell and inter-cell inductors $L_a$ and $L_b$, are connected between the left node and ground, as illustrated in Figure~\ref{figVampLESuppMat}a. Although three inductors are connected in parallel at the left node, the scroll transition still occurs at the same effective inductance. As previously shown in the main text (Figure~1), the transition from single-scroll to double-scroll occurs at $L_c = 20.17\,\text{mH}$ when only one inductor is present. In Figure~\ref{figVampLESuppMat}b, we plot the scroll types while varying $L_a$ and $L_b$ independently, keeping $L_c = 24\,\text{mH}$ fixed, which we adopt as the default value throughout the main text. For small values of both $L_a$ and $L_b$, the circuit remains damped without sustained oscillations. As either $L_a$ or $L_b$ increases, the system first exhibits single-scroll chaos (blue region), followed by a transition to double-scroll behavior (green region) when one of the inductors exceeds approximately $200\,\text{mH}$.

The effective inductance at the left node, denoted by $L_E$, is simply given by 
\begin{equation}
	L_E = \left(\frac{1}{L_a}+\frac{1}{L_b}+\frac{1}{L_c}\right)^{-1}.
	\label{LEsupp}
\end{equation}
In Figure~\ref{figVampLESuppMat}c, we plot the scroll transition diagram as a function of $L_E$. The blue region, corresponding to single-scroll oscillations, transitions to the green region, corresponding to double-scroll oscillations, precisely when the effective inductance reaches $L_E = 20.17\,\text{mH}$. This is exactly the same inductance at which the scroll transition occurs when only $L_c$ is present. This result demonstrates that the scroll type is determined by the effective inductance at the node.

Remarkably, as the effective inductance increases, the voltage amplitude at the left node also increases, and there exists a well-defined $V_\text{Amp}$ threshold beyond which the system transitions from single-scroll to double-scroll, as shown in Figure~\ref{figVampLESuppMat}d. The qualitative correlation between these two quantities allows us to evaluate both the scroll amplitude and scroll type based on the effective inductance.

\subsection{Scroll transitions in our Chua-SSH circuit}

We now consider an array of Chua circuits and investigate the scroll types by evaluating $V_\text{Amp}$. In Figure~\ref{figVampLESuppMat}e, we display plots of $V_\text{Amp}$ alongside the corresponding effective node inductance $L_E$. These 3D plots share axes representing the intra-cell inductance $L_a$ and the node number. We fix $L_c = 24\,\text{mH}$ and present results for six different values of $L_b$. To determine the scroll types in our simulations, we utilize the Poincaré section method, coloring each data point blue to indicate a single-scroll portrait or green for a double-scroll portrait.

Notably, the scrolls remain single-scroll state below a threshold of $V_\text{Amp} = 3\,\text{V}$, corresponding precisely to the breakpoint voltage $E_p$, set to 3\,V by default. For visual clarity, we include a red reference plane at $3\,\text{V}$ in each voltage amplitude plot. As exemplified in Figure~\ref{figVampLESuppMat}e for cases where $L_b < 75\,\text{mH}$, scrolls below this threshold consistently exhibit single-scroll behavior, while amplitudes exceeding $3\,\text{V}$ transition into double-scroll states. Note that this threshold is inherently tied to the breakpoint voltage, and would thus shift accordingly if a different breakpoint voltage were employed.

Interestingly, the spatial profile of scroll amplitudes demonstrates qualitative agreement with the corresponding effective node inductance profile. This correlation allows us to infer general trends, including the evolution of bulk-edge scroll behavior. Physically, this correlation arises because higher effective inductances typically reduce coupling strength between adjacent nodes, leading to enhanced local energy confinement and thus increased amplitude at the node voltage $V(l)$. Consequently, one can reasonably estimate the scroll amplitude distribution by examining the effective inductance profile.

\section{Visual illustrations of scroll profiles across varying lattice sizes}
\begin{figure}
	\centering
	\includegraphics[width=0.95\linewidth]{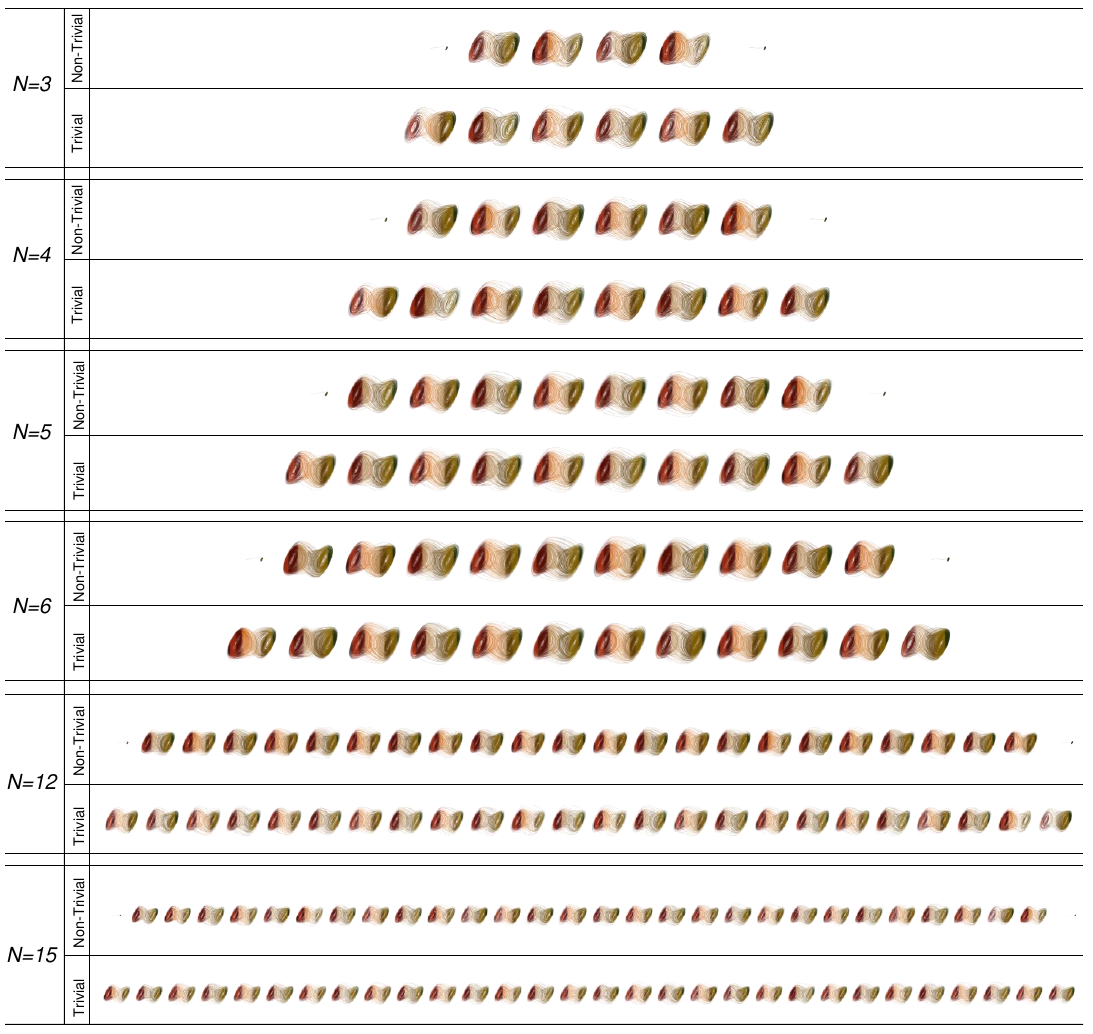}
	\caption{\textbf{Scroll profiles illustrating the effect of lattice size on chaotic dynamics for non-trivial and trivial phases.} Scroll dynamics for varying lattice sizes $N=(3,4,5,6,12,15)$ demonstrate topological edge suppression, even in smaller lattices ($N=3,4$), within the non-trivial phase. This indicates that edge suppression is robust and consistent, confirming the universality of topologically protected chaos across different system sizes. The parameters used here are $(L_c,C_1,C_2,R,b)=(24\,\text{mH},10\,\text{nF},100\,\text{nF},1.85\,\text{k}\Omega,-0.71)$, with inductance values set at $(L_a,L_b)=(80\,\text{mH},8\,\text{mH})$ for the non-trivial phase; these values are switched for the trivial phase.}
	\label{figSizeSuppMat}
\end{figure}

Here, we provide visual illustrations of scroll profiles corresponding to the phase diagrams presented in Figure~12 of the main text. We examine both non-trivial and trivial phases, specifically choosing parameters $(L_a,L_b)=(80\,\text{mH}, 8,\text{mH})$ for the non-trivial phase and $(L_a,L_b)=(8\,\text{mH}, 80\,\text{mH})$ for the trivial phase. As shown in Figure~\ref{figSizeSuppMat}, topological suppression of edge scrolls remains prominent even for very small lattices comprising only 3 or 4 unit cells, particularly when the ratio of inductances clearly distinguishes the non-trivial from the trivial phase.

\newpage
\section{Linear Discriminant Analysis (LDA) and Support Vector Machines (SVMs)}

LDA operates under the assumption that classes are linearly separable and that each class follows a multivariate normal distribution with identical covariance matrices. Mathematically, LDA seeks to find a projection vector $\mathbf{w}$ that maximizes the ratio of between-class variance to within-class variance:

\begin{equation}
	J(\mathbf{w}) = \frac{\mathbf{w}^\top \mathbf{S}_B \mathbf{w}}{\mathbf{w}^\top \mathbf{S}_W \mathbf{w}},
	\label{eq:lda_objective}
\end{equation}
where:
\begin{align*}
	\mathbf{S}_B &= \sum_{i=1}^{C} N_i (\boldsymbol{\mu}_i - \boldsymbol{\mu})(\boldsymbol{\mu}_i - \boldsymbol{\mu})^\top \quad \text{(Between-class scatter matrix)}, \\
	\mathbf{S}_W &= \sum_{i=1}^{C} \sum_{x \in \mathcal{C}_i} (x - \boldsymbol{\mu}_i)(x - \boldsymbol{\mu}_i)^\top \quad \text{(Within-class scatter matrix)}.
\end{align*}
Here, $C$ represents the number of classes, $N_i$ is the number of samples in class $i$, $\boldsymbol{\mu}_i$ is the mean vector of class $i$, and $\boldsymbol{\mu}$ is the overall mean vector of all classes.

\begin{figure}[b]
	\centering
	\includegraphics[width=1\linewidth]{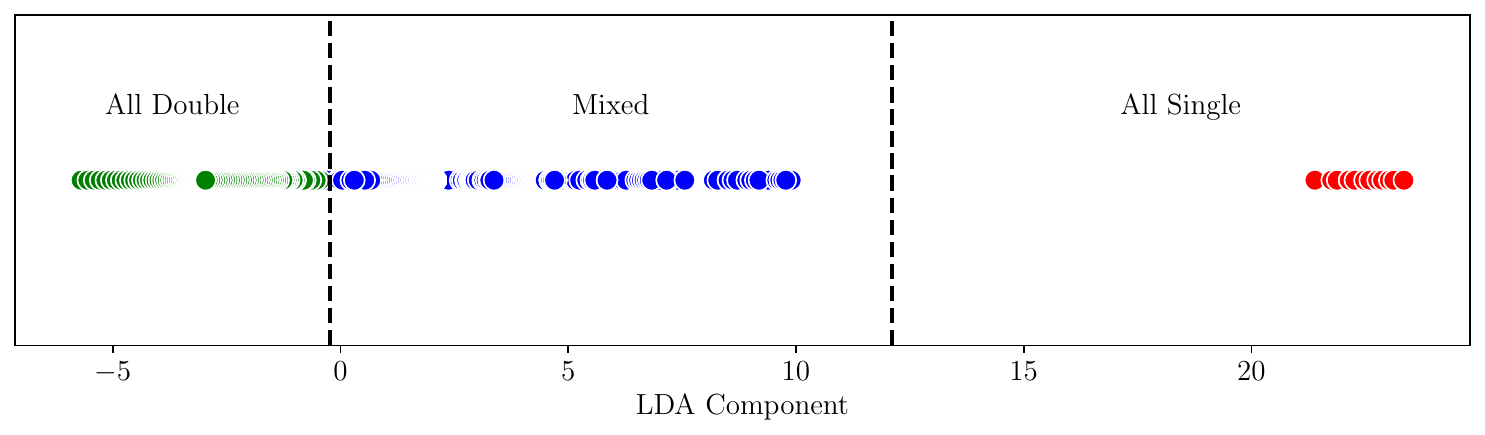}
	\caption{\textbf{LDA Reduced 1D Space} This figure demonstrates how Linear Discriminant Analysis (LDA) distinguishes classes by maximizing the ratio of between-class variance to within-class variance. As shown, three regions are well-separated, resulting in three distinct clusters. This indicates that, overall, the class space is governed by well-defined linear equations.}
	\label{fig:lda_separation}
\end{figure}

In this study, LDA was applied to reduce the dimensionality of the four dimensional space $(L_a, L_b, L_E(\text{bulk}), \Delta L_E)$ to the one dimensional projection shown in Figure~\ref{fig:lda_separation}. The figure illustrates that the classes form three well-separated clusters in the reduced 1D space. This clear separation indicates that the classes are governed by well-defined linear decision boundaries, validating LDA's assumptions. Additionally, the high classification accuracy achieved further supports the notion that the underlying classes are effectively modeled by linearly separable equations within the feature space.

\begin{figure}
	\centering
	\includegraphics[width=1\linewidth]{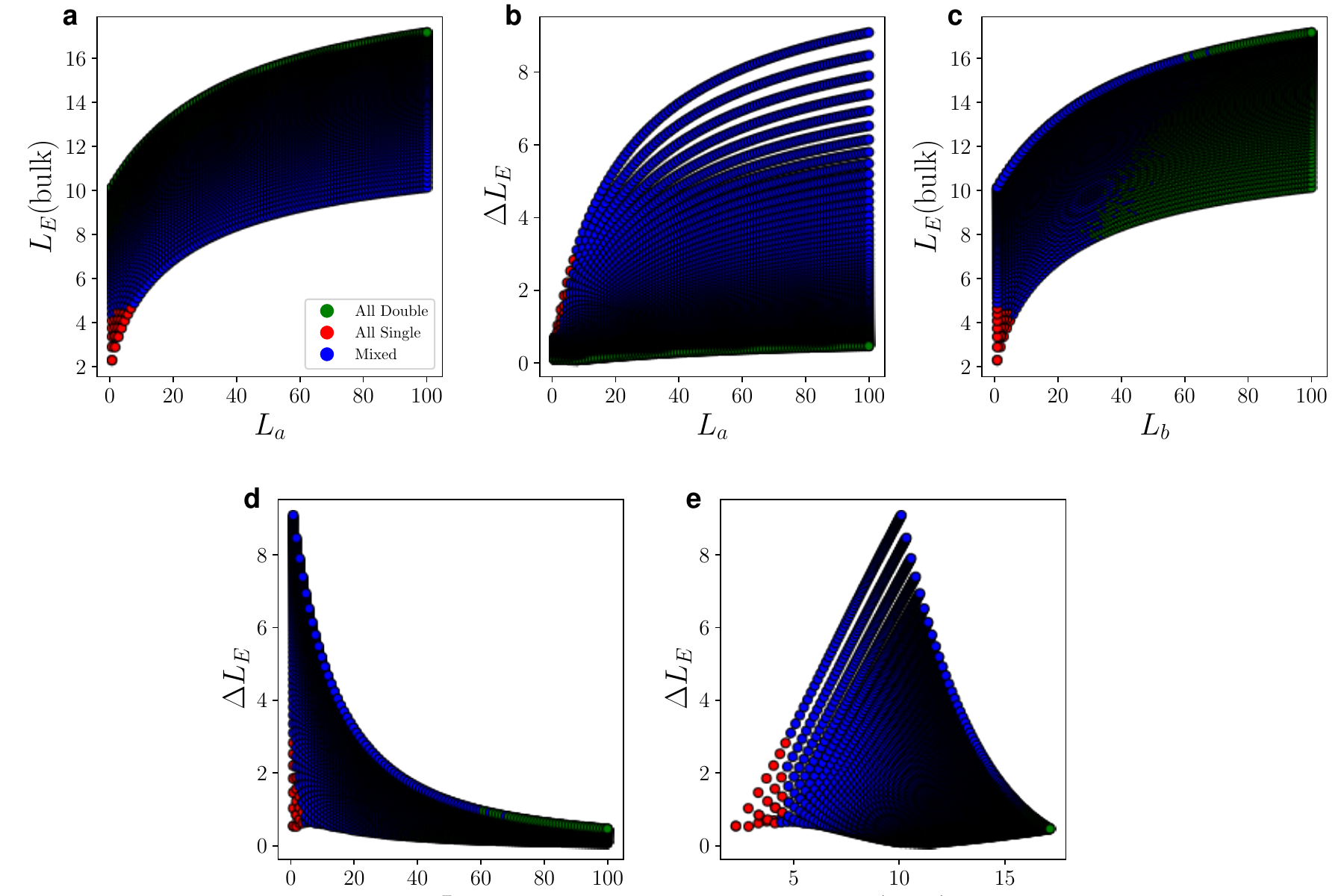}
	\caption{\textbf{Linearly Separable 2D Phase Space Projections.} Shown are two-dimensional slices of the $(L_a, L_b, L_E(\text{bulk}), \Delta L_E)$ phase space, where linearly separable boundaries emerge across distinct subspaces. \textbf{a} $(L_a, L_E(\text{bulk}))$ separates all double-scroll cases. \textbf{b} $(L_a, \Delta L_E)$ separates all double-single cases. \textbf{c} $(L_E(\text{bulk}), L_b)$ clearly separates all three classes. \textbf{d} $(\Delta L_E, L_b)$ separates all double cases. \textbf{e} $(\Delta L_E, L_E(\text{bulk}))$ distinctly separates all three scroll types. Each of these subspaces defines an equation that governs the phase space and different combinations can be selected.}
	\label{fig:svm_separation}
\end{figure}

By verifying the existence of linear separability, we can identify the governing linear equations of the phase space through the application of Support Vector Machines (SVM). Figure~\ref{fig:svm_separation} showcases how classes are separated in various two-dimensional projections of the phase space. Each subspace represents an equation governing the phase space and the whole phase space can be represented by combination of these conditions. It should also be noted that 3- and 4-dimensional slices exist with their own governing equations. However, due to the increased complexity with additional variables, we focus on high-accuracy two-dimensional equation combinations.

The presence of clear linear boundaries in these two-dimensional projections indicates that the classes are indeed linearly separable within these subspaces. Linear decision boundaries separating the classes are observed, demonstrating that these classes can be effectively modeled using different variables. The variables with the highest classification accuracy were selected for Equation~\eqref{ClassEquations}. To formalize this separability, we derive the governing linear equations using SVM, which define the hyperplanes that best distinguish between the classes. Although a two-dimensional space is sufficient for distinguishing classes in our study, the general form of the separating hyperplane in higher dimensions is expressed as:

\begin{equation}
	\label{ClassEquations}
	\mathbf{w}^\top \mathbf{x} + b = 0,
\end{equation}
where $\mathbf{w}$ is the weight vector perpendicular to the hyperplane, $\mathbf{x}$ is the feature vector, and $b$ is the bias term.

\section{Robust machine learning equations under realistic experimental constraints}
	In the previous sections, we presented the experimental conditions includes various uncertainties in real-world implementations. Here, we extend our SVM formulation to ensure robustness against these perturbations commonly encountered in experimental settings. Specifically, we investigate the use of adversarial training to enhance robustness in practical applications.

\begin{figure}
	\centering
	\includegraphics[width=1\textwidth]{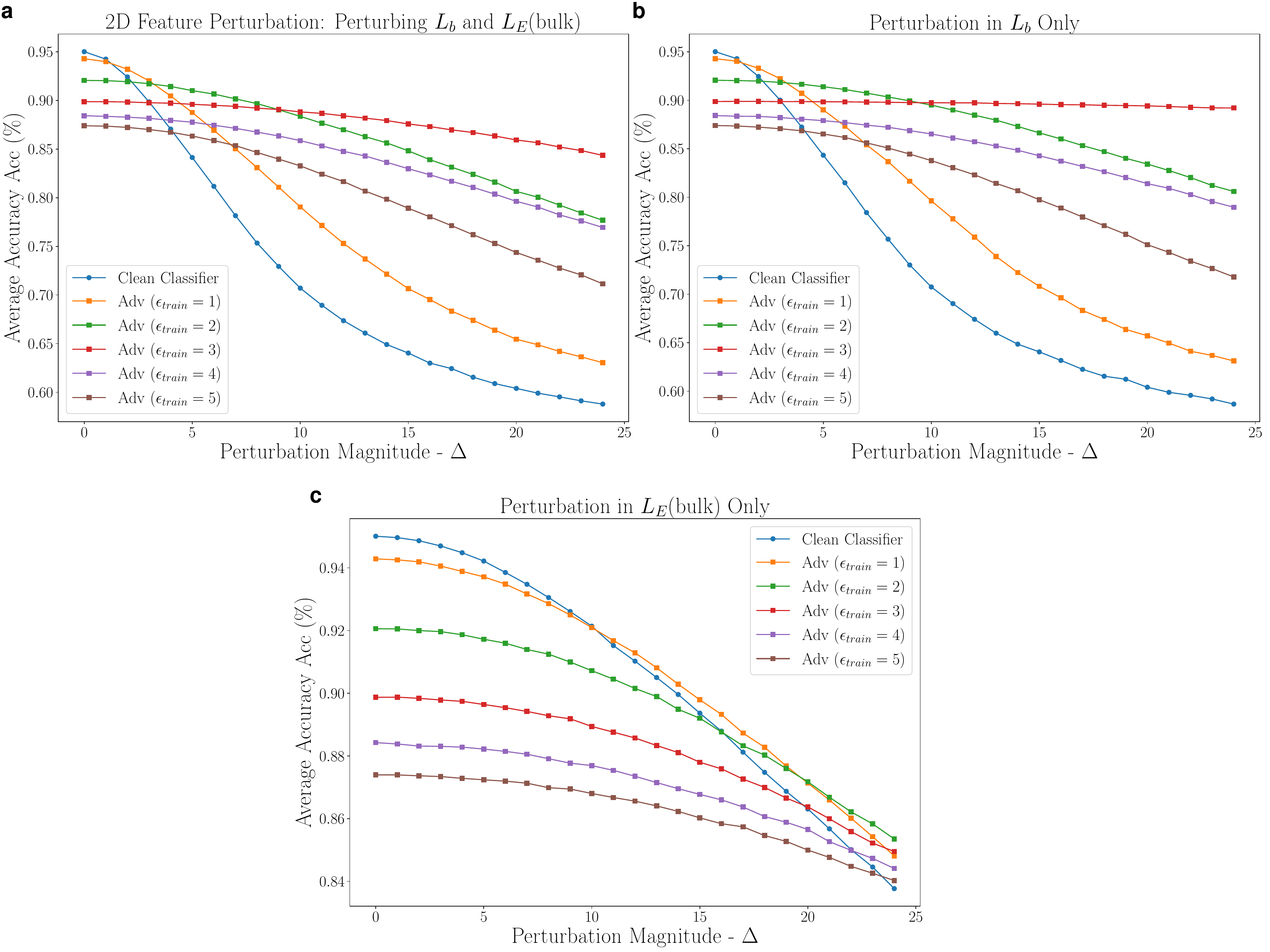}  
	\caption{\textbf{Robustness of adversarially trained support vector machines (SVMs) against perturbations.} 
			\textbf{a)} Average classification accuracy of SVMs trained with varying adversarial perturbation amplitudes, evaluated under simultaneous perturbations in both dimensions ($L_E(\text{bulk}), L_b$). Each curve corresponds to a different adversarial training regime, where during training, 25 perturbed versions of each original data point were generated using noise sampled independently from $\text{Uniform}(-\epsilon_{\text{train}}, \epsilon_{\text{train}})$, and the one yielding the highest SVM loss was selected. At test time, 50 new perturbations were applied to each data point in Figure~\ref{fig:AdversarialTraining}a of the main text, and accuracy was averaged across these realizations. Adversarially trained SVMs—such as the red curve trained with $\epsilon_{\text{train}} = 3$—maintain accuracy above 85\% even under test noise drawn from $\text{Uniform}(-\Delta=25, \Delta=25)$, indicating that adversarial training effectively enhances model robustness. 
			\textbf{b)} and \textbf{c)} Classification accuracy under perturbations applied individually to $L_b$ and $L_E(\text{bulk})$, respectively. These results demonstrate that aligning the adversarial training range with anticipated experimental uncertainties enables an optimal trade-off between accuracy and robustness, ensuring reliable classification in realistic, noisy environments.}
	\label{fig:AdversarialTraining}
\end{figure}

In the adversarial training scheme, SVMs were trained on perturbed samples. Each training configuration involved perturbations along specific axes—$L_E(\text{bulk})$, $L_b$, or both—and their robustness was quantified under external noise conditions, as shown in Figure~\ref{fig:AdversarialTraining}. Figure~\ref{fig:AdversarialTraining}a depicts the average classification accuracy as a function of simultaneous perturbations to both $L_E(\text{bulk})$ and $L_b$. While the horizontal axis denotes the perturbation magnitude ($\Delta$), the vertical axis shows the resulting accuracy. Each curve corresponds to a different adversarial training regime. For example, the red curve represents an SVM trained using perturbations sampled from $\text{Uniform}(-\epsilon_{\text{train}}, \epsilon_{\text{train}})$ with $\epsilon_{\text{train}} = 3$.

For each data point \(\mathbf{q}_i\) shown in Figure~4a, we perform the following steps during training:

\begin{enumerate}
	\item Generate a set of 25 perturbed samples:
		\[
		\mathbf{q}_{i,j} = \mathbf{q}_i + \boldsymbol{\epsilon}_{i,j}, \quad \boldsymbol{\epsilon}_{i,j} \sim \text{Uniform}\big(-\epsilon_{\text{train}}, \epsilon_{\text{train}}\big), \quad j = 1, 2, \dots, 25.
		\]
	
	\item Identify the adversarial sample (\(\hat{\mathbf{q}}_i\)) by selecting the perturbation that maximizes the SVM loss \( \ell_{\mathrm{SVM}} \) for the corresponding label \( y_i \):
		\[
		\hat{\mathbf{q}}_i = \underset{1 \leq j \leq 25}{\operatorname{arg\,max}} \; \ell_{\mathrm{SVM}}(\mathbf{q}_{i,j}, y_i)
		= \underset{1 \leq j \leq 25}{\operatorname{arg\,max}} \; \left\{\max\Bigl(0,\, 1 - y_i\, (\mathbf{w}^\top \mathbf{q}_{i,j} + b)\Bigr)\right\}.
		\]   
	
	\item Use \(\hat{\mathbf{q}}_i\) as the training sample for the original data point \(\mathbf{q}_i\) to enhance model robustness.
\end{enumerate}

During testing, for each original data point \(\mathbf{q}_i\), we sample 50 new perturbations:
	\[
	\mathbf{q}_{i,k}' = \mathbf{q}_i + \boldsymbol{\epsilon}_{i,k}', \quad \boldsymbol{\epsilon}_{i,k}' \sim \text{Uniform}\big(-\Delta, \Delta\big), \quad k = 1, 2, \dots, 50,
	\]
	where \(\Delta\) denotes the test perturbation magnitude. The final classification accuracy is computed as the average performance over 50 perturbed instances for each data point.

While accuracy generally decreases as perturbation strength increases, models trained with these worst-case scenarios demonstrate significantly greater resilience---indicating that noise-injected training enhances robustness without a substantial loss in performance. For example, the red curve illustrates that classification accuracy remains consistently high---between 90\% and 85\%---across a broad range of perturbations sampled from \(\text{Uniform}(-25,25)\), highlighting the model's robustness in the presence of large perturbations. Moreover, by aligning the training perturbation range with the expected experimental uncertainty in each individual dimension, classification performance can be optimized for specific conditions, leading to more reliable phase-profile determinations. For instance, although the red curve demonstrates the highest robustness over an extended perturbation range (\(\text{Uniform}(-25,25)\)), if the experimental uncertainty is known to be \(\text{Uniform}(-10,10)\), the green curve---trained with adversarial perturbations in the narrower range \(\text{Uniform}(-2,2)\)---emerges as the best choice.

Similarly, Figure~\ref{fig:AdversarialTraining}b and Figure~\ref{fig:AdversarialTraining}c illustrate how model performance responds to perturbations applied individually to \(L_b\) and \(L_E(\text{bulk})\), respectively. These results demonstrate that adversarial training can be effectively tailored to match the distinct noise characteristics of each dimension. For example, when \(L_E(\text{bulk})\) is relatively stable---perturbed by less than \(\text{Uniform}(-5,5)\)---the clean classifier (blue curve in Figure~\ref{fig:AdversarialTraining}c) already performs optimally, so no additional adversarial training is required in that dimension. In contrast, when \(L_b\) is subject to more substantial noise drawn from \(\text{Uniform}(-10,10)\), the most effective strategy is to apply adversarial training exclusively along the \(L_b\) dimension using perturbations from \(\text{Uniform}(-2,2)\), since the green curve in Figure~\ref{fig:AdversarialTraining}b (\(\epsilon_{\text{train}} = 2\)) achieves the highest accuracy in this regime. This analysis allows for the optimal choice of adversarial training regime when the uncertainties in each dimension are known. Lastly, a comparison between Figure~\ref{fig:AdversarialTraining}b and Figure~\ref{fig:AdversarialTraining}c reveals that perturbations in \(L_b\) degrade model performance more significantly than those in \(L_E(\text{bulk})\), highlighting the importance of dimension-specific training strategies.

In summary, adversarial training provides a systematic approach for enhancing model robustness to experimental uncertainties. This framework not only improves classification reliability but also enables the design of classifiers specifically optimized for the noise profiles encountered in different experimental settings.

\section{Evolution of Phase Boundaries}

\begin{figure}
	\centering
	\includegraphics[width=0.8\textwidth]{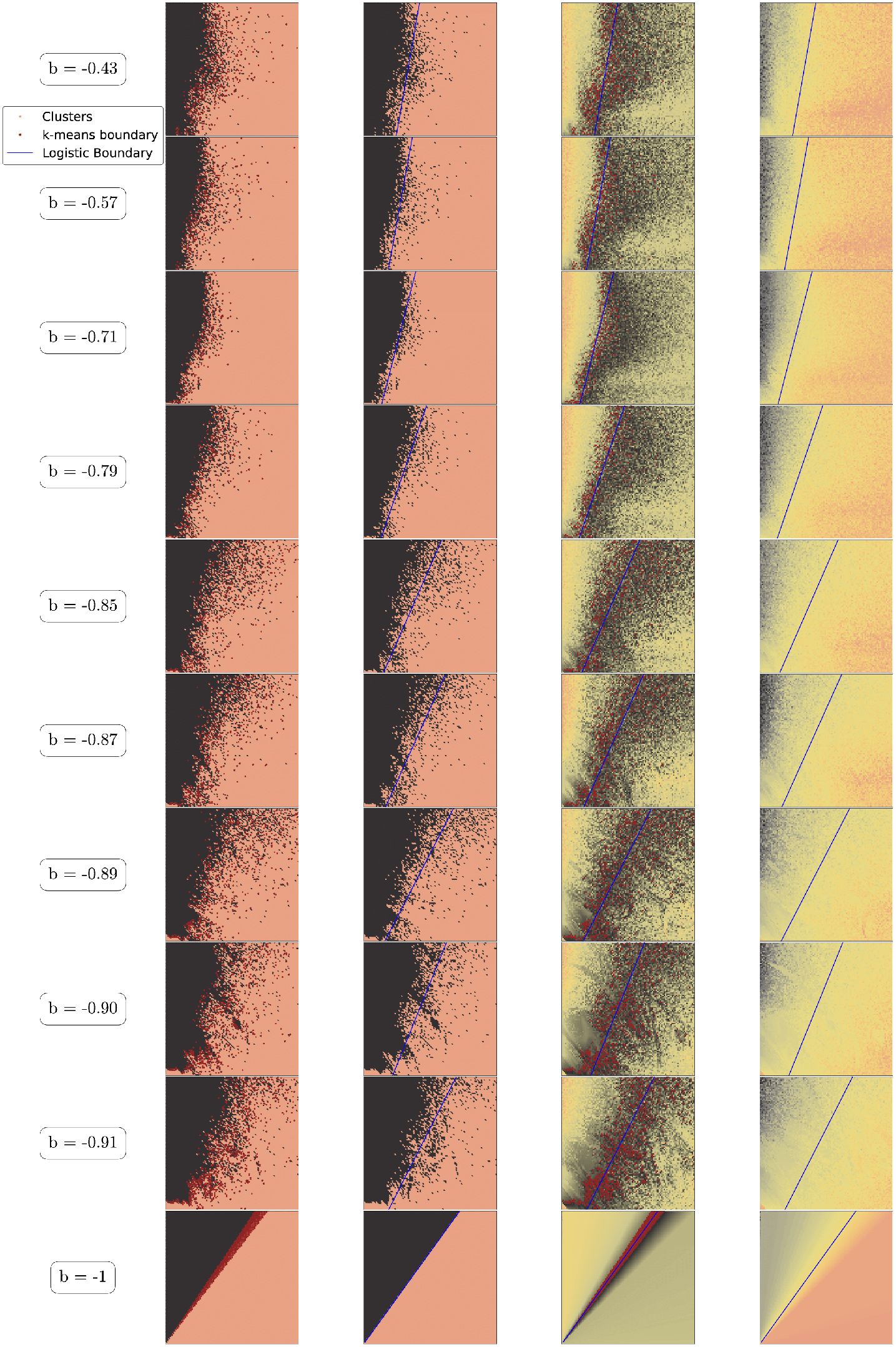}
	\caption{\textbf{Determining the phase boundaries.} This figure presents various processed representations of phase spaces for different \( b \) values. The leftmost column depicts the phase boundaries and clusters identified by k-means clustering. The second column shows the linear decision boundaries derived from logistic regression, marking the optimal separation between clusters. The third column overlays the resulting linear decision boundary and k-means boundary on the high-pass filtered phase space. Finally, the fourth column shows how the decision boundary maps onto the real phase space.}
	\label{fig:phase_spaces}
\end{figure}
The acquired linear boundaries provide a robust quantification of the evolving k-means clusters. Additionally, the consistent alignment between the high-pass filtered space, the k-means boundaries, and the linear decision boundaries further supports the validity of the applied methodology. The final column of Figure~\ref{fig:phase_spaces} highlights the real space evolution, corresponding to the results shown in Figure~14e.

\bibliography{TopoChaos_SuppMatAccepted}